\documentclass[onecolumn]{emulateapj}

\def\mo{M$_\odot$}

\def\kms{\,km\,s$^{-1}$}

\def\sles{\lower2pt\hbox{$\buildrel {\scriptstyle <}
   \over {\scriptstyle\sim}$}}

\def\sgreat{\lower2pt\hbox{$\buildrel {\scriptstyle >}
   \over {\scriptstyle\sim}$}}
\def\sharpnull#1{}

\begin{document}

\slugcomment{\bf }
\slugcomment{Accepted to Ap.J. April 23, 2007}

\title{Simulations of Magnetically-Driven Supernova and Hypernova Explosions 
in the Context of Rapid Rotation}

\author{A. Burrows\altaffilmark{1},  
L. Dessart\altaffilmark{1},
E. Livne\altaffilmark{2},
C.D. Ott\altaffilmark{1},
J. Murphy\altaffilmark{1}}
\altaffiltext{1}{Department of Astronomy and Steward Observatory, 
                 The University of Arizona, Tucson, AZ \ 85721;
                 burrows@as.arizona.edu,luc@as.arizona.edu,cott@as.arizona.edu,jmurphy@as.arizona.edu}
\altaffiltext{2}{Racah Institute of Physics, The Hebrew University,
Jerusalem, Israel; livne@phys.huji.ac.il}

\begin{abstract}

We present here the first 2D rotating, multi-group, radiation
magnetohydrodynamics (RMHD) simulations of supernova core 
collapse, bounce, and explosion.  In the context of rapid 
rotation, we focus on the dynamical effects of magnetic stresses 
and the creation and propagation of MHD jets. We find that a 
quasi-steady state can be quickly established after bounce, during which
a well-collimated MHD jet is maintained by continuous pumping of power from
the differentially rotating core.  If the initial spin period of the progenitor
core is $\sles$ 2 seconds, the free energy reservoir in 
the secularly evolving protoneutron star is adequate 
to power a supernova explosion, and may be enough for a hypernova.
The jets are well collimated by the infalling material and 
magnetic hoop stresses, and maintain a small opening angle.  
We see evidence of sausage instabilities in the emerging jet stream. 
Neutrino heating is sub-dominant in the rapidly rotating models we explore,
but can contribute 10$-$25\% to the final explosion energy.  
Our simulations suggest that even in the case of modest or slow rotation,
a supernova explosion might be followed by a secondary, weak MHD jet
explosion, which, because of its weakness may to date have gone unnoticed in supernova debris. 
Furthermore, we suggest that the generation of a non-relativistic 
MHD precursor jet during the early protoneutron star/supernova
phase is implicit in both the collapsar and ``millisecond magnetar" 
models of GRBs. The multi-D, multi-group, rapidly rotating RMHD 
simulations we describe here are a start along the path 
towards more realistic simulations of the possible role of magnetic fields
in some of Nature's most dramatic events.

\end{abstract}

\keywords{supernovae, multi-dimensional radiation magneto-hydrodynamics, MHD}

\section{Introduction}
\label{intro}

There is an emerging consensus that the core-collapse
supernova mechanism is essentially aspherical and that instabilities and the 
breaking of spherical symmetry are keys to explosion 
(Buras et al. 2006ab; Burrows et al. 2006,2007; Blondin, 
Mezzacappa, \& DeMarino 2003).  In modern simulations, due to
copious neutrino losses, the tamp of accretion, and
photodissociation, the direct hydrodynamic bounce mechanism 
always fails. The delayed neutrino-driven mechanism
works in spherical models only for a small subset of the lightest
massive star progenitors (e.g., $8.8$ M$_{\odot}$; Kitaura, Janka, \& Hillebrandt 2006). It 
works, if only marginally, in 2D simulations when aided by post-shock convection 
and overturn only for slightly more massive progenitors (e.g, 11.2 M$_{\odot}$;
Buras et al. 2006b).  The neutrino mechanism does work in the context 
of accretion-induced collapse (AIC; Dessart et al. 2006a),
but these systems must be rare.  In all cases explored to date, when neutrino heating is
the driving agency the explosion energies are not the canonical supernova values
near 10$^{51}$ ergs ($\equiv$ one Bethe), but half a dex or more smaller.  The low accretion densities
that enable explosion also inhibit the deposition of neutrino energy.  
However, it is indeed possible that when credible 3D simulations 
are available, we will find that a robust neutrino mechanism 
obtains for the majority of the relevant progenitor mass range 
($8 < M < 50$? M$_{\odot}$).  

Recently, Burrows et al. (2006,2007) have found that if the delay
to explosion is long, perhaps 0.5 to 1.0 seconds, the inner core can
be excited to oscillate in an $\ell = 1$ g-mode which damps by the anisotropic
emission of sound.  At this late stage, acoustic power can rival or exceed 
neutrino deposition power and can explode the envelope.  The excitation of core g-modes
in this manner is unavoidable.  The question is whether the pulsation amplitudes 
that can be achieved are adequate and whether the accretion/no-explosion 
phase lasts long enough to allow the core modes to excite to significance.
Given these desiderata, Burrows et al. (2007) witness the explosion
of all the progenitors they studied.  Whether the acoustic mechanism, or a hybrid neutrino/acoustic
mechanism, is relevant to supernova explosions remains to be seen, but it
has several intriguing features, including the production of environments
in which the r-process can occur and in which pulsar kicks are naturally imparted.

A third general class of explosion mechanisms, magnetohydrodynamic (MHD),
has a long history (LeBlanc \& Wilson 1970; Bisnovatyi-Kogan et al.
1976; Meier et al. 1976; Symbalisty 1984).  With the observation of collimated jets in the context
of gamma-ray bursts (GRBs) and hypernovae (Nomoto et al. 2005; 
Mazzali et al. 2006; Woosley \& Bloom 2006), the association between GRBs and a small
subset ($< 0.5$\%) of supernovae, and the detection of signatures of asphericity
in the generic supernova (Wang et al. 2002,2003), the MHD mechanism is 
experiencing a comeback (Ardeljan et al. 2000ab,2005; Wheeler et al. 2000; 
Wheeler, Meier, \& Wilson 2002; Akiyama et al. 2003; Yamada
\& Sawai 2004; Kotake et al. 2004; Mizuno et al. 2004; Takiwaki et al. 2004; Akiyama \& Wheeler 2005; 
Ohnishi, Kotake, \& Yamada 2005; Sawai, Kotake, \& Yamada 2005; Thompson, 
Quataert, \& Burrows 2005 (TQB); Proga 2005; Wilson, Mathews, \& Dalhed 2005; 
Moiseenko et al. 2006; Matt, Frank, \& Blackman 2006; Masada, Sano, \& Shibata 2007; 
Shibata et al. 2006).  In addition, the possible role of magnetic winds in 
powering secondary explosions and in
spinning down newly-born neutron stars or magnetars is being actively explored
(Thompson,~Chang,~\&~Quataert 2004; Bucciantini et al. 2006; Metzger, Thompson, \& Quataert 2006).
While modern models of whatever stripe of supernova explosion 
are very asymmetrical, MHD models are fundamentally so.

However, all credible MHD models require rapid 
rotation, which the current crop of solar-metallicity supernova progenitor 
models do not have (Hirschi,~Meynet,~\&~Maeder 2004,2005; Heger et al. 2000; 
Heger, Woosley, \& Spruit 2005).  Pre-collapse spin periods of a few seconds
may be needed, but Heger, Woosley, \& Spruit (2005), for example, derive
spin rates at collapse ten times slower for the cores of solar-metallicity 
stars. The rotational kinetic energies at bounce are consequently 
one hundred times smaller than required and are totally inadequate to power 
a supernova explosion. However, are these progenitor calculations correct?  Could
the progenitor cores of massive stars before collapse actually be 
rotating fast?  More conservatively, is there a subset of supernova progenitors
that are spinning fast enough to make MHD processes relevant on explosion timescales?
Importantly, the low-metallicity models of Woosley \& Heger (2006), 
some of which they suggest are GRB progenitors, have much faster spinning 
cores at collapse.

In this paper, we explore the physics and phenomenology of MHD-driven explosions 
when and if suitably rapid initial rotation rates obtain and leave open the
question of the progenitor systems in which they might.  Unlike all past investigations,
we incorporate 2D multi-group neutrino radiation transport, a realistic equation of state,
and realistic progenitor density, temperature, and composition structures. Our current 
supernova code, VULCAN/2D, has been upgraded to include MHD in 2.5D (Livne et al. 2007).
We start at the onset of core collapse and follow the evolution from collapse, 
through bounce, to explosion, and all the way to many hundreds of milliseconds
after explosion.  We observe and document bipolar jet explosions due to
magnetic stresses and buoyancy in magnetic towers (Uzdensky \& MacFadyen 2006ab),
accretion and magnetic collimation of the jets, sausage/neck instabilities (Lampe 1991), and the 
growth to quasi-steady state of the MHD engine power.  The jets are powered and 
maintained by the kinetic energy in differential rotation, which for these calculations is 
often significantly more than one Bethe. For the strongest jets, we observe
core spindown by magnetic torques. 

Akiyama et al. (2003) introduced into supernova theory the possibility that 
whatever the initial seed magnetic fields, the magneto-rotational instability (MRI; 
Balbus \& Hawley 1991; Pessah \& Psaltis 2005; Pessah, Chan \& Psaltis 2006),
driven by the negative gradients in the angular velocity ($\Omega$) after bounce that are
unavoidable in the context of core collapse (Ott et al. 2006), could foster the growth
to saturation values of poloidal ($B_{P}$) and toroidal ($B_{\phi}$) 
fields that are dynamically important.  The recent 
calculations of Obergaulinger, Aloy, \& M\"uller (2006), 
Obergaulinger et al. (2006), Etienne, Liu, \& Shapiro (2006),
and Shibata et al. (2006) make a good case that this is so.  Hence, if the shear
in $\Omega$ in the protoneutron star (PNS; Burrows \& Lattimer 1986) between $\sim$10 and $\sim$100 kilometers
is large, we conclude that strong, if transient, magnetic fields with a
dynamical role are inevitable in core collapse.  In all situations with 
rapid rotation, the growth of B-fields due to compression, 
winding, and the MRI (and, perhaps, due to classical dynamo action) will play an essential 
role in the explosion mechanism. This paper is a preliminary exploration 
of this scenario and its consequences. 

However, our simulations have a number of limitations.  The calculations are Newtonian, are done in 2D
(axisymmetric, plus rotation), and the transport is flux-limited (not multi-angle).
Importantly, Etienne, Liu, \& Shapiro (2006) and Shibata et al. (2006) have shown
that 2D spatial grids with thousands-squared of points are 
required to properly resolve the MRI in core-collapse simulations.
Moreover, Obergaulinger, Aloy, \& M\"uller (2006) and Obergaulinger et al. (2006)
have documented the dependence of the MRI growth rates on the initial, unknown, magnetic 
fields.  Since we can't yet afford both high spatial resolution and multi-group neutrino 
transport, we have opted for the latter, since no calculation performed to date has 
had this central feature.  Given our choice, we are still able to perform evolutions
for not only tens of milliseconds, but for $\sim$1 second and to witness the full
panoply of dynamical phases.

In \S\ref{general}, we discuss the overall physical context, motivate our
approach to the simulations portrayed, and review
the simple physics of flux compression and winding, as well as the basics of the MRI.
We also describe many of the relevant timescales of the problem, all of which we
argue are short compared with the time available in the explosion phase.  This
motivates the idea that a quasi-steady state engine is established and maintained.
Furthermore, we present qualitative discussions of the anticipated dependence of the 
jet power on spin rate and of the free energy from differential motion available 
to power the MHD engine.  The latter is continuously fed by equatorial accretion 
during the early phases of bipolar explosion. In \S\ref{setup}, we present
our computational setup and the parameters of the simulations we 
describe.  In \S\ref{results}, we go on to describe and detail our simulation
results and the phenomenology and morphology of the resultant blasts.
We discuss the magnetic tower phenomenon, the poloidal and toroidal field 
distributions, the emergence of the jet and its collimation, the weak sausage instability, 
and the power density distributions.  In \S\ref{period}, we discuss the magnitude and evolution
of the spin rates and the free energy of differential rotation.  We also explore
the initial spin-rate dependence of the latter.  Section \ref{energy} presents
estimates of the derived jet powers and explosion energies 
and we conclude in \S\ref{conclusion} with both a summary of 
our central results and speculations on their wider astrophysical import.

\section{The Overall Physical Context}
\label{general}

A progenitor core will have initial rotational and magnetic field 
profiles. Magnetic torques and wind mass loss during the pre-collapse 
evolution of the star will help determine both these distributions. 
Yet despite recent detailed efforts to model supernova progenitors 
(e.g., Heger, Woosley, \&  Spruit 2005; Hirschi,~Meynet,~\&~Maeder 2004,2005), 
we still have no reliable constraints on the spin or magnetic structures of the 
white-dwarf-like core that collapses when achieving the Chandrasekhar mass.
For instance, the predictions for the spin period, $P_0$, of the Chandrasekhar core 
vary from $\sim$1.5 seconds to $\sim$50 seconds.  This involves a range in the possible 
initial rotational kinetic energies of $\sim$$10^3$.  There is a similarly wide range of
predictions for the magnitudes of the pre-collapse toroidal and poloidal
fields and there are no predictions for 3D structures that are consistent
with Maxwell's equations.  

However, if the initial spin rate and angular momentum 
are large, and the MRI (or dynamo action in 3D) occurs in 
the protoneutron star after bounce, then the resulting magnetic
energy will be expected to exponentiate to saturation values that are in rough 
equipartition with the rotational free energy in the post-bounce differential motion.
The growth of B-field energy due to the MRI instability feeds off
the free energy available in differential rotation, which is the difference
between the rotational kinetic energy and the kinetic energy for solid-body
rotation at the same total angular momentum.
For example, Shibata et al. (2006) find that the MRI saturates in the core-collapse 
context at magnetic energies that are very roughly 10\% of the
rotational energy. The degree to which the free energy is close to the
total rotational energy itself depends upon the magnitude and distribution
of the angular-velocity and density gradients.  In particular, if the post-bounce 
core, which may contain a lot of the rotational energy, is in near solid-body 
rotation, and most of the shear is in an outer fraction of the mass (albeit with 
larger moment arms), then a smaller fraction of the total rotational energy will
be available to pump up the magnetic energy by the MRI.  This fraction could
be only 10\%, but might be 10\%$-$50\%.  For instance, the work of Ott et al. (2006) 
suggests that progenitor cores with a fast $P_0$ of 2.34 seconds, can have rotational
free energies at $\sim$200 milliseconds (ms) after bounce that range from $\sim$2 to $\sim$12
Bethes, depending upon progenitor mass and model.  Progenitor cores with slower $P_0$s
of $\sim$10 seconds can have free energies (at the same epoch) 
that range from $\sim$0.1 to $\sim$0.2 Bethes, while those with 
a slow $P_0$ of $\sim$50 seconds can have rotational free energies of
only $\sim$0.03 Bethes.   Hence, depending upon the initial spin profile, the free energy
available in differential rotation to be tapped to generate strong magnetic
fields and MHD jets varies from insignificance to supernova and hypernova magnitudes. 

In this paper, we explore the consequences for supernova theory of rapid initial
rotation that should lead to strong magnetic fields and MHD-jet-driven explosions.
We envision initial $P_0$s less than $\sim$4 seconds, but focus for specificity
on a $P_0$ of 2 seconds.  Such rapid spin rates may or may not obtain, or may obtain 
rarely and/or only in the hypernova context, but this is not our concern here. 
Since with our modest number of grid points (\S\ref{setup}) we can not fully resolve the MRI 
(see \S\ref{intro}), we hypothesize (and calculate) that the MRI growth 
rates are sufficiently fast that within 100-200 milliseconds
after bounce the resulting MHD jet structure will stabilize on longer-term 
secular evolution times, such as the accretion and neutrino cooling times 
(0.5-10 seconds), and will lose track of the initial B-field configuration.
In this way, to a reasonable approximation, the structure, hydrodynamic flow,
rotational profiles, and jet power will settle into a long-term evolution
that depends most centrally only on the initial spin and thermodynamic 
profiles, and not on the initial B-fields (at least, not very much). Such should be  
a consequence of quick exponential growth to saturation.  This ansatz motivates our choices 
of the initial poloidal and toroidal fields (\S\ref{setup}); these are
set to lead by flux compression and rotational field winding to 
final B-fields whose magnitudes roughly
approximate those expected due to the MRI at and around saturation. 
For a $P_0$ of 2 seconds, the resulting toroidal fields between 
$\sim$10 kilometers (km) and $\sim$100 km turn out to be very
roughly $10^{15\pm0.5}$ Gauss (G).  In \S\ref{saturation}, we 
provide a very approximate analytic connection between period and saturation B-field.  

Since flux freezing, field winding on infall, field winding after infall,
and amplification due to the magneto-rotational instability (MRI), of
both poloidal and toroidal field components, are of interest, we 
next provide short discussions of each in turn, with a focus on the relevant
timescales and approximate magnitudes.  None of this is original, 
but we include it for general context.  Note that the various 
timescales all scale with the rotational period.
Hence, in the case of rapid rotation, all the relevant processes  
are accelerated and should lead to a jet-driven explosion that 
achieves a quasi-steady state on a ``short" timescale. 
After the asides in \S\ref{amp}, \S\ref{mri}, 
and \S\ref{saturation} on order-of-magnitude estimates, 
we describe in \S\ref{setup} our computational 
and initial model setup, and then transition in \S\ref{results} 
to a detailed discussion of our simulation results.

\subsection{Field Amplification and Timescales}
\label{amp}

Rotational winding converts/stretches poloidal field into toroidal field.
The basic equation:
\begin{equation}
\frac{\partial B_{\phi}}{\partial{t}} \sim B_P\left(\frac{\partial\Omega}{\partial\ln{r}}\right)
\end{equation}

leads to

\begin{equation}
\Delta{B_{\phi}}  \sim B_P\Omega t \\
 \sim B_P\left(\frac{2\pi t}{P}\right) \, .
\label{wind_after}
\end{equation}
In eq. (\ref{wind_after}), $t$ is the time, $P$ is the local spin period, 
and $\Omega$ is the corresponding angular frequency.
Eq. (\ref{wind_after}) suggests that each turn leads to a boost in $B_{\phi}$ near
$2\pi$ and that growth is linear in $t$.

On infall, winding alone yields
\begin{equation}
\frac{\Delta{B_{\phi}}}{B_P} \sim \Omega_0\int \left(\frac{r_0}{r}\right)^2 dt \\
 \sim 3\Omega_0 t_0 \left(\frac{r_0}{r}\right)^{1/2}\\
 \sim 60\pi\left(\frac{t_0}{P_0}\right) \, ,
\label{wind}
\end{equation}
where we have used $r^{3/2} \sim r_0^{3/2}(t_0 - t)/t_0$ and angular momentum
conservation: $\Omega = \Omega_0 (\frac{r_0}{r})^2$.  $t_0$ is the time to
bounce (``collapse time"), $r_0$ is the initial radius of the collapsing shell, and $P_0$
is the initial rotation period. With $t_0 = 0.2$ seconds, $P_0 = 20$ seconds,
$r_0 \sim 2000$ km, and $r \sim$ 20 km,
$\frac{\Delta{B_{\phi}}}{B_P} \sim$ 2, with a reasonable range between 1 and 10.    
However, if $P_0 = 2$ seconds, this ratio is ten times larger.

Core collapse leads to matter compression. Magnetic flux conservation would
then amplify both the toroidal and the poloidal fields.
The magnitude of the compression is $\sim\!(\frac{\rho}{\rho_0})^{2/3}$.
With $\rho \sim 3\times 10^{14}$ g cm$^{-3}$ and $\rho_0 \sim 7\times 10^{9}$ g cm$^{-3}$,
$(\frac{\rho}{\rho_0})^{2/3} \sim 10^3$.  If we combine the prescription for winding
on infall with flux compression, assuming that the collapse is homologous
($\frac{\rho}{\rho_0} \sim (\frac{r_0}{r})^3$), we find
\begin{equation}
\frac{\Delta{B_{\phi}}}{B_P}
\sim \frac{3}{5}\Omega_0 t_0 \left(\frac{r_0}{r}\right)^{5/2}\, .
\label{wind2}
\end{equation}
This expression is within a factor of two of the product of flux compression and
winding (eq. \ref{wind}) on infall taken separately.  Given the parameters above, flux compression
yields a factor of $\sim$$10^3$ and winding a factor of a few (again, for an initial period
of $\sim 10-20$ seconds).  Note that the winding amplifies only the toroidal component, while flux compression
amplifies both $B_{\phi}$ and $B_P$.  Flux compression alone takes a toroidal field
from $\sim 10^{10}$ G to $\sim 10^{13}$ G and a poloidal field from
$\sim 10^{7}$ G ($B_P^0$) to $\sim 10^{10}$ G ($B_P$). Using eq. (\ref{wind_after}), and starting with
a $B_P$ of $10^{10}$ G, we can derive the time to achieve a given change in the toroidal
field due solely to flux winding:
\begin{equation}
t_{\rm wind} \sim 32\ {\rm ms} \left(\frac{P}{2\ {\rm ms}}\right)
\left(\frac{\Delta B_{\phi}}{10^{12}\ {\rm G}}\right)\left(\frac{10^{10}\ {\rm G}}{B_P}\right)
\label{wind3}
\end{equation}
If the initial fields are not very large, winding by itself, and not aided by the MRI or
dynamo action in 3D, can require a long time to achieve dynamically interesting
field strengths ($\sim$10$^{15}$ G).  It is, nevertheless, a very important process
in rotational supernovae, particularly if the MRI or a dynamo can deliver after bounce 
and on exponential timescales high values of the poloidal field.

\subsection{Amplification by Magneto-Rotational Instability}
\label{mri}

In local linear mode analysis, one can derive a characteristic dispersion relation
that depends on the rotational shear ($\frac{\partial\Omega}{\partial\ln{r}}$),
the magnitude of the B-field, and the Brunt-V\"ais\"al\"a frequency 
(Balbus \& Hawley 1991; Pessah \& Psaltis 2005; Akiyama et al. 2003).
Ignoring the Brunt-V\"ais\"al\"a frequency, the maximum 
rate of MRI modal amplification occurs for a given wavenumber, $k_{mri} \sim \Omega/v_A$,
where $v_A$ is the Alfv\'{e}n speed, and has an associated growth time ($\tau_{mri}$) of
\begin{equation}
\tau_{mri}  \sim 4\pi\left(\frac{\partial\ln{r}}{\partial\Omega}\right)\\
 \sim 2 P\, ,
\label{mri_tau}
\end{equation}
where $P$ is the spin period at and after bounce.  
Note that $\tau_{mri}$ is proportional to
$P$, and that if $P$ $\sim$ 2 ms, $\tau_{mri}$ is also on millisecond timescales.

The Alfv\'{e}n speed is given by
\begin{equation}
v_A  = \frac{B}{\sqrt{4\pi\rho}} \\
 \sim 10^9\ {\rm cm\ s^{-1}}\ \frac{B_{15}}{\rho_{11}^{1/2}}  \, ,
\label{alfven}
\end{equation}
where $\rho_{11}$ is the mass density in units of $10^{11}$ g cm$^{-3}$
and $B_{15}$ is the field strength in units of $10^{15}$ G.
With eq. (\ref{alfven}), we see that the physical scale of the mode with the 
largest MRI growth rate is approximately:
\begin{equation}
\lambda_{mri}^{max} \sim \frac{2\pi v_A}{\Omega} \\
 \sim v_A P \\
 \sim 10^4\ {\rm cm}\ P_{10}\frac{B_{12}}{\rho_{11}^{1/2}}   \, ,
\label{lambda_max}
\end{equation}
where $P_{10}$ is the rotation period in units of 10 milliseconds.
Eq. (\ref{lambda_max}) shows that for ``small" initial seed fields, 
$\lambda_{mri}^{max}$ can, in a computational sense,  be quite small.

Etienne et al. (2006) note that the 
characteristic time for tapping differential rotation, 
transferring angular momentum by magnetic torques,
and saturating the MRI is the Alfv\'{e}n crossing time:
\begin{equation}
t_A  = \frac{R}{v_A}\\
 \sim 10\ {\rm ms}\ \frac{R_{7}}{B_{15}}\rho_{11}^{1/2}  \, ,
\label{time_sat}
\end{equation}
where $R_7$ is the radius in units of $10^7$ cm.
Hence, when high fields are achieved, $t_A$ is quite short in evolutionary terms.

Finally, using eq. (\ref{mri_tau}), exponential growth of the MRI yields
the time, $t_{mri}$, it takes to achieve a field of $B_f$, starting from a field
of $B_i$, of
\begin{equation}
t_{mri}  \sim 20\ {\rm ms} \left({P}_{10}\right)\ln{\frac{B_f}{B_i}}  \\
 \sim \left[92 + 20 \ln{\left(\frac{B_f}{10^{15}\ {\rm G}}\right)\left(\frac{10^{13}
{\rm G}}{B_i}\right) }\right] \left({P}_{10}\right)\ {\rm ms}\, .
\label{mri_time}
\end{equation}
Eq. (\ref{mri_time}) suggests that rapid spin rates can deliver fields of $\sim$$10^{15}$ G
within 100-200 milliseconds, with a weakish dependence on $B_i$.

The approximate formulae from eq. (\ref{mri_tau}) to eq. (\ref{mri_time}) suggest that all the relevant
timescales scale with $P$, that they can be quite short, and that if the MRI obtains
it can generate dynamically interesting fields within 100-200 milliseconds. Furthermore, 
eq. (\ref{wind3}) demonstrates that subsequent rotational winding 
can act to maintain high fields as long as the rotational energy of the core is maintained.

\begin{figure}
  \plottwo{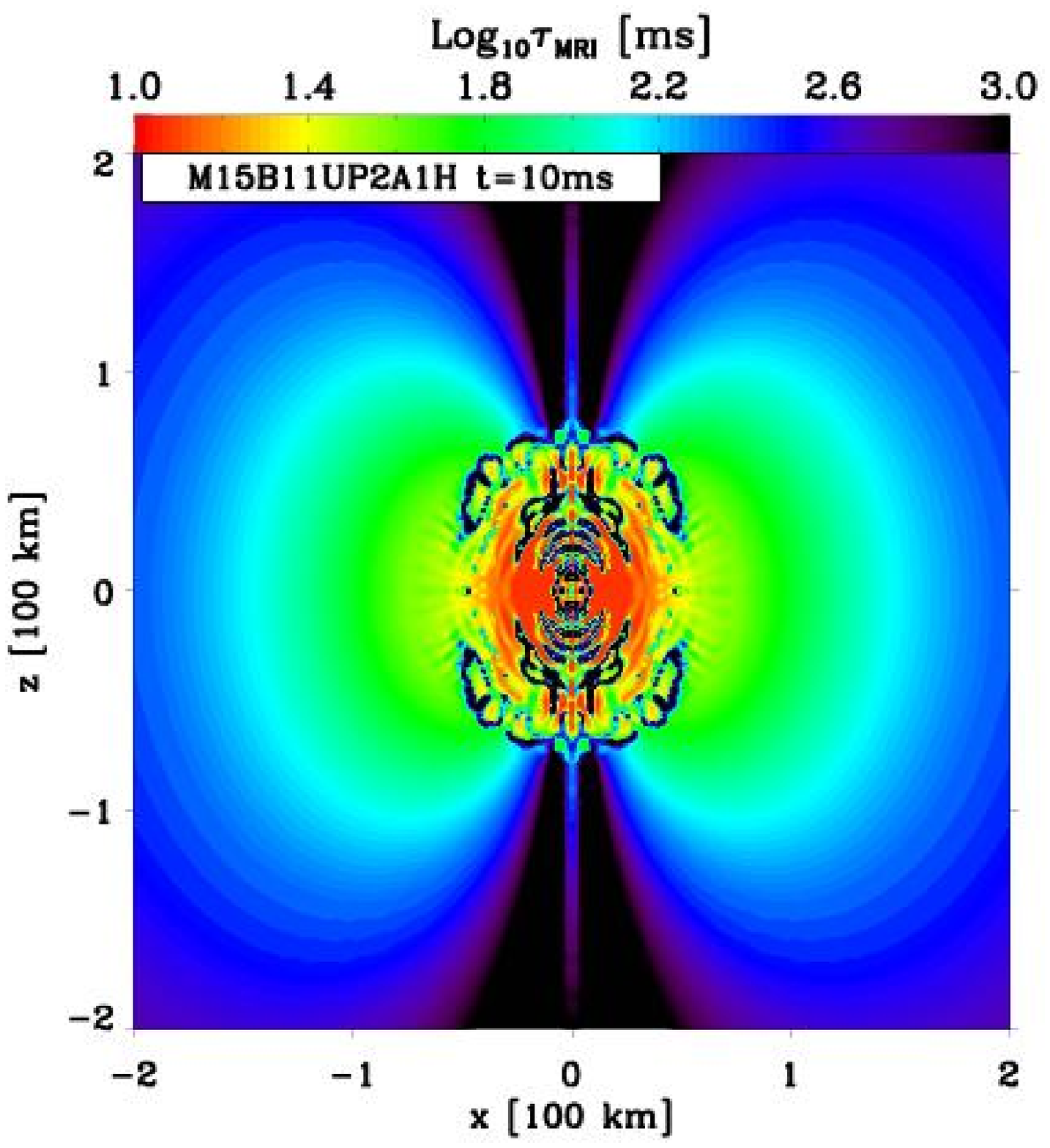}{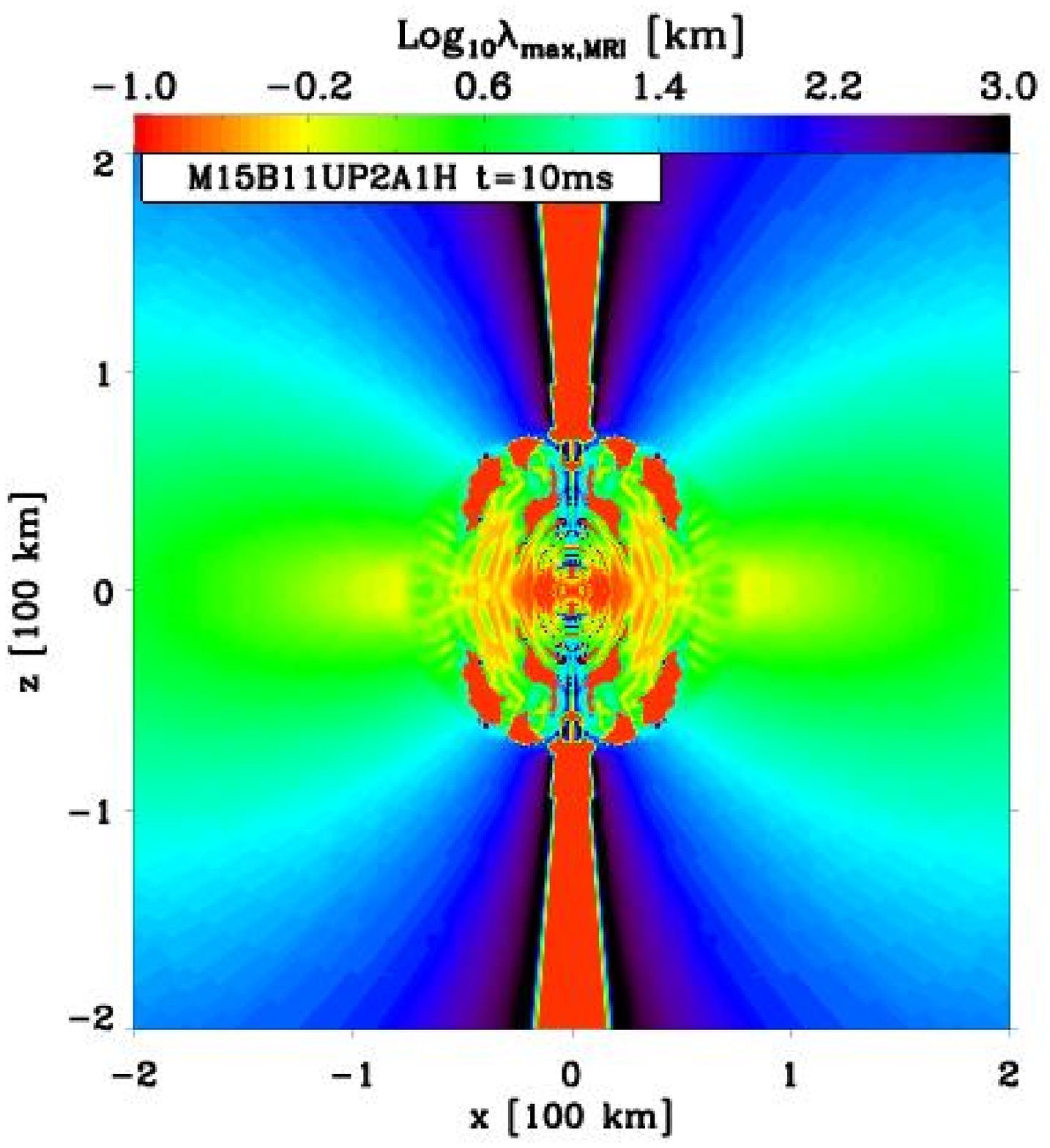}
  \caption{Colormap of the growth times (left) and spatial scales (right)
for the most unstable mode associated with the magnetorotational instability,
obtained using the perturbation formalism of Pessah \& Psaltis (2005) at every
interior location in our simulation for model M15B11UP2A1H. 
(See text for a discussion.)}
  \label{fig_mri}
\end{figure}

We end this subsection by providing in Fig. \ref{fig_mri} representative colormaps of the peak 
linear MRI growth timescale, $\tau_{mri}$ (left-hand side), and the associated wavelength, 
$\lambda_{mri}^{max}$ (right-hand side).  These numbers were obtained using the output 
of our 2D radiation/hydro simulation, M15B11UP2A1H (see \S\ref{setup}), 10 milliseconds 
after bounce and the formalism of Pessah \& Psaltis (2005).  Figure \ref{fig_mri} 
shows that in the inner 50 km, $\tau_{mri}$ in our calculations is only 10-30 ms, 
but that $\lambda_{mri}^{max}$ is 0.1 to 0.5 km. In these regions, our spatial 
resolution is on the order of 0.5 km, and since Etienne et al. (2006) have 
shown that $\sim$10 resolution elements over $\lambda_{mri}^{max}$ are needed to fully 
resolve the growth of the MRI, our calculations fall short in this regard.  Were we to
attempt the resolution required, we would need to increase the total number of zones 
employed in our calculations by more than an order of magnitude, rendering our 2D, multi-group radiation
hydrodynamic calculations extremely expensive and time-consuming.  Worse still,
model M15B11UP2A1H incorporates a large initial poloidal field, which yields an overlarge
$\lambda_{mri}^{max}$.  Had we used an even smaller, more realistic, initial $B_{\phi}$,
$\lambda_{mri}^{max}$ would have been correspondingly smaller (eq. \ref{lambda_max}), putting 
its resolution even further out of reach for our simulations with multi-group radiation that are evolved
for many hundreds of milliseconds past bounce. Since each simulation presented in this paper already
consumes about a month on modern supercomputers, one can see why we have opted for
our current strategy.  That strategy involves using high initial poloidal fields ($B_{P}$)
that nevertheless result in toroidal and poloidal fields after bounce that are
of similar magnitude to what could be expected given the saturation of a fully functioning MRI.
We did calculations for a spectrum of initial conditions, but highlight only models
M15B11UP2A1H and M15B11DP2A1H which very approximately result in the requisite fields.  
We now turn to back-of-the-envelope estimates of those saturation fields as a function of
$\Omega$.

\subsection{Expected Systematics with $\Omega$ of Saturation Fields and Jet Power}
\label{saturation}

At saturation, the energy in the magnetic field in the post-bounce region experiencing
significant rotational shear should scale with the kinetic energy in that differential
motion (the ``free energy" of rotation).  The inner core, with a radius of $\sim$10 km, 
is generally in solid-body rotation, and, hence, is not a factor in generating 
new field by the MRI (Akiyama et al. 2003).  The region of interest is between $\sim$10
km and $\sim$100 km.  Following the arguments of Shibata et al. (2006) and Hawley, Gammie, \& Balbus (1996),
to determine saturation fields we can set either a fraction ($\varepsilon \sim 0.1$) of the 
energy density in shear motion equal to the magnetic energy density, or set
the square of the Alfv\'{e}n speed equal to that same fraction 
of $(r\Omega)^2$, where $r$ is the cylindrical radius.
Approximately, this yields:
\begin{equation}
B  \sim \sqrt{4\pi\varepsilon\rho r^2 \Omega^2} \\
 \sim 10^{15}\ {\rm G}\ \left(\frac{\varepsilon}{0.1}\right)^{1/2} 
\left({\rho_{11} r_{30}^2}\right)^{1/2} \left(\frac{\Omega}{10^3\ {\rm rad}\ {\rm s}^{-1}}\right) \, ,
\label{satur}
\end{equation}
where $r_{30}$ is the radius in units of 30 kilometers.  Of course, the fields generated
will have spatial distributions, as do $\rho$ and $\Omega$, and $\Omega$ generally decreases
with $r$.  Eq. (\ref{satur}) merely hints at the scalings and 
magnitudes expected.  Importantly, it shows that $B$ is expected to be linear
with $\Omega$ and that $\Omega$s near $\sim$10$^3$ translate into fields near 10$^{15}$ G.
Ott et al. (2006) have shown that the ratio of initial spin period to spin period at and after bounce
is roughly $\sim$10$^3$.  So an $\Omega$ of 10$^3$ 
${\rm rad}\ {\rm s}^{-1}$ after bounce implies a $P_0$ near $\sim$5 seconds.
Therefore, for such initial progenitor periods, saturation fields of around $\sim$10$^{15}$ G 
at radii of tens of kilometers are reasonable.

The spindown torque due to a magnetic jet should be proportional to $B_{\phi}\times B_{P}$. 
Since field winding will ensure that $B_{\phi} > B_{P}$, the torque is proportional 
to $\frac{B_{P}}{B_{\phi}}$ times the total magnetic energy in the MRI/shear region.
As eq. (\ref{satur}) implies, the latter is proportional to $\Omega^2$ and since the jet
power ($\dot{E}$) is approximately equal to $\Omega$ times the torque, $\dot{E}$ 
is proportional to $\Omega^3$ (Shibata et al. 2006).  This is a very useful result.
More quantitatively,  
\begin{equation}
\dot{E}  \sim \varepsilon \left(\frac{B_{P}}{B_{\phi}}\right) F_{rot} \Omega \\
 \sim 10^{52}\ {\rm ergs\ s^{-1}}\ \left(\frac{\Omega}{10^3\ {\rm rad}\ {\rm s}^{-1}}\right)^3 \, ,
\label{powero}
\end{equation}
where $F_{rot}$ is the free energy in differential rotation and we have set $\frac{B_{P}}{B_{\phi}}$
equal to 0.1 (Shibata et al. 2006; Etienne et al. 2006).  Since the outer post-shock region 
is rotating more slowly than the inner region, if $\Omega$ equals 
$10^3$ ${\rm rad}\ {\rm s}^{-1}$ at $\sim$10-30 kilometers,
using 10$^3$ in eq. (\ref{powero}) is likely to be a slight overestimate.  Nevertheless,
jet powers of 10$^{51-52}$ ergs\ s$^{-1}$ for $\Omega$s near 10$^3$ ${\rm rad}\ {\rm s}^{-1}$ are sensible.  Such
powers will pump supernova-like energies into the ejecta within 0.1 to 1.0 seconds.
This conclusion is the essence of our study and our simulations bear it out.

It should be noted that Thompson, Quataert, \& Burrows (2005), using 
the analogy of $\alpha$-disk accretion, have explored the local heating
rate due to field dissipation in a region experiencing the MRI and derive
a $\Omega^3$ dependence for the associated heating power. In fact, eq. (\ref{powero}),
representing large-scale power and the TQB formula, representing the associated small-scale 
dissipational heating, yield numbers that are within an order of magnitude of one another.  
So, our modest-resolution simulations may be underestimating the driving power of a saturated MRI.

Finally, it is instructive to compare eq. (\ref{powero}) with the corresponding 
formula for the magnetic dipole radiation of a pulsar into a vacuum:
\begin{equation}
\dot{E}_{pul} \sim \left(\frac{B_{\perp}^2R^3}{6}\right) \left(\frac{R\Omega}{c}\right)^3 \Omega \, ,
\label{pulsar}
\end{equation}
where $B_{\perp}$ is the surface field perpendicular to the rotation axis.  The first term
in eq. (\ref{pulsar}) is near the total energy in the magnetic field.  That, multiplied 
by $\frac{B_{P}}{B_{\phi}}$ and $\Omega$, is an approximate form for eq. (\ref{powero}),
without the assumption that $B \propto \Omega$.  However, eq. (\ref{pulsar}) 
contains the term $\left(\frac{R\Omega}{c}\right)^3$, which
is the cube of the ratio of the radius of the acceleration region 
to the light-cylinder radius and is a small number ($\sim$10$^{-3}$) for $\Omega$s near 10$^3$
${\rm rad}\ {\rm s}^{-1}$. So, the power of the MHD-driven jets 
we investigate in this paper, $\dot{E}$, generally trumps,
at least in the first seconds after bounce, $\dot{E}_{pul}$. Note that the characteristic
dipole spindown time (time to ``halve" the initial period) is only days for a magnetar 
with $B_{\perp} = 10^{15}\ {\rm G}$ and an initial spin period of 10 ms.

Eq. (\ref{powero}) indicates that $\dot{E} \propto F_{rot}$. Since $F_{rot}$
might be considered the energy reservoir being tapped for the MHD jet, their ratio
is a measure of the characteristic time ($t_{jet}$) during which the jet operates.   
Remarkably, $t_{jet}$ is approximately equal to $\frac{P}{0.1\varepsilon(2\pi)}$,
demonstrating that yet another time scales with period. In fact, $t_{jet}$ 
is roughly 10-20$\times P$ and, if $P = 4\ {\rm ms}$, $t_{jet}$ $\sim$60 ms.  This is after
saturation is achieved, which might take 100-200 ms, but is still rather 
short.  However, as we will see, $F_{rot}$ is constantly replenished 
by equatorial accretion during polar explosion.  
In summary, \S\ref{amp}, \S\ref{mri}, and \S\ref{saturation} collectively demonstrate that
rapid rotation leads to magnetic field strengths, energies, and characteristic timescales that 
are very relevant in the context of collapse dynamics and the associated supernova.

\section{Numerical Technique and Computational Setup}
\label{setup}

We perform the calculations presented in this paper with
the code VULCAN/2D (Livne et al. 2004; Walder et al. 2005; Ott et al. 2006; Dessart et
al. 2006ab; Burrows et al. 2006,2007), newly enhanced 
with an ideal MHD capability (Livne et al. 2007). VULCAN/2D solves
the multi-energy-group, 2D radiation-magnetohydrodynamic 
(RMHD) equations of fluid motion with rotation.  The toroidal  
component of the magnetic field ($B_{\phi}$) is axisymmetric.
Otherwise, the three B-field components in cylindrical 
coordinates ($B_{\phi}$, $B_{z}$, and $B_{r}$) are evolved using
the corresponding Maxwell's equations in the MHD approximation 
with infinite conductivity.  The magnetic stresses are fully incorporated 
into the matter momentum equations.  The divergence-free condition
of the B-field is maintained to machine precision. 

For this study, we use 16 energy groups for each of three neutrino 
species ($\nu_e$, $\bar{\nu}_e$, and ``$\nu_{\mu}$")
and a 2D flux-limiter (Burrows et al. 2006,2007).  Hence, 
the calculations are not multi-angle, though VULCAN/2D 
has this capability (Livne et al. 2004).  Redistribution 
due to neutrino-electron scattering, a subdominant process with
respect to the charged-current URCA processes involving free nucleons,
is ignored and velocity-dependent terms in the transport equations
have been dropped.  The latter approximation can underestimate the
degree of neutrino heating in the gain region by $\sim$10\% 
(Hubeny \& Burrows 2007).

We use the hybrid grid that we have employed in previous VULCAN/2D 
multi-D radiation-hydrodynamics simulations (Ott et al. 2006; Dessart et 
al. 2006ab; Burrows et al. 2006,2007). The grid
is spherical-polar outside of a transition radius, here set equal to
25\,km, and switches smoothly to a pseudo-Cartesian grid
inside, extending right down to the center.
Using 221 grid points, we adopt a constant radial increment in the log between the transition
radius at 25\,km and the maximum radius, 5000\,km, of the computational domain.
We use 71 equally-spaced angular zones over a 90$^{\circ}$ angular domain from the rotation axis
to the equator (but see M15B10DP2A1F \footnote{F means that the calculations 
were done on the full 180$^{\circ}$ semicircle.} below) to ensure that 
cells have an aspect ratio of about unity throughout the spherical-polar region
\footnote{Note that limiting the computational domain to 90$^{\circ}$ suppresses 
the $\ell =1$ SASI (standing accretion shock instability) and core g-modes (Burrows 
et al. 2006,2007).}. The total number of zones is 17909. The highest resolution
is achieved at the center. Because of the Cartesian gridding there,
resolution is limited to $\sim$350 meters to guarantee reasonably 
``large" Courant times of no less than about a microsecond.
The lowest resolution is at the outer radius of 5000 km and is $\sim$110\,km.

The progenitor employed for the simulations reported in the
present study is the rotating 15-\mo\ model (m15b6) of Heger, Woosley, \& Spruit 
(2005, HWS). The cores of the HWS suite of models experience
significant magnetic spindown during stellar evolution to collapse
and generally leave very slowly rotating cores, 
with central periods from $\sim$30 to $\sim$50 seconds.
The HWS 15-\mo\ model has an interior period near $\sim$30 seconds.
Although this family of progenitor models was evolved by HWS in 1D with a 
well-motivated prescription for rotation, magnetic fields, 
and their evolution, we use instead our own analytical description 
for the initial angular-velocity and magnetic field profiles and spin this
initial model up significantly in order to explore the effects of rapid rotation. The 
initial rotation law is given by $\Omega(r) = \frac{\Omega_0}{1+(r/A_0)^2}$, where
$r$ is the cylindrical radius,
for which we choose $\Omega_0=\pi$ s$^{-1}$ (initial inner core period of 2\,seconds) and
$A_0=1000$\,km. We adopt two initial 
divergence-free field configurations: 1) the poloidal and toroidal fields 
are uniform throughout the computational domain (this is 
our ``uniform" [U] configuration) and 2) the poloidal and toroidal fields are
uniform in the cylindrical $z$ direction within a sphere of radius $A_0$ and dipolar (falling off
as $1/r^3$) outside.  This is our ``D" configuration.  In the latter setup, the toroidal component 
is taken to have the same radial dependence as the magnitude of the 
poloidal component. 

We reiterate that our simulations are 
axisymmetric and, thus, that no variable has a $\phi$-angle dependence.
In the full study (a subset of which is reported here), the magnitude 
of the initial poloidal and toroidal fields was varied.
We focus here on models with initial $B_z = 10^{11}$ Gauss (G) 
in the center, which we refer to as models M15B11DP2A1H and M15B11UP2A1H
and the initial toroidal magnetic field ($B_{\phi}$) is chosen to be 
3 orders of magnitude smaller than the polodial component.
The model naming convention we follow is M15BA$^{\prime}$A$^{\prime}$B$^{\prime}$PCAD$^{\prime}$H,
where A$^{\prime}$A$^{\prime}$ is the log of the initial poloidal field in Gauss (e.g., 10, 11, or 12),
B$^{\prime}$ is the B-field configuration given above (either ``U" or ``D"), C
is the initial central spin period ($= 2\pi/\Omega_0$) in seconds (here either 2 or 4 seconds),
D$^{\prime}$ is $A_0/1000 {\rm km}$, and H means that the calculations were done on a 90$^{\circ}$ quadrant.
We have performed many simulations with a variety of parameters, but 
focus here on A$^{\prime}$A$^{\prime}$ = 11. 
   Most previous purely MHD simulations (see \S\ref{intro}) focussed on
   models with initial fields at the high end, in fact near $10^{12-13}$ Gauss,
   in order to obtain some dynamical effect on the short timescales (tens of
   milliseconds) they could address.
Table \ref{tab_model} lists the simulations that have informed 
our investigation and conclusions, and includes general information on 
each run.  Note that model M15B0DP2A1H is the non-magnetic model we include   
to gauge the effects of rapid rotation alone and employs a slightly 
altered naming convention. Model M15B11DP4A1H explores
the dependence on $P_0$ and model M15B12DP2A1H explores the effects of 
very large fields that might perhaps be too large for an assumed $P_0$ 
of 2 seconds, but nevertheless demonstrates several interesting, and dramatic, 
features.  Model M15B10DP2A1F is calculated over the full 180$^{\circ}$ angular realm,
thereby liberating the $\ell = 1$ mode of the standing accretion shock instability 
(SASI; Blondin, Mezzacappa, \& DeMarino 2003; Foglizzo, Scheck, \& Janka 2006; 
Buras et al. 2006ab; Burrows et al. 2007). With model M15B10DP2A1F, we can 
determine the influence of rotation and magnetic fields on the SASI. Note 
that all these models explode before the core g-modes would have had 
a chance to achieve significant amplitudes (Burrows et al. 2006,2007).

\section{Simulation Results}
\label{results}

\begin{figure}
\includegraphics[height=.3\textheight]{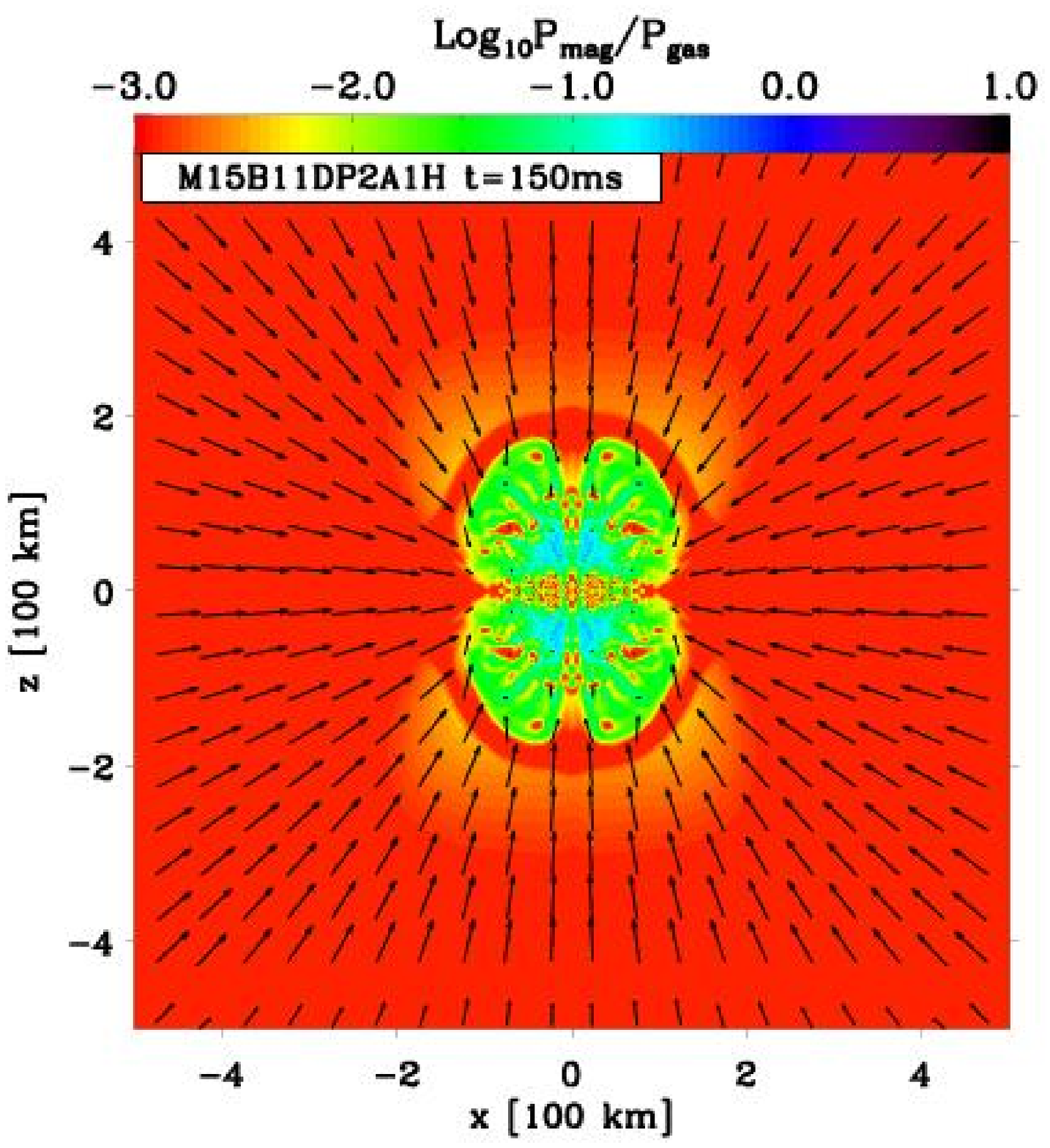}
\includegraphics[height=.3\textheight]{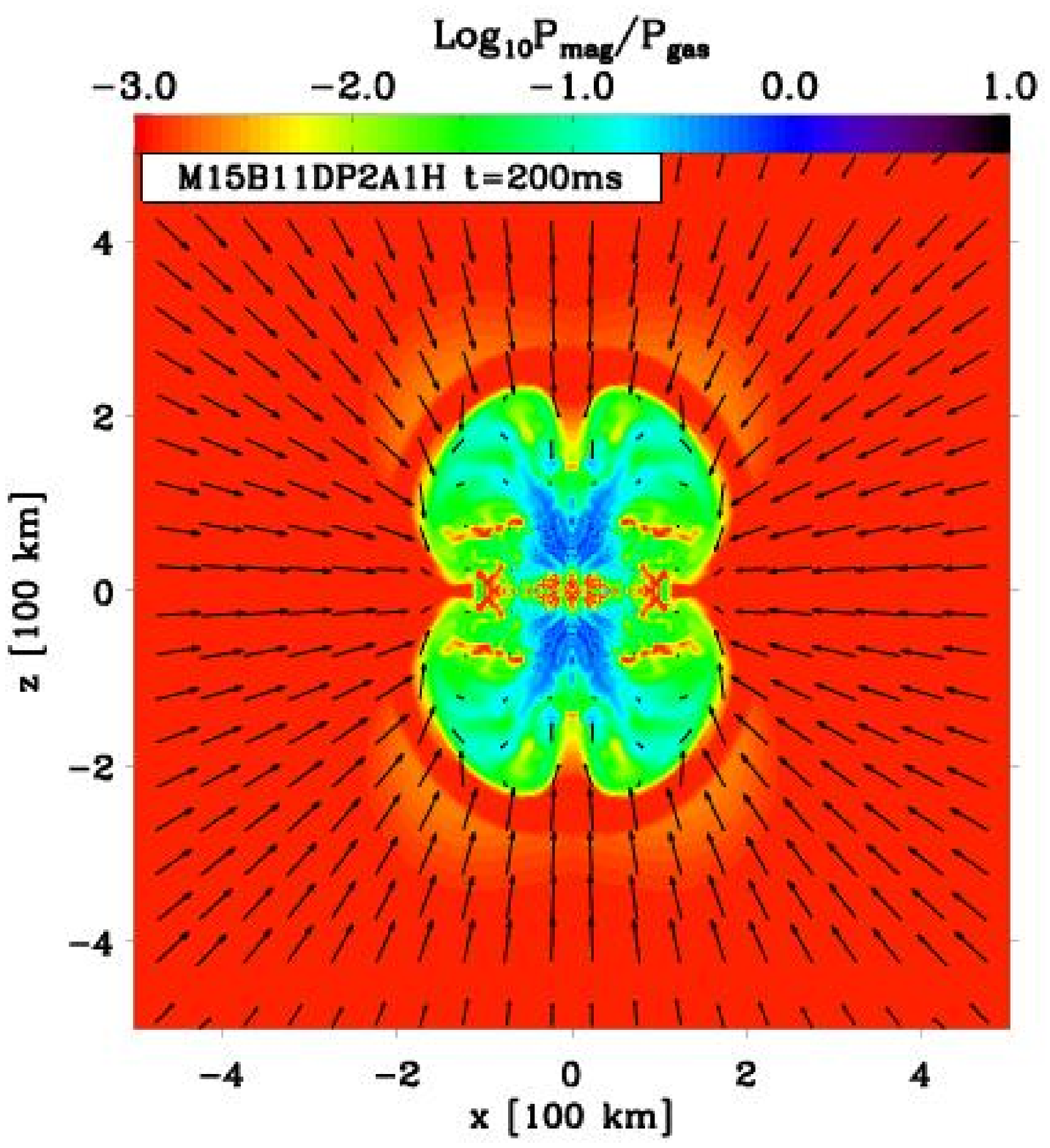}
\includegraphics[height=.3\textheight]{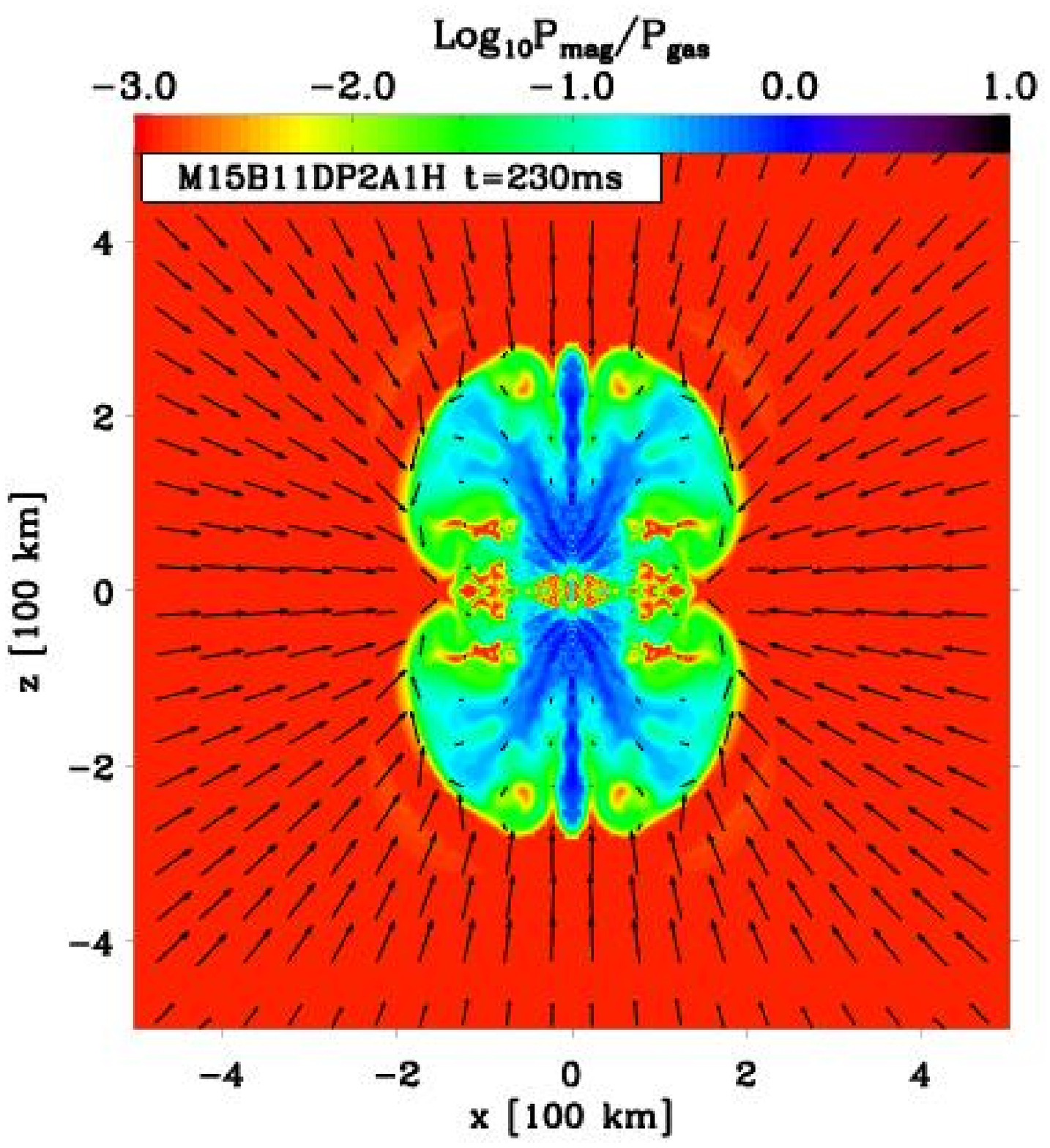}
\includegraphics[height=.3\textheight]{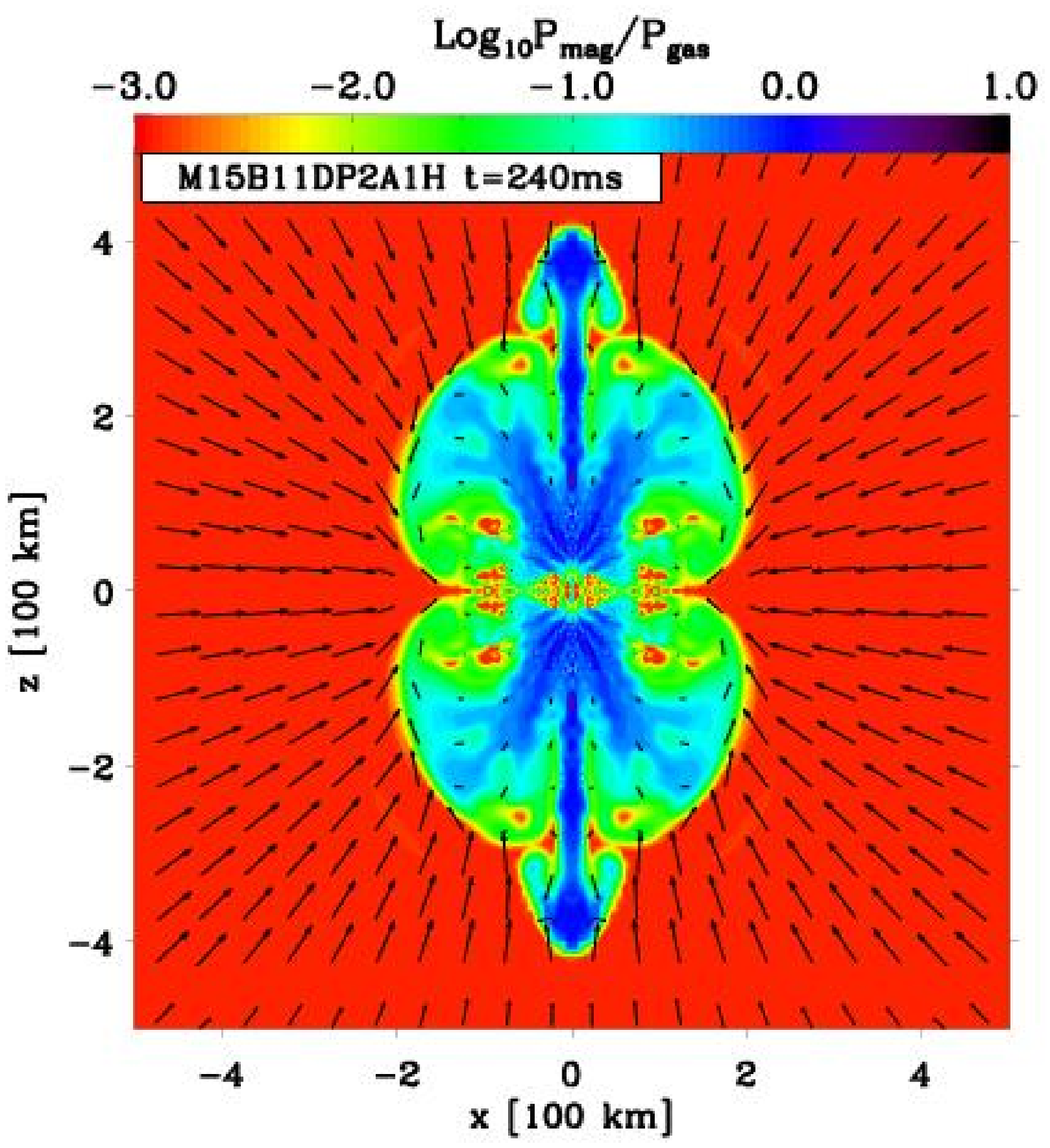}
\includegraphics[height=.3\textheight]{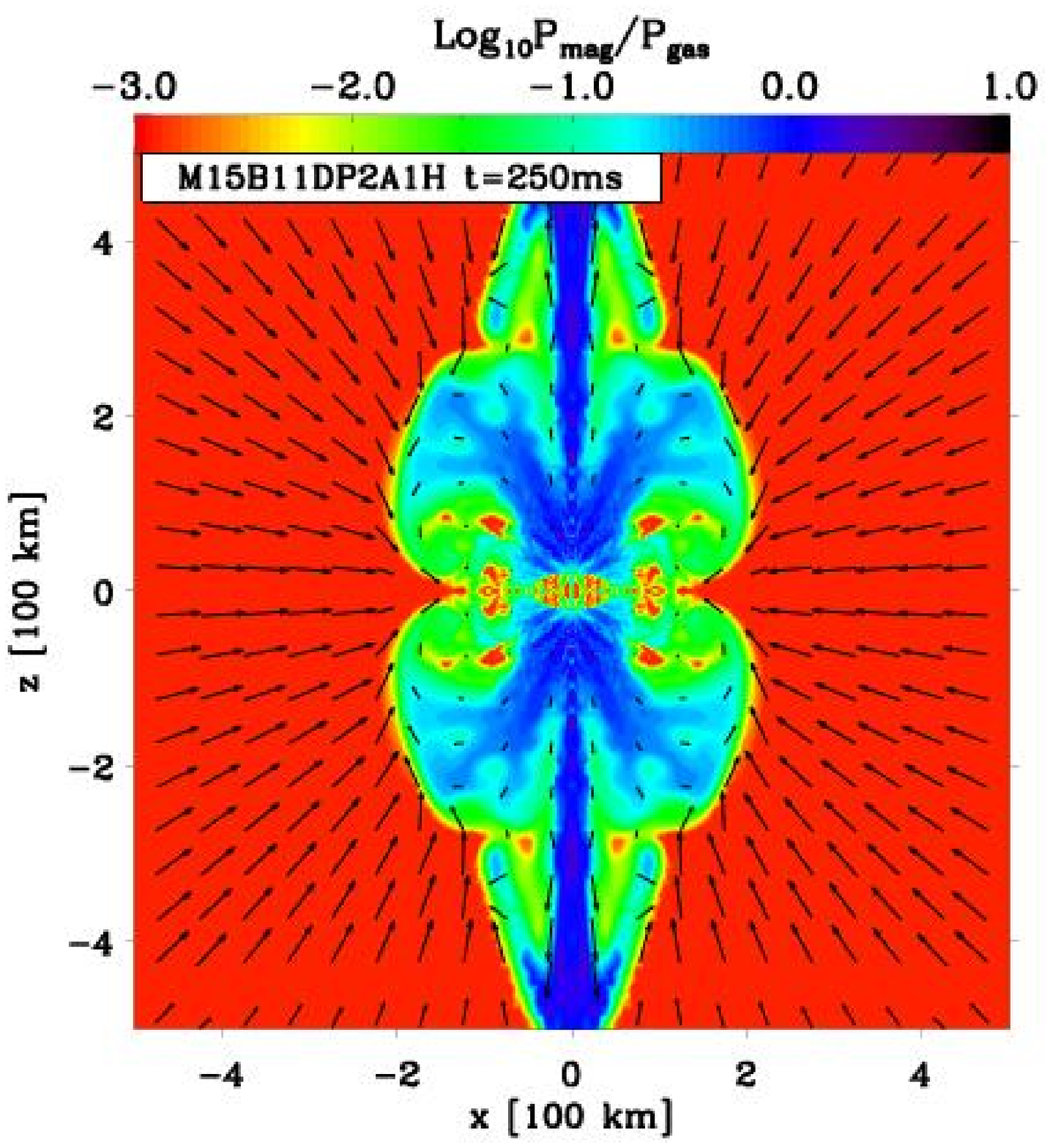}
\includegraphics[height=.3\textheight]{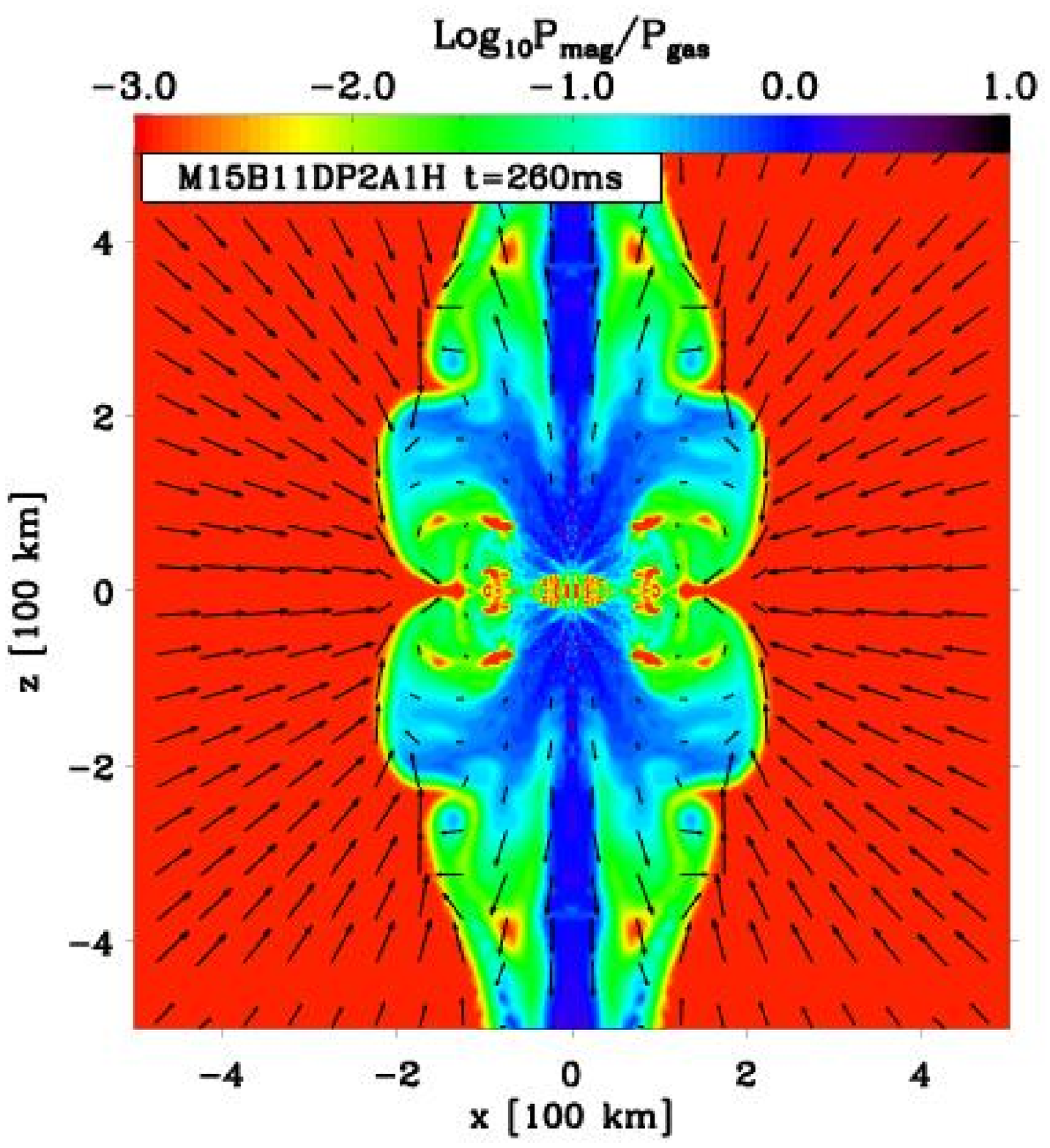}
\caption{
Colormap of the ratio of the magnetic pressure to the gas pressure for model
M15B11DP2A1H at 150\,ms (top left), 200\,ms (top right), 230\,ms (middle left),
240\,ms (middle right), 250\,ms (bottom left), and 260\,ms (bottom right)
after bounce, in the inner 1000$\times$1000\,km$^2$ region.
This series of six stills shows the approach to and onset of explosion
as the magnetic pressure behind the shock becomes more prelevant.
We overplot black velocity vectors, with a length
saturated at 10000\kms that corresponds to 7\% of the width of the display.
Note how the magnetic pressure behind the shock (near $\sim$100-200\,km early on)
goes from being subdominant to being comparable to the gas pressure.  For
this model, this happens $\sim$200\,ms after bounce. Notice that though the last three stills
are each separated by only 10\,ms, once the ratio of the magnetic pressure to the gas pressure
near the shock wave is near unity the polar explosion quickly commences. In all our models,
the approach of this ratio to unity always signals the onset of a jet explosion,
but at different physical times.}
\label{fig_pmagopgas_sequence}
\end{figure}

\begin{figure} 
\includegraphics[height=.3\textheight]{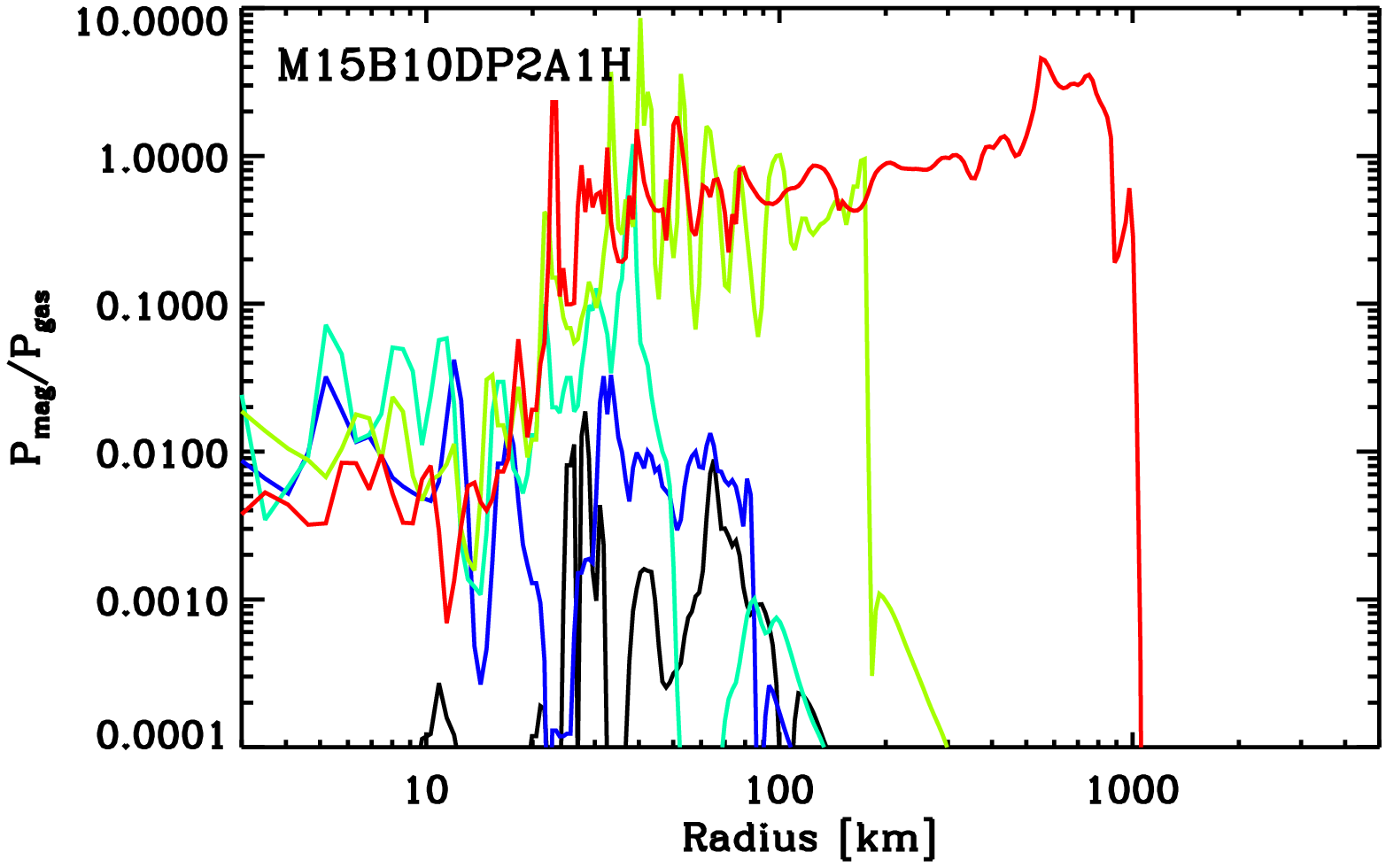}
\includegraphics[height=.3\textheight]{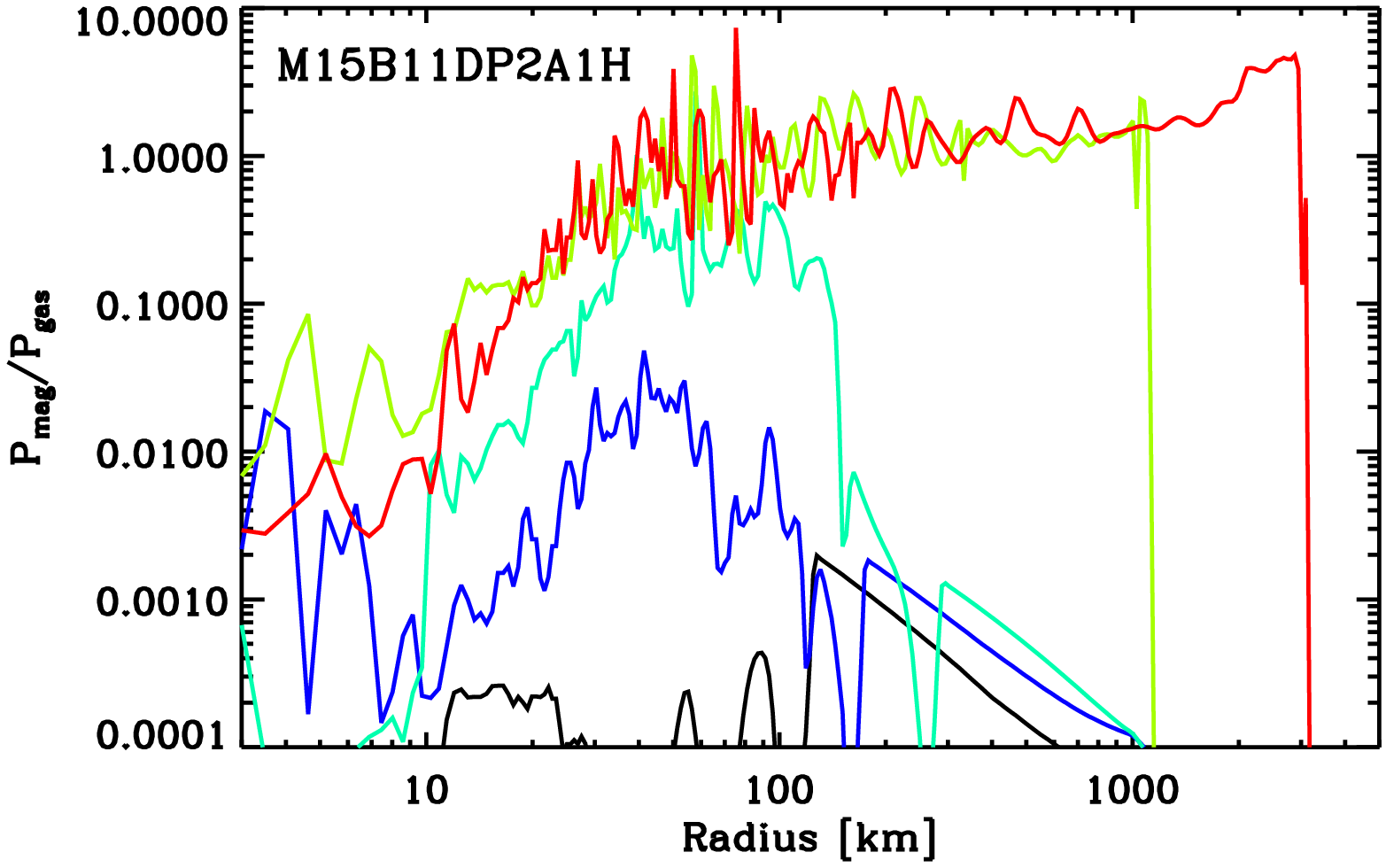}
\includegraphics[height=.3\textheight]{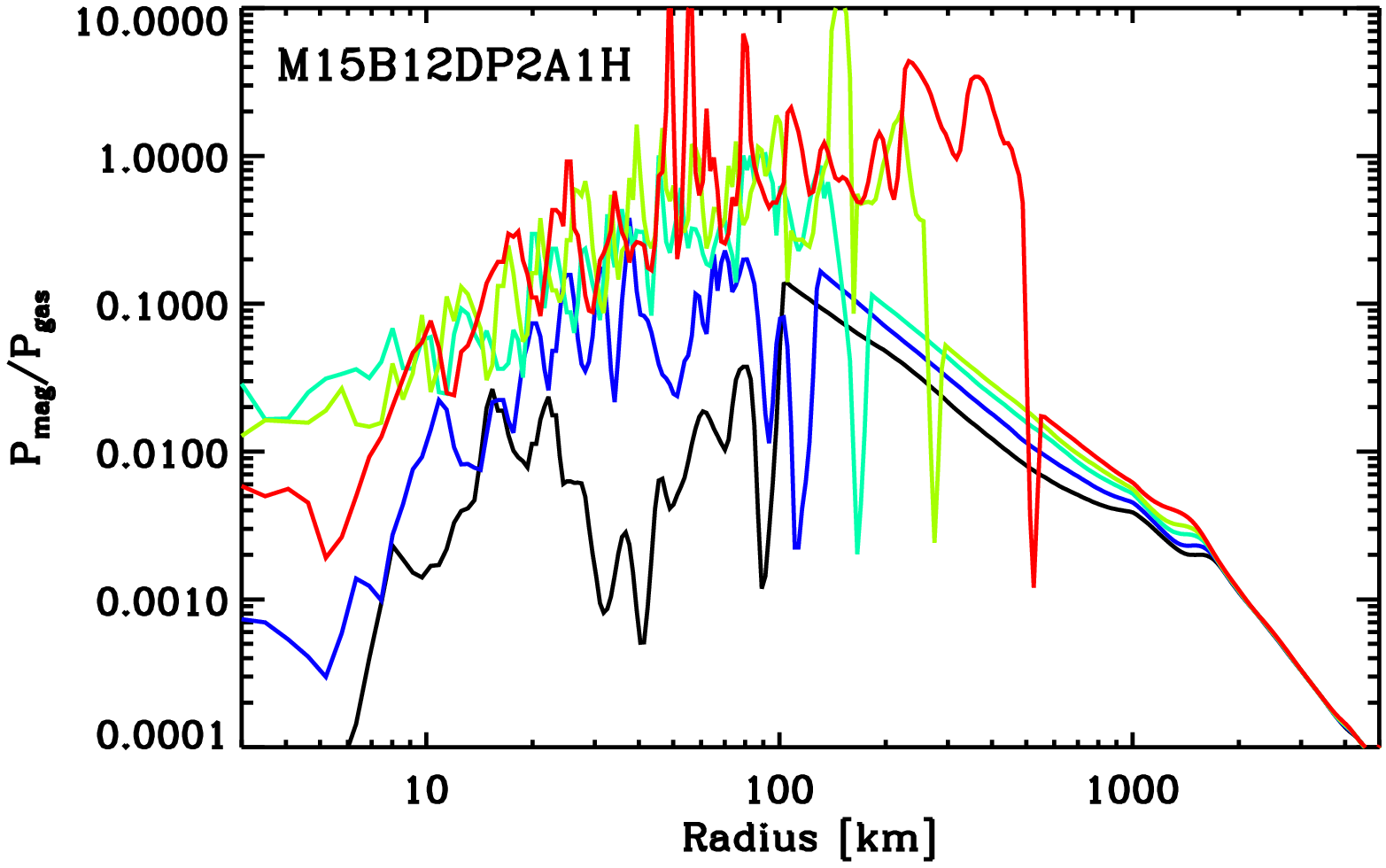}
\caption{
Sequence of snapshots of the radial variation along the polar direction of
the ratio of magnetic pressure to gas pressure for model
M15B10DP2A1H (top left), M15B11DP2A1H (top right), and M15B12DP2A1H (bottom left).
The colors corresponds to time, increasing in the order: black, blue, turquoise, green,
and red. The turquoise color corresponds to the time when the displayed ratio first exceeds unity,
i.e., the time that signals the launching of the jet, and varies from model to model
(see Table~1).
For model M15B10DP2A1H, the snapshot times are 218, 334, 451, 568, and 685\,ms after bounce.
For model M15B11DP2A1H, the snapshot times are 46, 125, 204, 283, and 362\,ms after bounce.
For model M15B12DP2A1H, the snapshot times are 21, 35, 49, 63, and 76\,ms after bounce.
See text for a discussion.
}
\label{pmax.pgas.profile}
\end{figure}

All the RMHD collapse models we have simulated go through the same general stages
and manifest the same phenomena.  Within $\sim$10$-$30 ms of bounce, the shock wave
launched at bounce stalls near 100$-$150 km in the canonical fashion (Bethe \& Wilson 1985; 
Wilson 1985; Burrows et al. 2006,2007; Buras et al. 2006ab).  Differential rotation in the cavity
between the stalled shock and the inner PNS core inexorably winds the toroidal component of
the B-field to higher and higher values (\S\ref{amp}).  Eventually, the magnetic stresses due
to the amplified $B_{\phi}$s achieve values in a quasi-cylindrical column along the poles
that begin to compete with the gas pressures ($P_{\rm gas}$).  In particular, magnetic stresses
($P_{\rm mag}$) just behind the shock start to rival the local $P_{\rm gas}$.  When this happens, a magnetically-driven
jet punches into the accreta along the poles, powering a bipolar explosion.  
Figure \ref{fig_pmagopgas_sequence} provides six colormaps of the ratio, $P_{\rm mag}$/$P_{\rm gas}$,
for model M15B11DP2A1H just before, during, and after the onset of explosion. Note
that the explosion commences only when this ratio reaches a value near unity (dark blue) along the
poles, at which time the jet emerges.  The last five stills are separated
from one another by only 10 milliseconds, indicating that when the pressure condition is
achieved the explosion does not long delay. Figure \ref{pmax.pgas.profile} portrays the 
evolution of this ratio along the poles for three representative models over a wider
time span.  The importance of the $P_{\rm mag}$/$P_{\rm gas}$ $\sim$ 1 condition
is manifest in all these as well, even though the epoch of explosion is very different for each
model (see caption).  For each line, a vertical drop traces the shock wave, 
which starts post-bounce life as an accretion shock, but then transitions into a 
propagating explosion when the $P_{\rm mag}$/$P_{\rm gas}$ $\sim$ 1 condition
is reached.

The structure of the magnetic field in the resultant cavity is that of a tightly coiled spring, and the explosion
resembles a ``magnetic tower" (Uzdensky \& MacFadyen 2006ab).  Radial gradients in the hoop
stresses along both rotation axes (top and bottom) are responsible for the jet thrust.  At the inauguration
of the jet, the magnetic stresses in the equatorial sector are not yet competitive 
with the local gas pressures.   Note that all that is required for a jet to emerge to reenergize
the shock and to overcome the local ram pressure of the accreting matter is the focussed 
application of magnetic pressure, such as occurs first along the rotation axis.   
Hence, even a ``small" MHD power (but over a small solid angle) can ``bore a hole."  The jet
is confined predominantly by the accreting matter through which it drills, and by the magnetic hoop
stresses. Since the jet is thereby inhibited from spreading laterally, 
it is not centrifugally flung out to large angles and,
hence, does not resemble a Blandford-Payne jet (Blandford \& Payne 1982).
For all our runs, which extend up to $\sim$1 second after bounce (Table \ref{tab_model}) 
and achieve jet penetrations of $\sim$5000 km, the jet remains tightly collimated and does not diverge. 

The jet-driven explosions we obtain are not energized by
the impulsive deposition of energy, but by the continuous injection of power
due to magnetic stresses in the inner core that are continuously replenished 
by winding of poloidal field into ``new" toroidal field.   The ultimate source of
energy and power is the gravitational energy of infall, converted into differential
rotational motion of the shocked outer core, and then transformed into magnetic energy.
The post-bounce time needed to launch the jet varies in our suite of models from $\sim$80 ms
(M15B12DP2A1H) to $\sim$550 ms (M15B10DP2A1H), but in all cases is much shorter than 
the time it would take to inject into the explosion at the jet powers achieved all the free 
energy that resides in core shear motion. In other words, the jets achieve a 
quasi-steady state of roughly constant power that changes only on secular 
evolution timescales.  An ``engine" is established.  Furthermore,
jet explosion along the poles is simultaneous with continuing accretion
along the equator.  Such accretion brings in ``fresh" rotational energy that maintains
the store of kinetic energy being tapped by the magnetic jet.  The free energy available
to the MHD jet actually continues to grow after bounce.  Hence, accretion does not inhibit
explosion, but powers it.  This is reminiscent of the situation with the acoustic 
mechanism (Burrows et al. 2006,2007), in which there is also simultaneous 
accretion and explosion.  As in that case, the breaking of spherical
symmetry into dual matter streams is the dynamical key.

\begin{figure}
\vspace{1in}
 \begin{center}
   \includegraphics[height=.5\textheight]{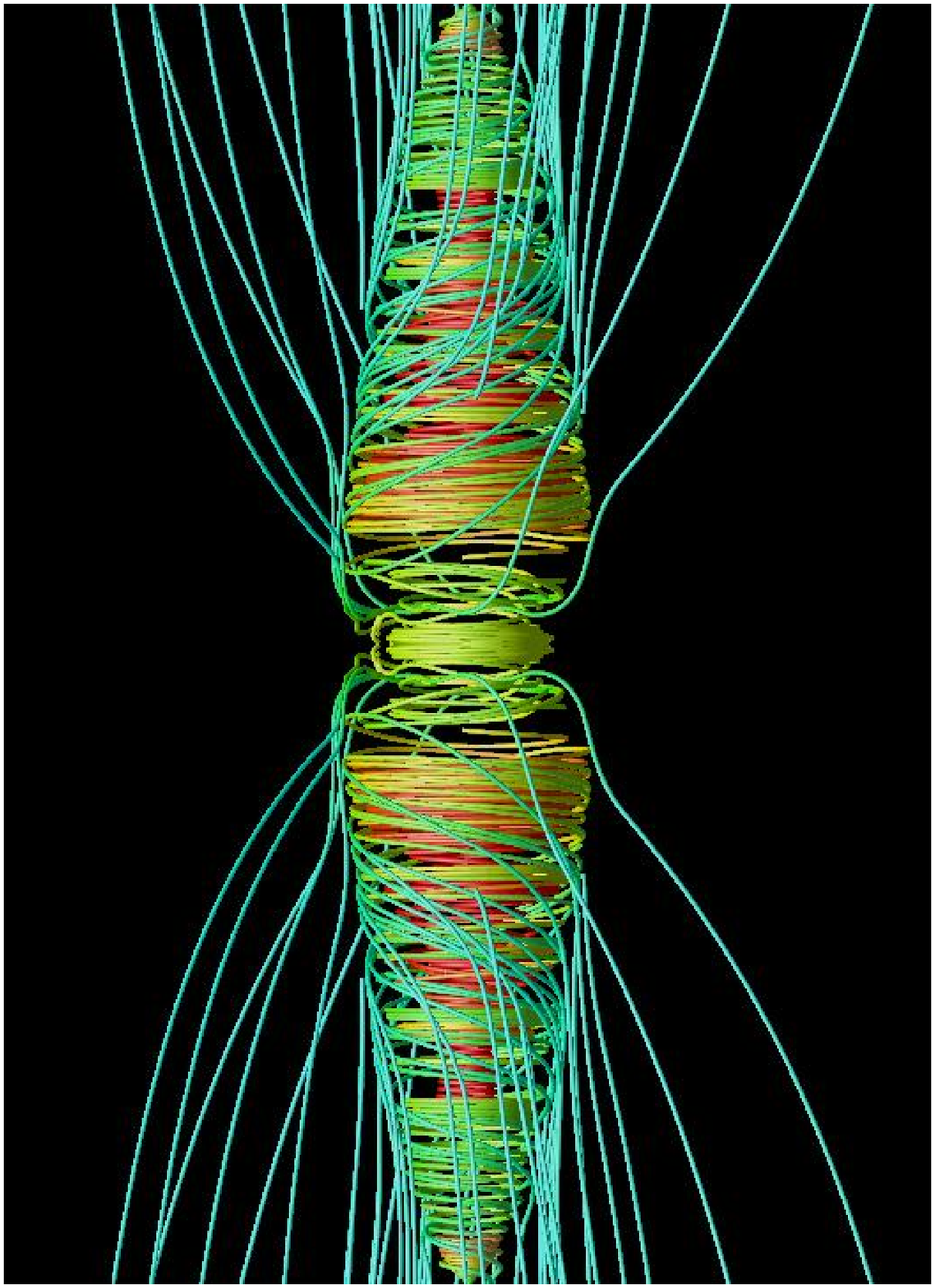}
   \includegraphics[height=.5\textheight]{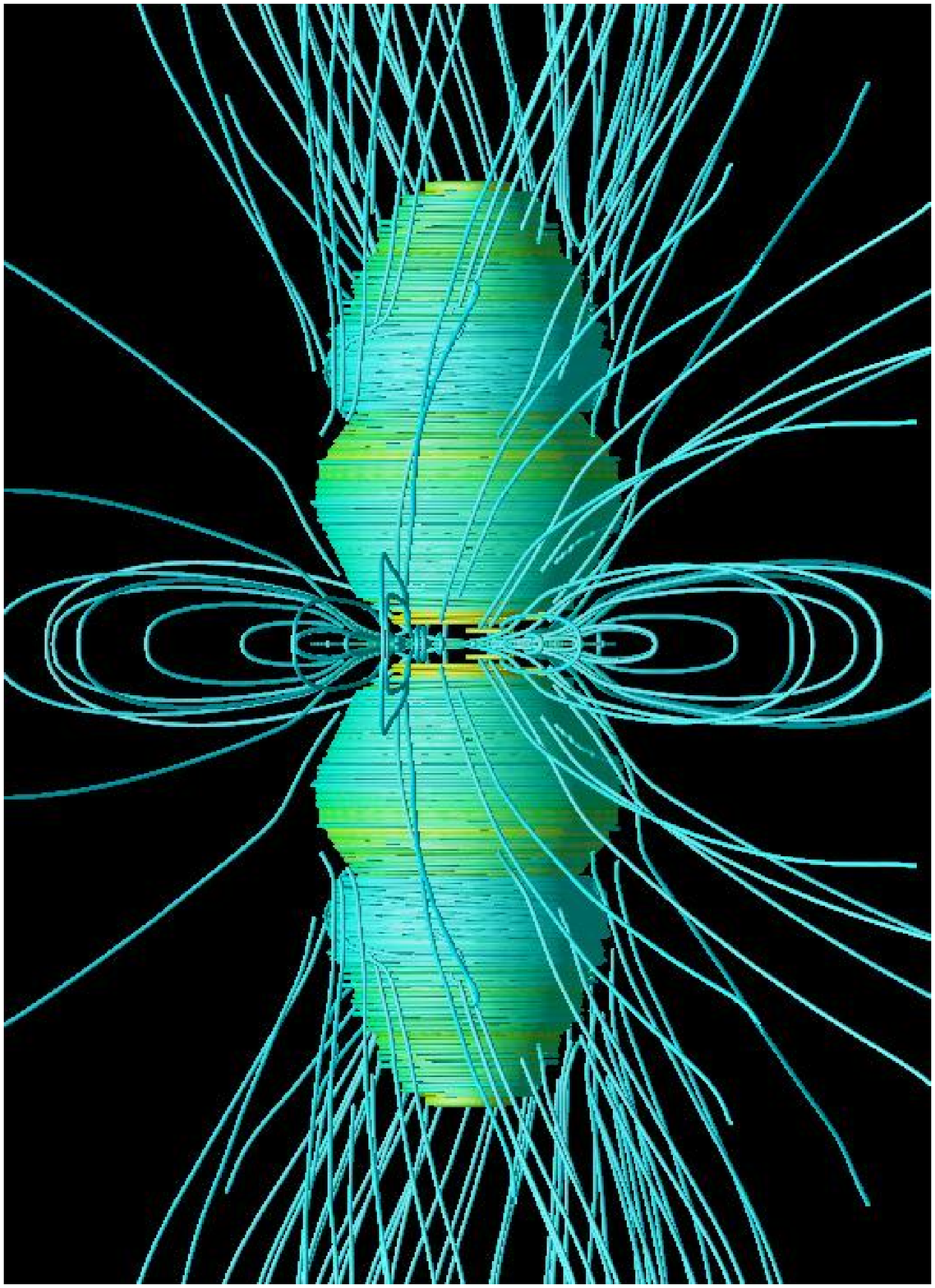}
\vspace{1in}
\caption{{\it Left:} Magnetic field lines for model M15B11UP2A1H at 264.5\,ms
after bounce. The size of the displayed region is 3000$\times$4000\,km$^2$.
``Footpoints'' for the field lines are randomly distributed in the inner
500-1000\,km, with a denser distribution along the polar
axis to probe the region of larger magnetic energy where the explosion takes
place in our simulations. Hence, the crowding of field lines does not correspond
directly and accurately to regions of larger magnetic fields. {\it Right:}
Same as on the left, but for model M15B10DP2A1H at 855.5 ms after bounce and
on a scale of 6000 km$\times$8000 km.  Notice how much more tightly the
B-field is wound. (See text for a discussion.)}
\label{mod2p_12_field_lines}
\end{center}
\end{figure}

Figure \ref{mod2p_12_field_lines} depicts snapshots of the distribution of 
magnetic ``streamlines" anchored in footpoints at arbitrary positions in the 
inner regions.  The left panel shows model M15B11UP2A1H at 264.5 ms after
bounce and the right panel shows model M15B10DP2A1H at 855.5 ms after bounce.
The lines are colored in entropy, with the yellow/red lines tracing the 
shocked regions and the light blue lines tracing the unshocked regions.
The severe twisting of the field lines due to rotational winding behind the shock
is clearly seen, as is the magnetic tower morphology\footnote{The vertical coiled 
spring structure depicted in Figure \ref{mod2p_12_field_lines}, and in 
Figure \ref{marker_streaklines_mod2p} below, makes clear why this
field distribution is referred to as a ``magnetic tower."}.
Moreover, the ``auger" or ``punch" shape of the confined jet is manifest.  Model 
M15B10DP2A1H has a more tightly wound field structure because its lower 
initial poloidal component (M15``B10"DP2A1H) necessitated more rotations 
to achieve the requisite magnetic pressures and because the resulting 
explosion power and ejection speeds were lower.  Therefore, the spring unwound 
more slowly.  Nevertheless, models M15B11UP2A1H and M15B10DP2A1H are 
qualitatively, if not quantitatively, the same.

Figure \ref{marker_streaklines_mod2p} portrays streaklines of the trajectories
of marker particles covering the last 200 ms of the M15B11UP2A1H (left) and 
M15B11DP2A1H (right) simulations.  The lines are the paths followed by 
representative parcels of matter that are co-extensive with the magnetically-driven jet.
Figure \ref{marker_streaklines_mod2p} clearly shows the lift-off of the cork-screwing 
Lagrangian parcels as rotation transitions into spiraling ejection, and then, 
at larger radii, into a directed jet.  In addition, in model M15B11UP2A1H the radius of the shock
in the equatorial regions is larger.  This is because the equatorial magnetic pressures 
achieved there at a given time are larger than in model M15B11DP2A1H.  This, in turn, is 
due to the fact that in model M15B11UP2A1H the uniform (``U") initial poloidal field results
in larger accreted fields at later times than in model M15B11DP2A1H,
for which the late-time accretion is of matter from the outer core where
the initial field decays in the $\frac{1}{r^3}$ dipolar manner (\S\ref{setup}). In fact, for 
model M15B11UP2A1H the equatorial regions join the explosion at later times. This outcome 
is expected eventually for all models, but due to the different magnetic field structures and magnitudes
for the models listed in Table \ref{tab_model} the times to equatorial 
explosion will vary greatly from model to model. 

The particle trajectories implied by Fig. \ref{marker_streaklines_mod2p} and magnetic 
flux freezing indicate that the ejected material stretches toroidal field into poloidal
field, in a reverse of what happens during rotational winding in the inner $\sim$20$-$150 km.
So, in the jet column at large radii the field has a significant poloidal component.

\begin{figure}
\plottwo{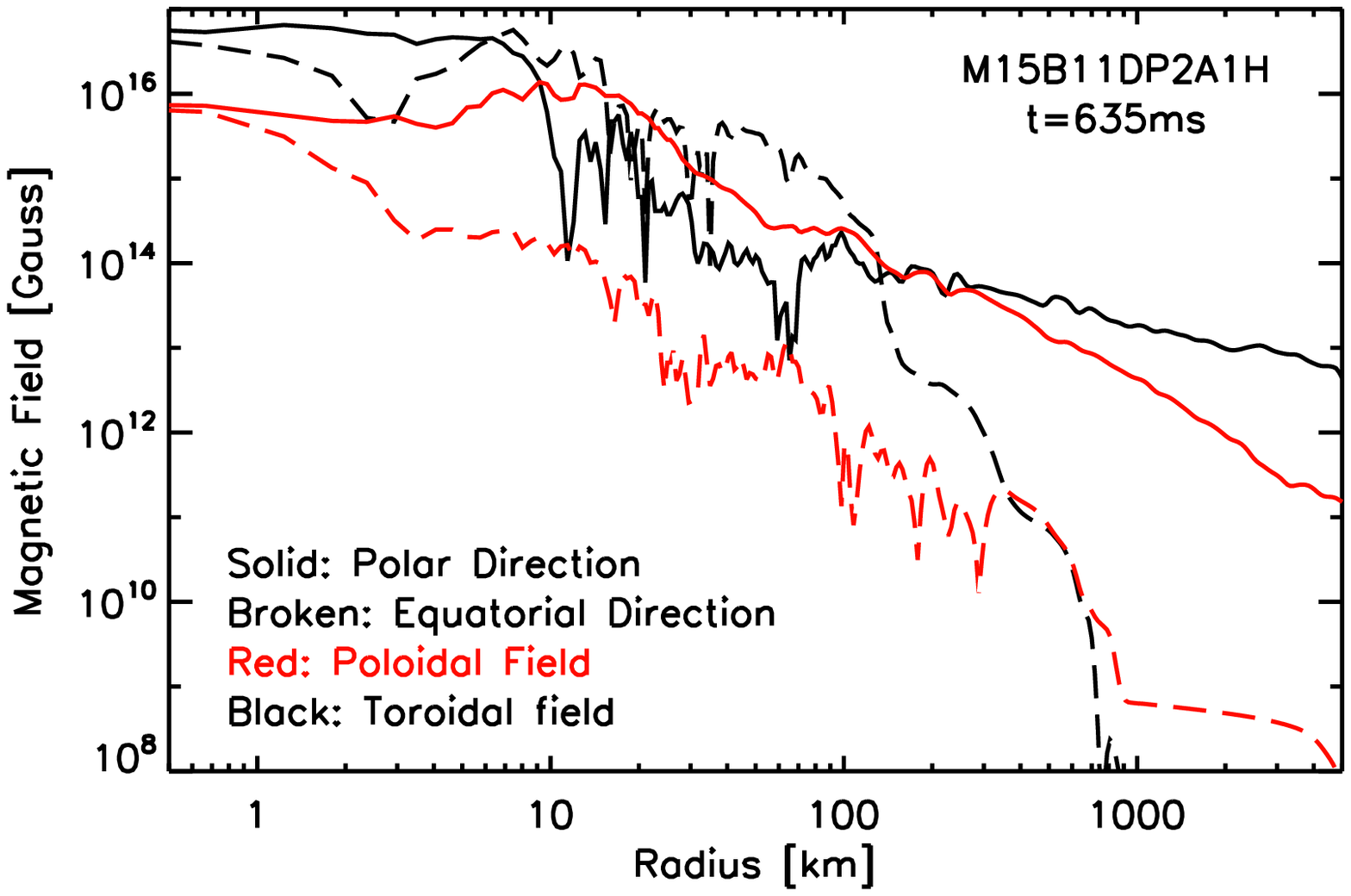}{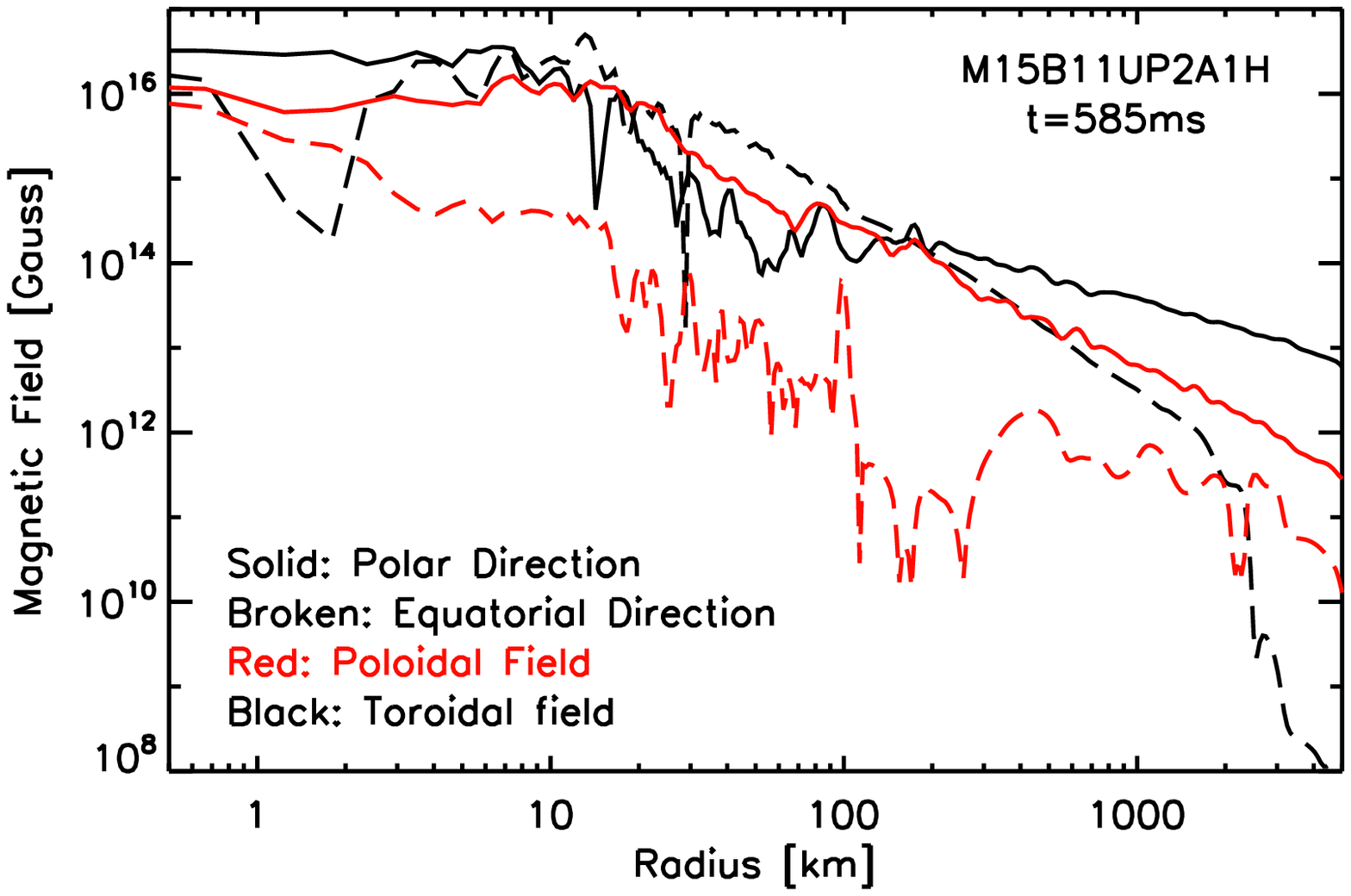}
\caption{Radial slices at the end of each simulation for models
M15B11DP2A1H (left) (at 635 ms after bounce) and 
M15B11UP2A1H (right) (at 585 ms after bounce) of the magnitude of the
toroidal (black) and poloidal (red) components of the magnetic field, along
the polar (solid) and equatorial (broken) directions.}
\label{mod2p_btor_pol}
\end{figure}

Figure \ref{mod2p_btor_pol} shows radial slices along the 
poles (solid lines) and along the equator (broken lines) 
of both the poloidal (red) and toroidal (black) fields for models M15B11DP2A1H 
(left) and M15B11UP2A1H (right) at 635 ms and 585 ms, respectively, after bounce. 
Since there is no appreciable rotational shear interior to $\sim$10 km,
the magnetic fields there have little dynamical effect.  It is the fields in the region
between $\sim$10 km and $\sim$150 km that are of consequence, since it is here 
that the magnetic tower is launched and maintained.  Figure \ref{mod2p_btor_pol} and 
\S\ref{saturation} indicate that the fields achieved in this region in these models are 
comparable to what is expected at saturation for a $P_0$ of 2 seconds ($\sim$10$^{15}$ G).
This justifies our focus on these models when assuming $P_0 = 2$ seconds, despite the 
fact that we underresolve the MRI.  

\begin{figure}
\plottwo{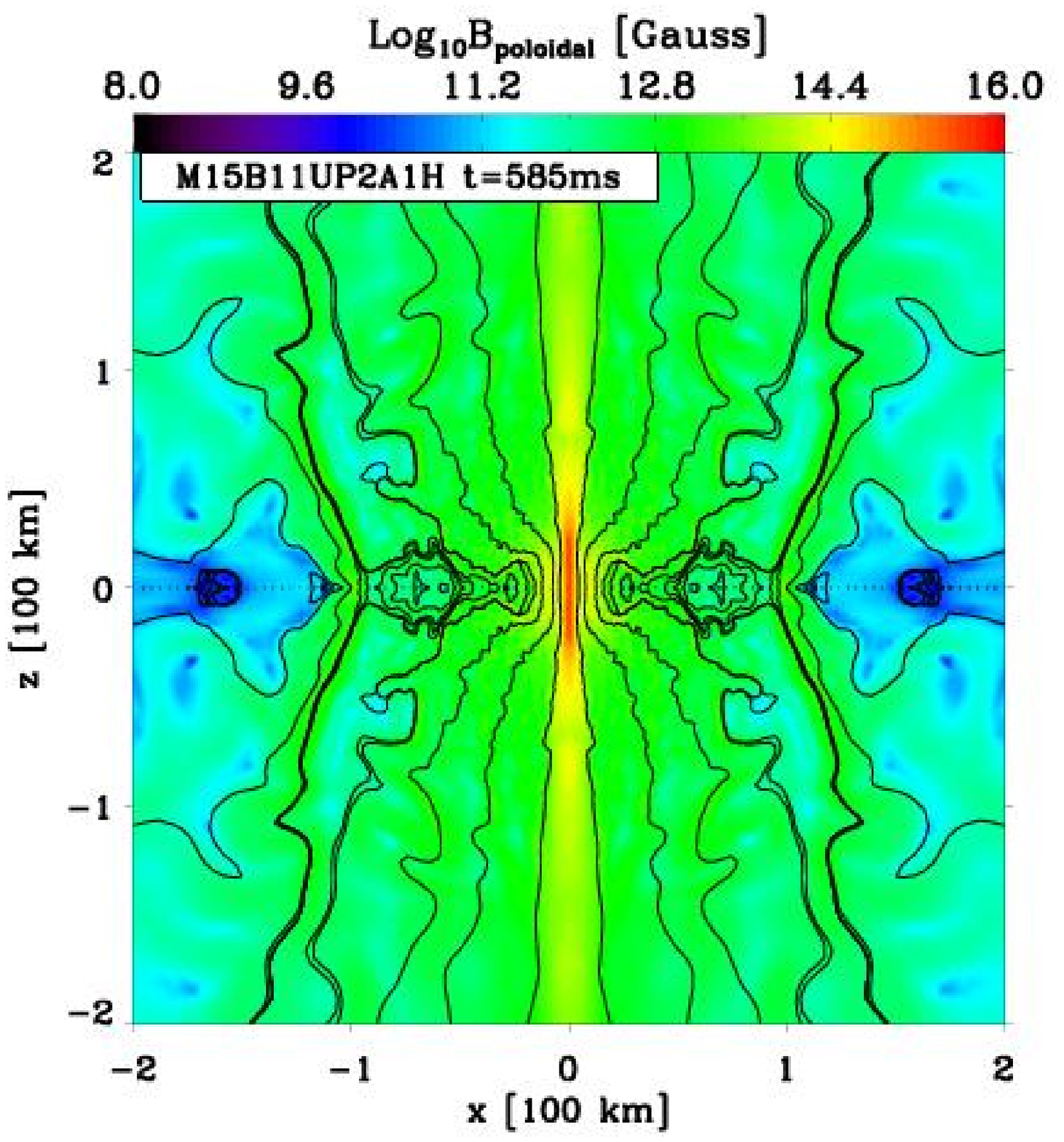}{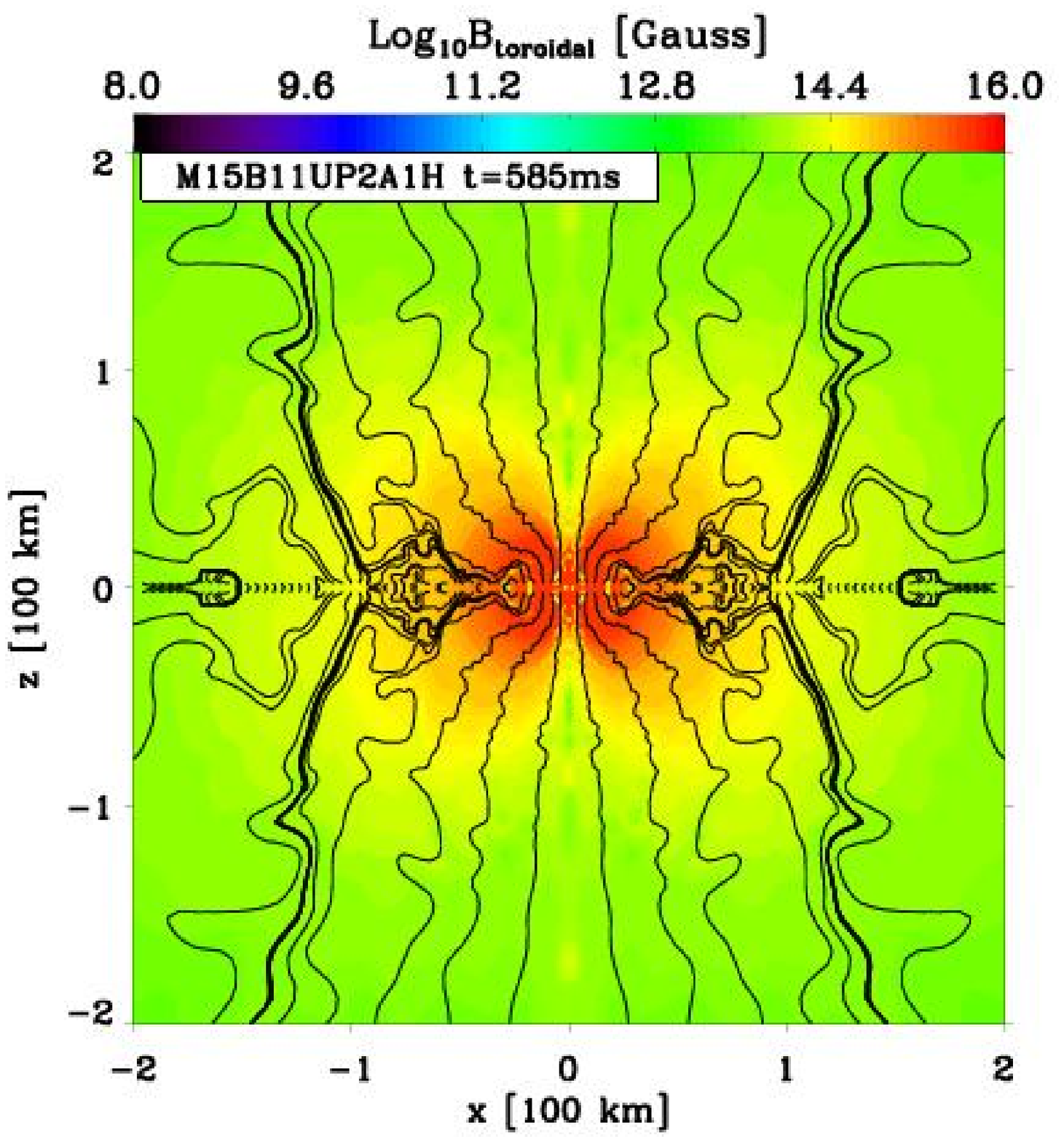}
\caption{Colormap of the magnitude of the poloidal (left panel) and the
toroidal (right panel) components of the magnetic field for model
M15B11UP2A1H at 585\,ms after bounce out to 200\,km. We also overplot 
poloidal field lines every 4\,km, starting along the equator within 200\,km of the polar axis.
}
\label{mod2p_r04k_bfield}
\end{figure}

Figure \ref{mod2p_r04k_bfield} depicts colormaps of the poloidal (left) and toroidal (right)
field distributions in model M15B11UP2A1H, 585 ms after bounce.  In both panels, the 
lines are iso-poloidal field lines and the inner 200 km on a side is shown.
The relative extents of the red and yellow regions demonstrate the dominance 
of the toroidal component in the inner zones at these late times well 
into the explosion, but the presence of a column of yellow/red (high-field) 
along the axis in the poloidal plot attests to the conversion due 
to stretching by ejected matter of toroidal into poloidal field 
(see also Fig. \ref{marker_streaklines_mod2p}).  Figure \ref{mod2p_r04k_bfield}
also demonstrates the columnar structure of this inner region due
both to equatorial accretion (and, hence, pinching) and rotation about the (vertical) axis.
However, it should be made clear that the actual field distributions
after saturation are likely to be different, and what they are in detail when the MRI
is fully enabled remains to be determined.

\begin{figure}
  \plottwo{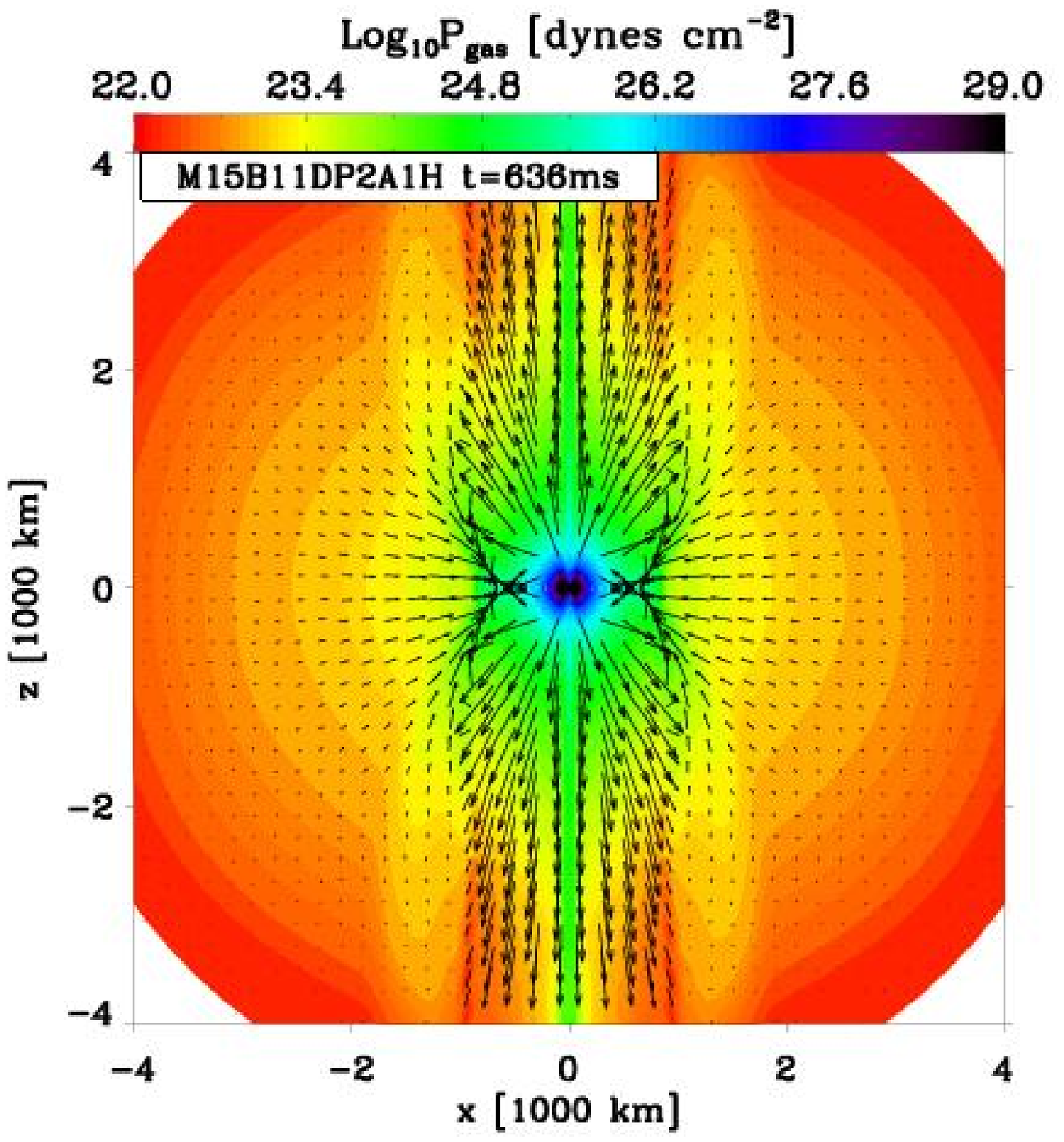}{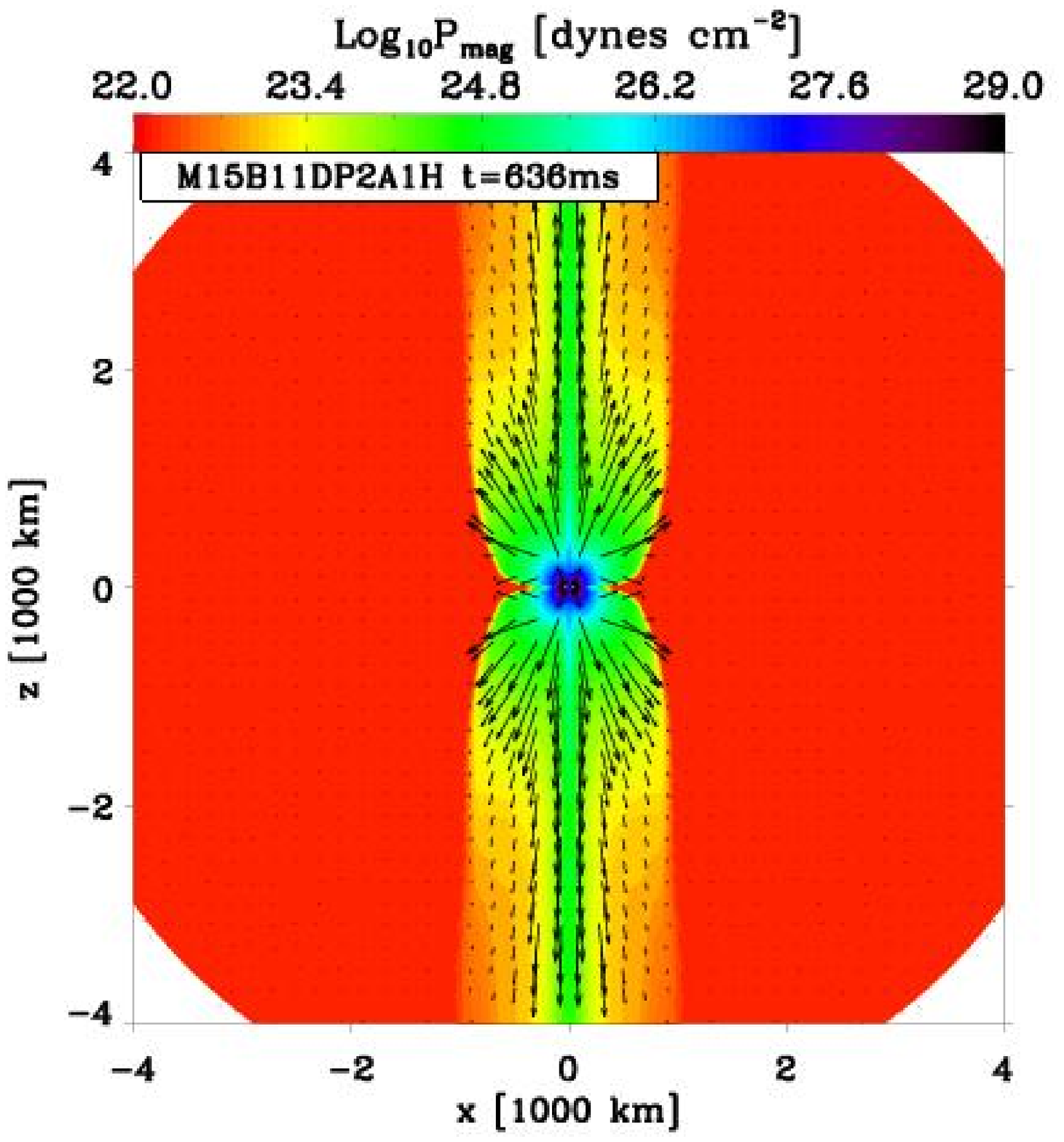}
  \plottwo{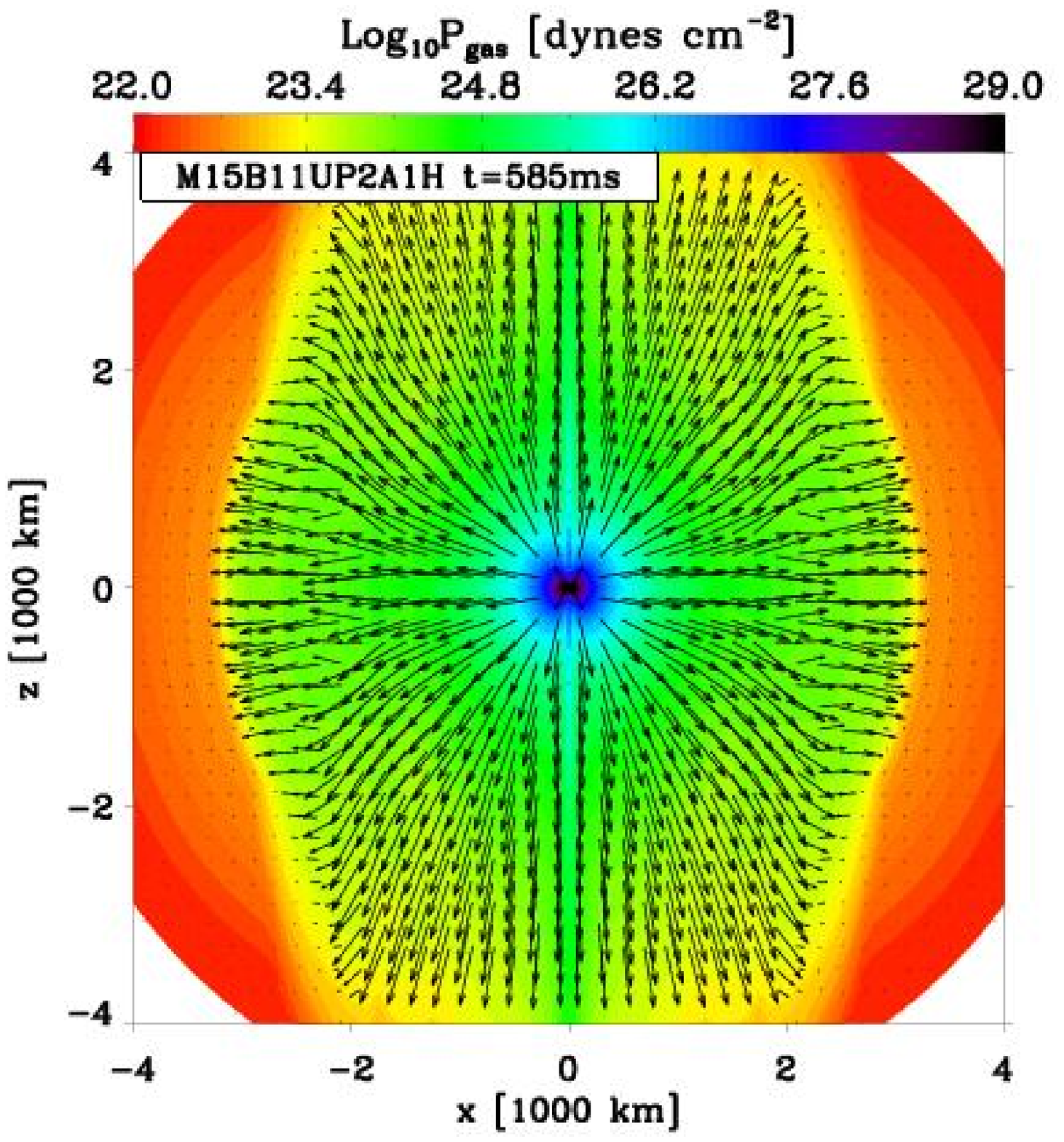}{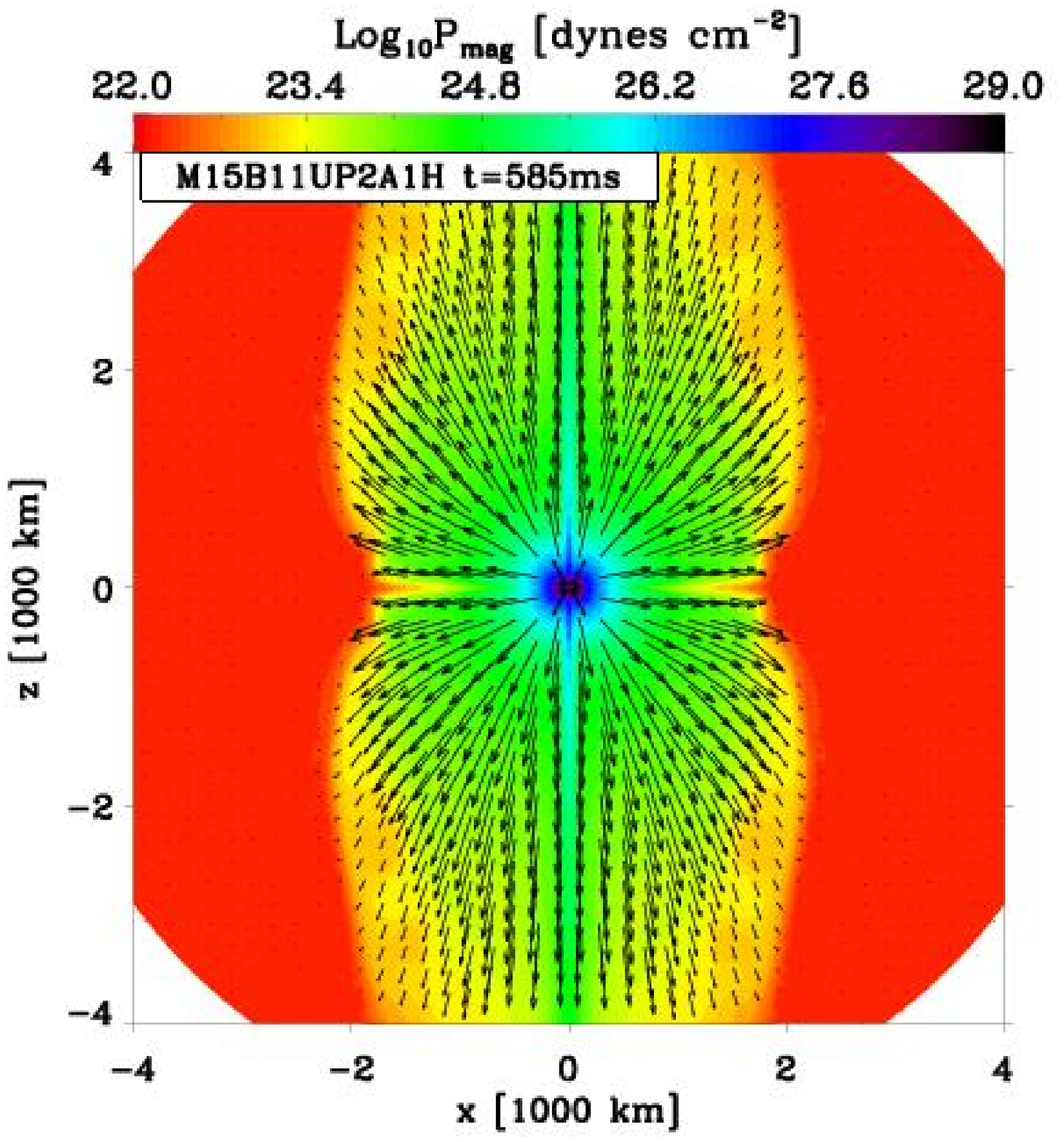}
\caption{{\it Top Left:} Colormap of the gas pressure for model M15B11DP2A1H,
at 636\,ms after bounce.
We also overplot ``Bernoulli'' vectors in black,
corresponding to $\rho v_R \times (e_{\rm th}+e_{\rm kin}
+P/\rho-GM_{\rm PNS}/R)$. This choice does not account for contributions from
the magnetic field. The vector length is saturated at
10$^{33}$\,erg\,cm$^{-2}$\,s$^{-1}$ and corresponds to 15\% of the side 
length of the display. {\it Top Right:} Colormap of the magnetic pressure, 
at 636\,ms after bounce, including the toroidal and the (sub-dominant)
poloidal components. We overplot vectors for the Poynting flux, with the 
same ratio between vector length and vector magnitude as in the
left panel. The Poynting flux has the following standard definition: $\vec S =
c/4\pi (\vec E \times \vec B)$, and measures the rate of electromagnetic energy flow,
here on this scale of comparable magnitude to the Bernoulli flux (though lower), shown in the
left panel. {\it Bottom Left and Right:} Same as above, but for model M15B11UP2A1H 
at 585 ms. (See text for discussion.)}
\label{mod2p_bernouilli_poynting}
\end{figure}

Figure \ref{mod2p_bernouilli_poynting} compares maps of the the gas 
pressures ($P_{\rm gas}$, left) with the magnetic pressures ($P_{\rm mag}$, right)
for models M15B11DP2A1H (top) and M15B11UP2A1H (bottom), 
at various times after their respective explosions commenced (see figure caption 
and Table \ref{tab_model}).  All four panels are on a large scale of 8000 km$\times$8000 km.
To facilitate comparison, the same colormaps are used for both $P_{\rm gas}$ and 
$P_{\rm mag}$.  The stronger equatorial explosion in model M15B11UP2A1H 
vis \`{a} vis model M15B11DP2A1H is obvious, even though model M15B11DP2A1H
has also started to expand laterally and at the times depicted 
the shock is approaching a radius of $\sim$1000 km in that 
direction. It is clear from the right panels that the magnetic stresses
are large only interior to the shock, even if it is elongated.  This
follows quite naturally from the facts that 1) winding is more significant
at small radii, 2) the shock flux-compresses accreted field, and 3) 
as it emerges the jet stretches the large toroidal component generated 
in the interior and, thereby, enhances the B-field along 
the axis at larger radii.  

Superposed on each left panel are vectors that represent the hydrodynamic flux 
at a given point.  This is the argument of the divergence term in the 
momentum equation (including the gravitational term), and we refer to it 
as the ``Bernoulli" flux.  It does not include the corresponding 
magnetic term.  This is one component in the jet power density and, quite
naturally, is pointing outwards.  The vectors on the right panels are the
Poynting fluxes, the MHD power densities, and they too point outwards and predominantly
along the axes.  Note, however, that at later times in model M15B11UP2A1H both the 
Poynting and Bernoulli fluxes have significant components in the equatorial direction. 
These vector maps detail the instantaneous driving terms of the jet phenomenon
in MHD-driven supernova explosions.

\begin{figure}
\includegraphics[height=.8\textheight]{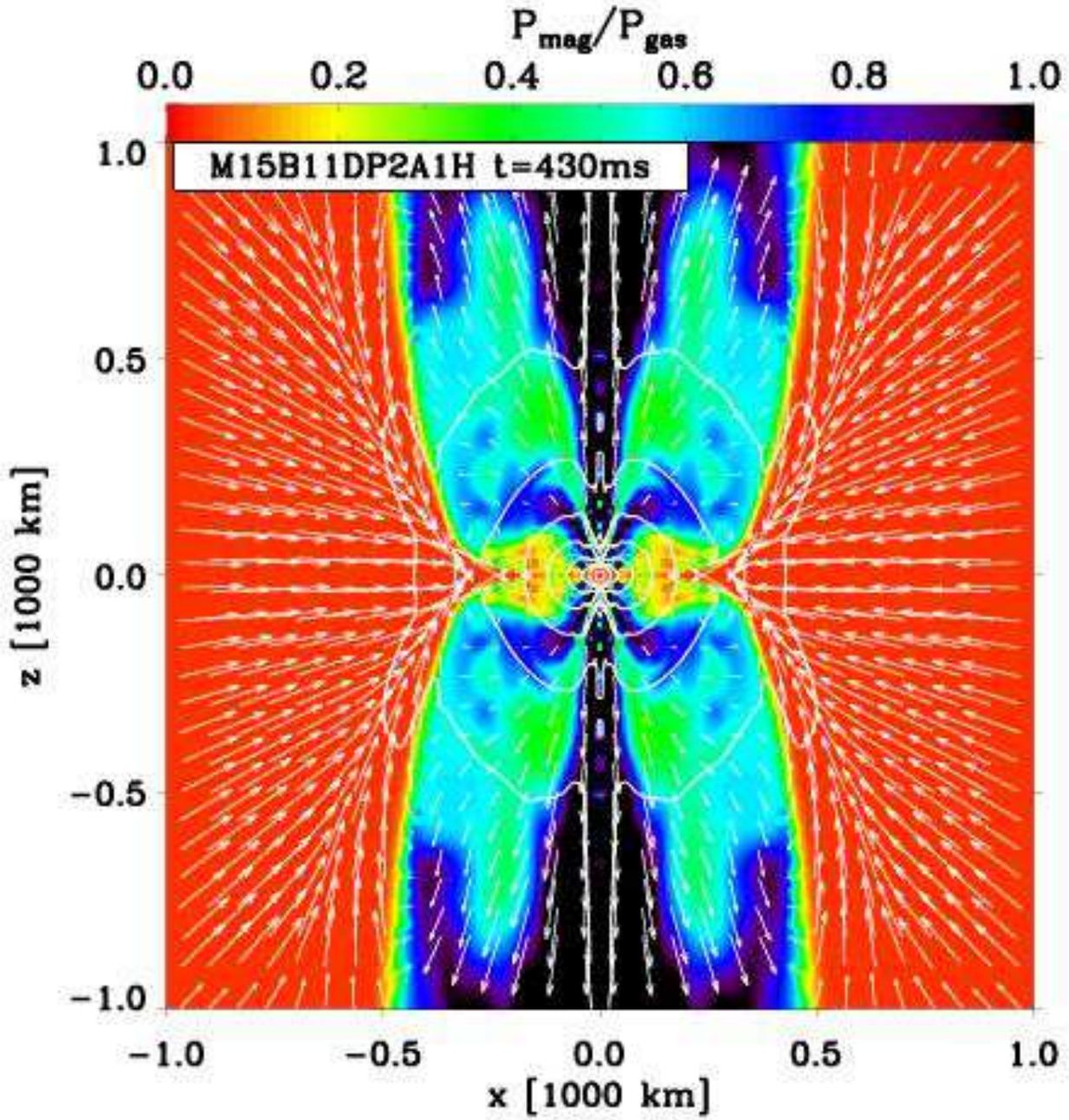}
\caption{
Colormap for model M15B11DP2A1H at 430\,ms after bounce of the
ratio of magnetic to gas pressure, overplotted with white isodensity contours
(every decade downward from 10$^{14}$\,g\,cm$^{-3}$) and velocity vectors
(length saturated to 15\% of the width of the figure and corresponding to a
velocity of 10000\kms).
}
\label{mod2p_pmopg}
\end{figure}

\begin{figure}
\includegraphics[height=.8\textheight]{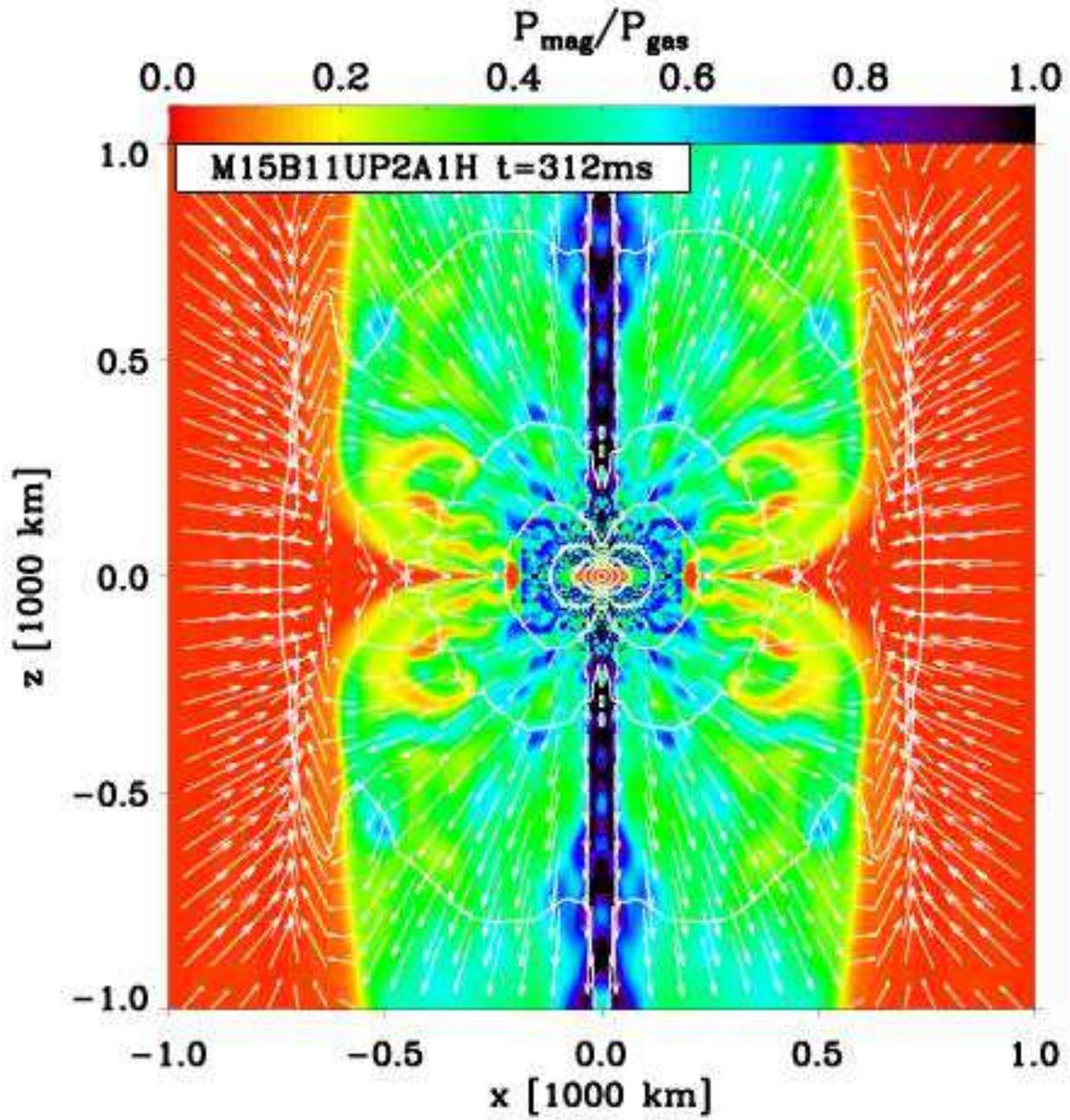}
\caption{Same as Fig.~\ref{mod2p_pmopg}, but for model M15B11UP2A1H at 312\,ms
after bounce.}
\label{mod2p_r04k_pmopg}
\end{figure}

Figures \ref{mod2p_pmopg} and \ref{mod2p_r04k_pmopg} show close-up views 
on a scale of 2000 km$\times$2000 km of the ratio of $P_{\rm mag}$ to 
$P_{\rm gas}$ for models M15B11DP2A1H and M15B11UP2A1H, respectively, at times 430 ms and
312 ms after bounce.  In these plots, the vectors are velocity vectors.  These
figures vividly communicate the fact that explosion along the poles and inflow/accretion
at the equator are simultaneous.  Note that the largest values of $\frac{P_{\rm mag}}{P_{\rm gas}}$
are found along the poles and are of order unity (as is also clearly demonstrated in 
Figs. \ref{fig_pmagopgas_sequence} and \ref{pmax.pgas.profile}).  Figure \ref{mod2p_r04k_pmopg}
also shows that along the poles the jet experiences oscillations as it emerges.
These are partially explained by magnetic sausage/neck instabilities.  The periods
of these instabilities are only a few milliseconds, and the wavelengths are only
$\sim$200 km, but we see them in all our simulations.  What consequences or signatures,
if any, these sausage instabilities may have remains to be seen.

\begin{figure}
\plotone{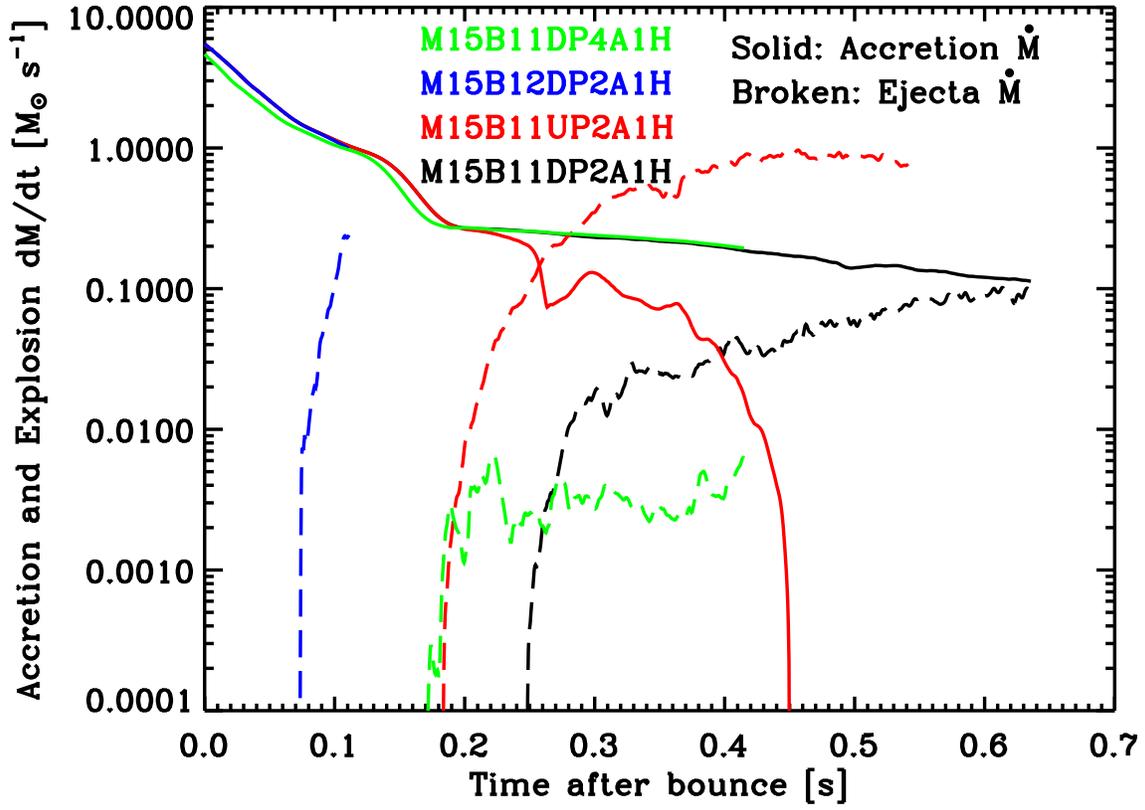}
\caption{Time evolution of the instantaneous integrated  mass flux
accreting (solid) or outflowing (broken) through a shell at a radius
of 500\,km, for four representative models. The mass flux for model
M15B0DP2A1H (not shown) is very similar to that for model 
M15B11DP2A1H.
}
\label{mdot}
\end{figure}

Given the fact that in the first many hundreds of milliseconds after bounce
equatorial accretion and polar explosion occur simultaneously, it is interesting
to compare the mass accretion rates with the mass ejection rates.  In Fig. \ref{mdot},
we provide both rates versus time after bounce for models M15B11DP2A1H, M15B11UP2A1H,
M15B12DP2A1H, and M15B11DP4A1H. The latter model is the same as model M15B11DP2A1H,
but with a slower $P_0$ of 4 seconds.  Comparing the broken and solid red curves
on Fig. \ref{mdot}, we see that the jets generated in model M15B11UP2A1H achieve
an $\dot{M}$ that exceeds the accretion $\dot{M}$ soon after the explosion commences.
In fact, the rapid turndown of the solid red curve indicates that the equatorial
explosion occurs rather early.  In contrast, the two black curves
depicting the corresponding behavior of model M15B11DP2A1H indicate that 
after $\sim$0.6 seconds the accretion and jet $\dot{M}$s are of 
comparable and modest magnitudes ($\sim$0.1 M$_{\odot}$ s$^{-1}$). 
We surmise that the actual behavior for a $P_0 = 2$ seconds may be roughly bracketed
by these two behaviors during the first second after bounce.  

On Fig. \ref{mdot}, we also show $\dot{M}$s for models M15B12DP2A1H (very explosive) and M15B11DP4A1H.
In fact, the simulation of model M15B12DP2A1H crashed early and we have not restarted it.
However, its rapid rise speaks for itself.  As our arguments in \S\ref{saturation} suggest,
we do not think that this behavior reflects what would actually happen for $P_0 = 2$ seconds,
but might reflect what would happen for $P_0 < 1$ seconds.  However, the anemic 
$\dot{M}$ seen in Fig. \ref{mdot} for model M15B11DP4A1H may indicate that $P_0$s of 4 seconds
or more would lead, at best, to weak MHD explosions. As we will show 
in \S\ref{energy} and Fig. \ref{e_expl}, though this model does explode 
via an MHD-driven jet, its explosion energy in the first second is less than a paltry  
$10^{49}$ ergs ($0.01$ Bethe).

\begin{figure}
  \includegraphics[height=.4\textheight]{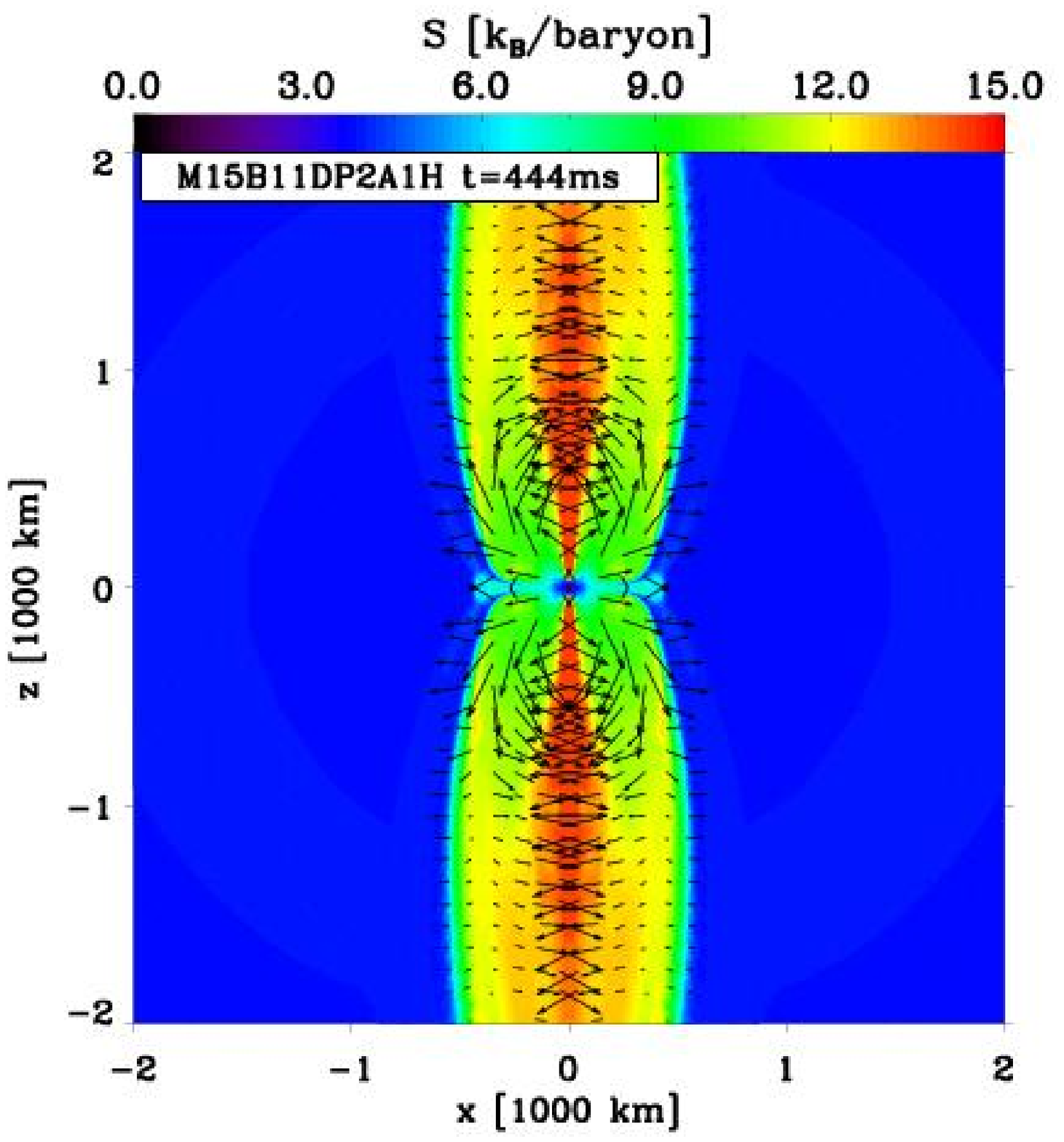}
  \includegraphics[height=.4\textheight]{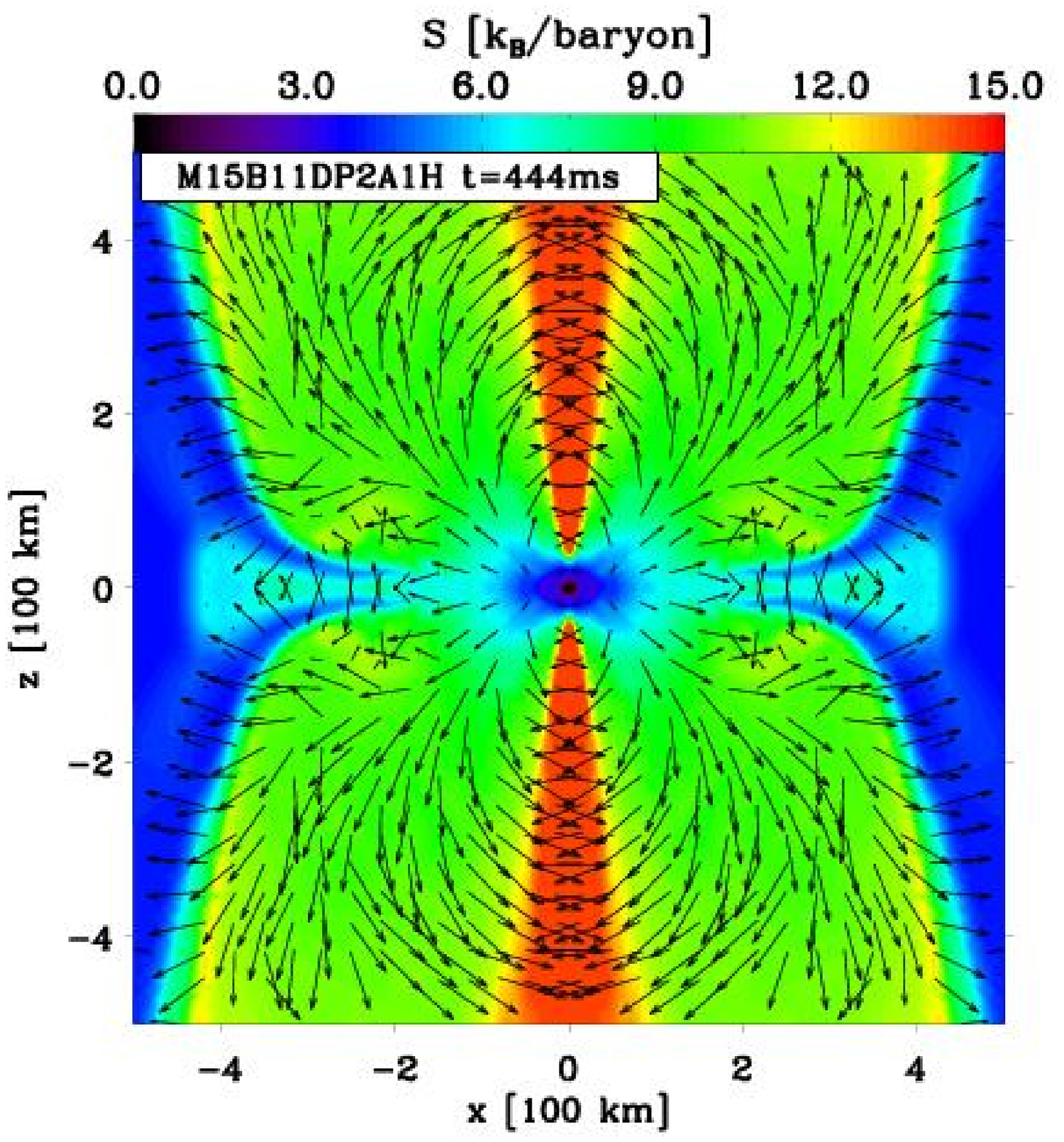}
\caption{
On the left-hand-side is a colormap of the entropy at 444\,ms after bounce for model M15B11DP2A1H on a
4000 km$\times$4000 km scale. We overplot the $(\nabla \times \vec B) \times \vec B$ 
vector field, with a length of 15\% of the width of the display corresponding to 
a saturation value of 10$^{18}$\,g\,cm$^{-2}$\,s$^{-2}$. This term enters the
momentum equation and thus represents the acceleration due to the magnetic field,
revealing here, in particular, the role of hoop stresses in confining the
jet as it moves to large distances. The right panel is the same as the 
left panel, but for the inner 1000 km$\times$1000 km region.
}
\label{mod2p_S_poynting_flux}
\end{figure}

\begin{figure}
  \includegraphics[height=.4\textheight]{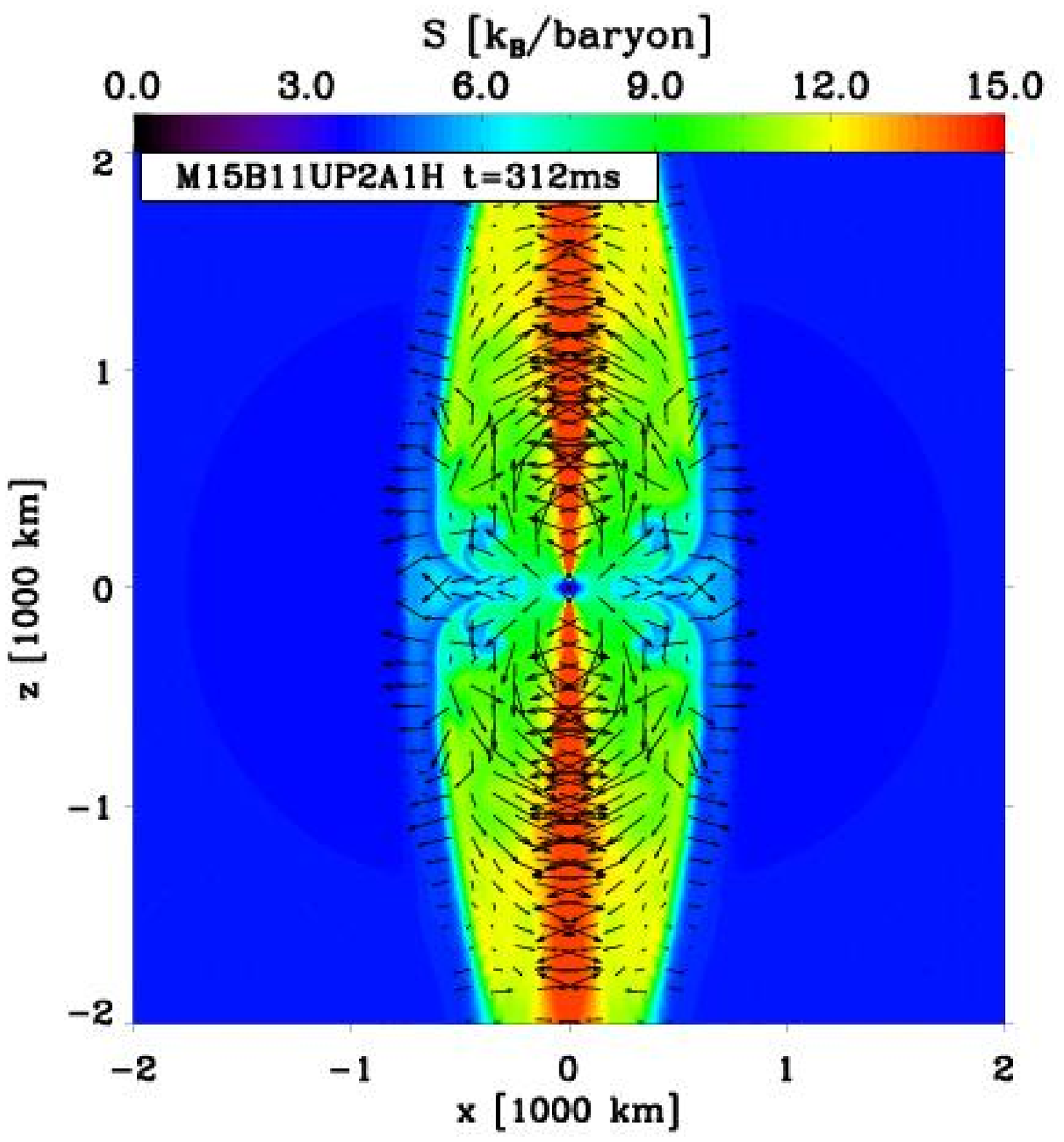}
  \includegraphics[height=.4\textheight]{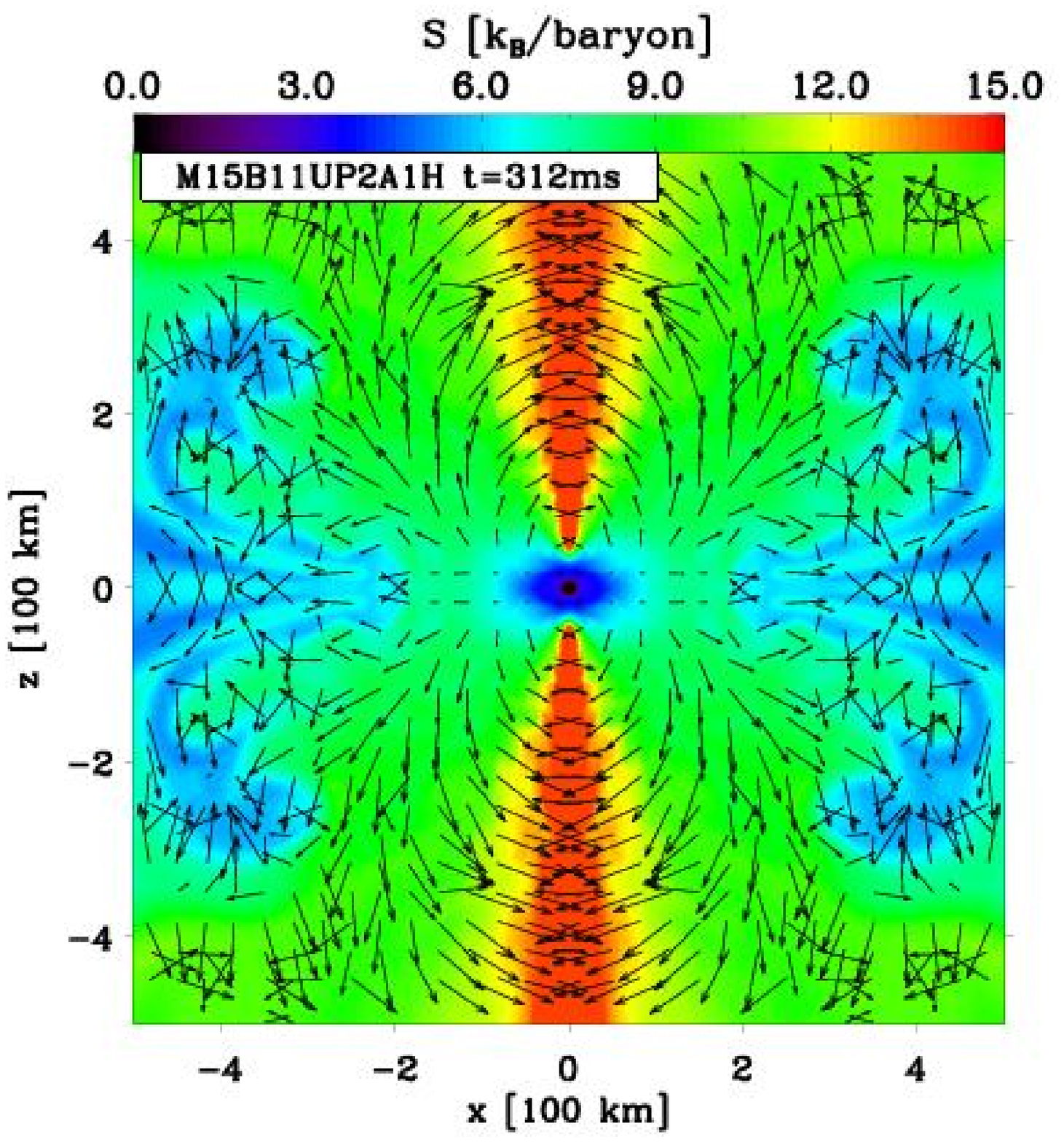}
\caption{
Same as for Fig.~\ref{mod2p_S_poynting_flux}, but for model M15B11UP2A1H, at
312\,ms after bounce. We use the same length-magnitude relationship for
the vectors.
}
\label{mod2p_r04k_S_poynting_flux}
\end{figure}

The magnetic stresses ($\propto$ $(\nabla \times \vec B) \times \vec B$) are the agencies
of jet propulsion and their distribution and direction explain the underlying
dynamics.  Figures \ref{mod2p_S_poynting_flux} and \ref{mod2p_r04k_S_poynting_flux} 
depict the entropy fields for models M15B11DP4A1H and M15B11UP2A1H, 
respectively, at post-bounce times of 444 ms and 312 ms. The right-hand
panels are finer-scale versions of the left-hand panels.
Superposed on Figs. \ref{mod2p_S_poynting_flux} and \ref{mod2p_r04k_S_poynting_flux}
are vector fields of the magnetic stress.  As the directions of the vectors
make clear, the vertical components of the magnetic stresses along the axis 
have positive signs.  This is what drives the jet.  In addition, the fact
that the vectors point inward towards the axes demonstrates that the jet
is partially confined in the shocked region by hoop stresses.  These hoop stresses 
are even more manifest, if indirectly so, in Fig. \ref{mod2p_12_field_lines}.

In Figs. \ref{mod2p_S_poynting_flux} and \ref{mod2p_r04k_S_poynting_flux}, 
the polar jet column is revealed to be the highest entropy region (red), but the peak
entropies achieved are only 15-20 units (per baryon per Boltzmann's constant).
This is much smaller that the corresponding values for the generic neutrino-driven
explosions ($\sim$20-40; Kitaura et al. 2006; Dessart et al. 2006a; Buras et al. 2006b) 
and acoustic explosions ($\sim$50-300; Burrows et al. 2006,2007).

\section{Angular Frequency and Period Evolution}
\label{period}

As we emphasize throughout this paper, rapid rotation is central to the 
phenomena we are calculating and describing.  Collapse transforms initial
rotation rates into spin rates at and after bounce that are $\sim$500-1000 times
faster (Ott et al. 2006).  Furthermore, all reasonable initial conditions result in 
protoneutron stars with differential rotation and shear (Akiyama et al. 
2003).  With a $P_0$ of 2 seconds, inner core periods 
of $\sim$2 ms and outer core periods of $\sim$10-30 ms arise.  Figure \ref{mod2p_omega}
provides the evolution of the profile of the angular frequency, $\Omega$, 
in the equatorial direction for the three models: M15B11DP2A1H (top)
M15B11UP2A1H (middle), and M15B11DP4A1H (bottom).  The colors and the 
colorbars indicate the time after bounce represented by each line.  We see 
from Fig. \ref{mod2p_omega} that for models M15B11DP2A1H and M15B11UP2A1H 
the inner $\sim$30 km rapidly achieves periods of $< 3$ ms, and that the PNSs
continue to spin up with time.  Continuing spin-up is due to continuing accretion, which compresses
the core, and to core deleptonization and cooling, which undermines pressure support for
the PNS mantle.  Note that nowhere in these models does matter achieve Keplerian
rotation rates (dashed lines in Fig. \ref{mod2p_omega}), though they can get close.

    As the bottom panel on Fig. \ref{mod2p_omega} shows, the early
    post-bounce evolution of the average period of the inner
    region of model M15B11DP4A1H is qualitatively similar to that of other
    models. However, the M15B11DP4A1H model explodes weakly, with little mass ejected,
    and is, therefore, not spun down due to the loss of rotational free energy
    to counteract the spin-up induced by the contraction of the protoneutron
    star. Interestingly, of all models, M15B11DP4A1H ends up with the
    shortest average rotation period in the core, despite being the slowest
    rotating model initially. We surmise that the reduced initial angular
    momentum leads to reduced centrifugal support, resulting in more efficient
    contraction and, therefore, enhanced spin-up. This convergence would
    explain the similar ``final" rotation period in models M15B11DP4A1H
    and M15B0DP2A1H. The weaker enhancement in magnetic pressure after
    bounce in model M15B11DP4A1H also leads to a more compact protoneutron
    star, and, thus, is less slowed-down by expansion than models M15B11DP2A1H
    and M15B11UP2A1H. Overall, accretion is not what distinguishes
    these models, since, as we see from Fig. \ref{mdot}, all models follow
    essentially the same path.

\begin{figure}
  \includegraphics[height=.3\textheight]{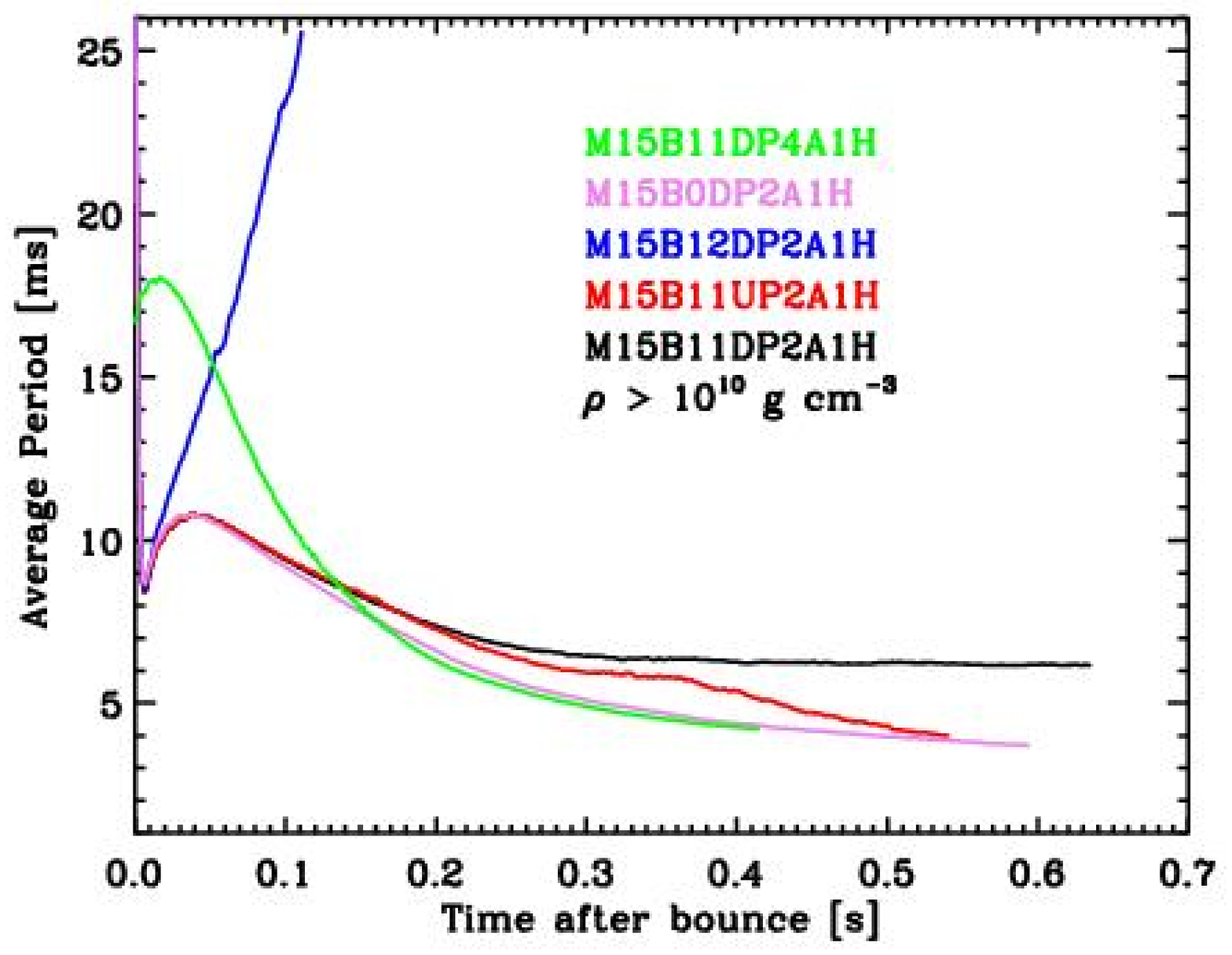}
  \includegraphics[height=.3\textheight]{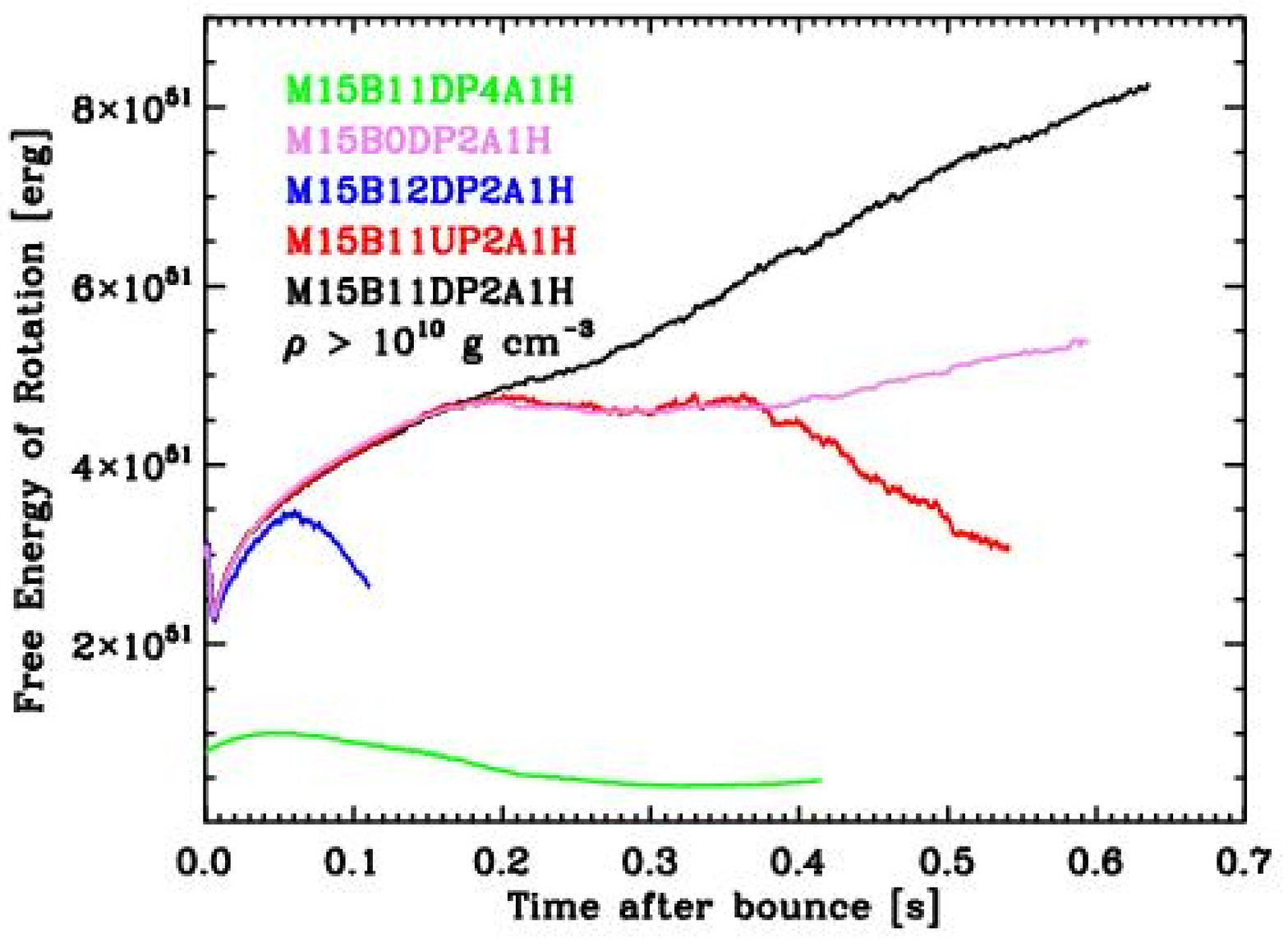}
\caption{{\it Left:} Time evolution for models of the average period, assuming
solid body rotation for the corresponding cumulative angular momentum,
for the region interior to the iso-density contour corresponding to
10$^{10}$\,g\,cm$^{-3}$. {\it Right:} Time evolution for models of the free
energy of rotation interior to the iso-density contour corresponding to
10$^{10}$\,g\,cm$^{-3}$. (See text for discussion.)
}
\label{av_period}
\end{figure}

The time histories of the corresponding mean periods in the central regions ($> 10^{10}$ g cm$^{-3}$)
of the PNS for five models, including model M15B0DP2A1H with no field, model M15B12DP2A1H
with the highest (unreasonable?) initial poloidal fields, and model M15B11DP4A1H
with a $P_0$ of 4 seconds, are portrayed in the left panel of Fig. \ref{av_period}.
We note the variety of behaviors.  The extreme model M15B12DP2A1H (dark blue) ``starts" at an
average period of $\sim$8 ms, but quickly spins down due to large magnetic torques to an 
average period of $\sim$25 ms within $\sim$100 ms.  This is not only dramatic, but indicates
the code's capability to follow the evolution of the entire core and its rotation.
The inner region is not excised and magnetic torques are given full rein to operate. 
Using the formula connecting torque ($\frac{dL}{dt}$) with magnetic jet power (Fig. 
\ref{power_comp}), $\dot{E} = \Omega\frac{dL}{dt}$, we can derive $\frac{dL}{dt}$ for 
this model at the end of the simulation (Table \ref{tab_model}):
$\frac{dL}{dt} \sim 4\times 10^{48}$ dyne$\cdot$cm.  Dividing by the PNS mass yields 
the ``specific" spindown torque, $\sim{1.5}\times 10^{15}$ [cm$^2/{\rm s}]$ s$^{-1}$. 

The evolution of the average period for baseline models M15B11DP2A1H (black) 
and M15B11UP2A1H (red) indicates that they spin up after bounce, but also that they evolve slightly
differently.  This is due in part to their different B-field (and, hence, torque density)
distributions, but is also due to their different mass fluxes (Fig. \ref{mdot}) and 
explosion histories, as well as the arbitrary density cut ($> 10^{10}$ g cm$^{-3}$) 
we apply.  Nevertheless, mean periods at the end of these simulations of 
$\sim$4$-$6 ms are instructive, even if they are smaller than the average inferred birth periods of
pulsars (see Ott et al. 2006 for a discussion).  Model M15B11DP4A1H with the higher initial 
$P_0$ bounces with a slower spin rate, but soon spins up due to 
inexorable accretion.  This model experiences a weak explosion at a later phase
and involves an $\dot{M}$ that is low (Fig. \ref{mdot}).  Curiously, compression and deleptonization
spin this model up to values that are comparable to those achieved by the models with 
faster initial spins.  

The right panel of Fig. \ref{av_period} depicts the 
corresponding values of the free energy in differential rotation versus time after bounce.
Here, the free energy is calculated by determining the kinetic energy of the PNS, and subtracting 
from it the kinetic energy for the same object if in solid-body rotation at the same total angular
momentum.  The density profiles in the two realizations are assumed to be the same,
so this procedure is only approximate.  Nevertheless, in this manner we obtain a useful measure of  
the energy available to the MHD jet.  As the figure indicates, the free energy available
to power the jets in all of the $P_0 = 2$-second
models exceeds $3\times 10^{51}$ ergs (3 Bethes).  Furthermore, models such as
M15B11DP2A1H that explode weakly evolve to increase the free energy available,
sometimes to large values that can approach the hypernova regime.  It 
would seem that at $P_0$ = 2 seconds, there is ample energy to power a supernova.

However, model M15B11DP4A1H has significantly less free energy
with which to maintain a respectable MHD jet explosion.  This energy is even less
than the 25\% one might have naively expected and is due to the fact that the feedback of
rotation on the dynamics of collapse and accretion results in a less differentially-rotating
PNS and in a slightly different density (mass) profile.  From this, we conclude 
that the $\Omega_0$-dependence of the MHD jet mechanism
may be a stronger function of $\Omega_0$ than simple scaling arguments would suggest;
the viability of the MHD jet mechanism  may be a stiff function of $\Omega_0$,
and may drop off quite fast with increasing $P_0$.  This possibility puts a premium
on precise (more precise) initial models and advanced simulation capabilities.

\section{Jet Powers and Explosion Energies}
\label{energy}

\begin{figure} 
\plotone{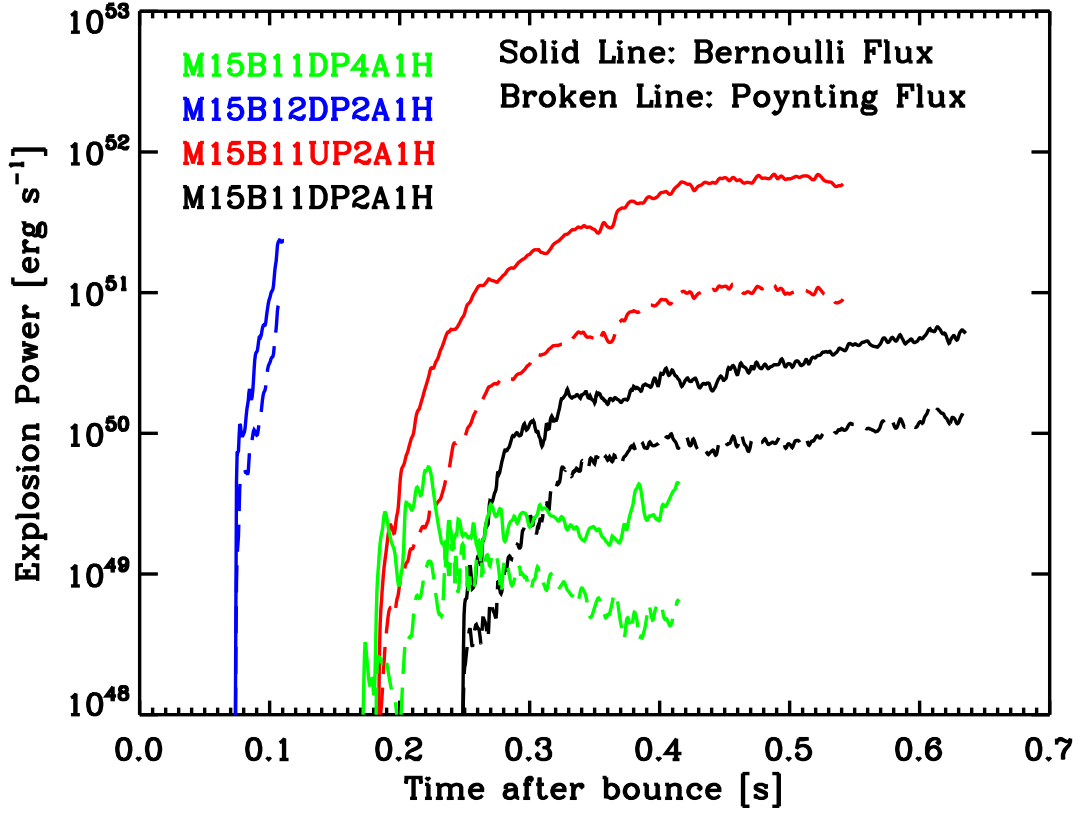}
\caption{Time evolution of the hydrodynamic/Bernoulli (solid
line) and Poynting (broken line) powers in the ejecta, integrated over a
radial shell at
500\,km, for four of the models simulated.
}
\label{power_comp}
\end{figure}

The most important information from our simulations is the energetics of
the explosions.  The jets carry kinetic, internal, gravitational, and magnetic
energies.  The total explosion power can be estimated by summing
the corresponding contributions for the ejecta.  We integrate the
Bernoulli and Poynting fluxes over the jet area at a 500-km radius ($\times$2 to account
for the bipolar jets) and plot in Fig. \ref{power_comp} the result versus 
time after bounce for four representative models.  In all cases,
at 500 km the hydrodynamic (Bernoulli) power is larger than the Poynting power,
even though in the inner regions the roles are reversed. 

Figure \ref{power_comp} shows that the extreme model M15B12DP2A1H 
achieves a total jet power (Bernoulli plus Poynting) in excess of 
$3\times 10^{51}$ ergs s$^{-1}$ within $\sim$25 ms of the onset of 
explosion, $\sim$75 ms after bounce.  It is clearly headed for much higher 
values above $10^{52}$ ergs s$^{-1}$ when the run crashes.
Model M15B11UP2A1H achieves a total jet power  
near $10^{52}$ ergs s$^{-1}$ within 0.5 seconds of bounce and evolves 
only slowly on secular timescales thereafter. Model M15B11DP2A1H boast lower
total powers, but still reaches values near $10^{51}$ ergs s$^{-1}$ by 0.6 seconds
after bounce.  On the basis of the arguments in \S\ref{saturation}, 
we believe that models M15B11UP2A1H and M15B11DP4A1H bracket
the true results for initial core spins of 2 seconds, so that quasi-steady-state jet 
powers for this $P_0$ are easily near (or greater than) $10^{51}$ ergs s$^{-1}$.  Such powers,
and the corresponding free energies plotted in Fig. \ref{av_period}, are 
well within the supernova regime. The slower rotator, M15B11DP4A1H, explodes, 
but musters total powers of only $\sim 0.05\times 10^{51}$ ergs s$^{-1}$.  

\begin{figure}
\plotone{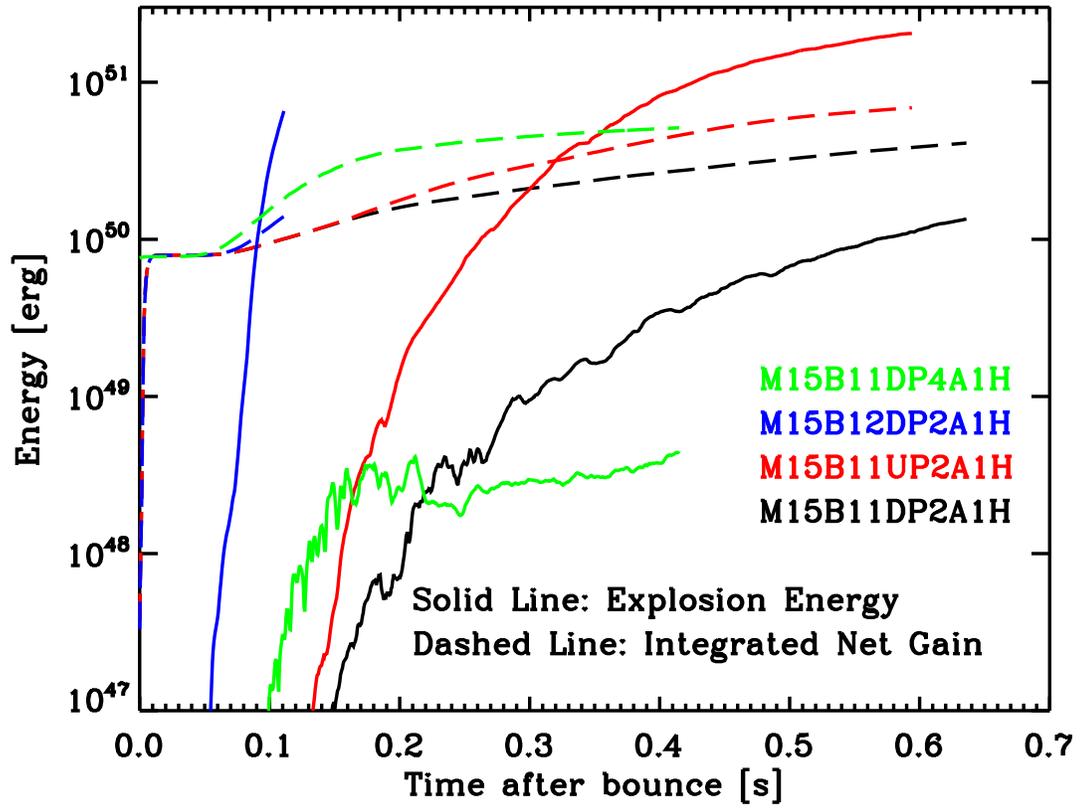}
\caption{Time evolution of the explosion energy (solid line)
and the net integrated neutrino gain (dashed) due to deposition outside the
high-density regions limited by the 10$^{10}$\,g\,cm$^{-3}$ contour, for four
models simulated in this study. (See text
for discussion.)
}
\label{e_expl}
\end{figure}

The rapid rise and near flattening of the explosion powers portrayed
in Fig. \ref{power_comp} lends credence to our scenario that an MHD jet engine,
energetically fed by continuing accretion and PNS settling, and lasting as long
as is necessary (perhaps many seconds) to produce a viable and vigorous explosion, 
is established in the context of rapid rotation. The total accumulated energy
versus time after bounce is plotted in Fig. \ref{e_expl}.  The hierarchy of behaviors
seen in Fig. \ref{power_comp} is replicated in this figure.  Model M15B11UP2A1H
takes only $\sim$0.45 seconds to achieve an explosion energy above $10^{51}$ ergs.
Model M15B11DP2A1H will take longer, but is headed in that direction.  As expected,
after $\sim$0.4 seconds model M15B11DP4A1H has achieved an explosion energy of only 
$\sim$0.005$\times 10^{51}$ ergs and is destined to remain anemic.

Figure \ref{e_expl} also portrays the integrated ``net gain" (Bethe \& Wilson 1985)
due to neutrino heating in the outer mantle. This number gives a sense of the 
contribution of neutrino heating to the explosive energy budget.  We see that
neutrinos can contribute some multiple of $10^{50}$ ergs, most of which in
model M15B11DP2A1H is used to lift the ejecta out of the deep 
gravitational well.  In these models, neutrinos do not,
without the MHD jet, lead to explosion in themselves.

\begin{figure} 
\plotone{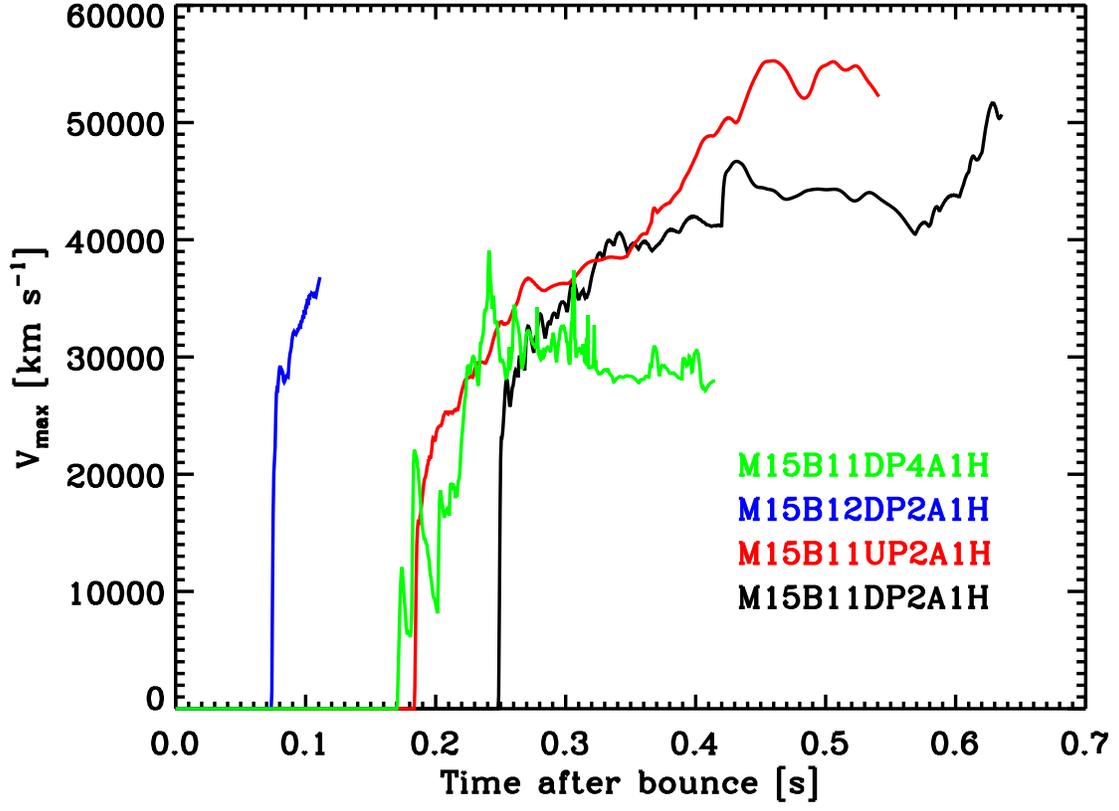}
\caption{Time evolution for four of the models we simulated in this study 
of the maximum (outward-oriented) velocity
along the polar axis and outside a radius of 500\,km.
}
\label{vmax}
\end{figure}

Finally, Fig. \ref{vmax} portrays the evolution of the maximum speed in the ejecta
as a function of time.  The maximum is $\sim$30000$-$55000 km s$^{-1}$. This is high, but
not near the speed of light.  If relativistic speeds are ever achieved
in the context of rapidly rotating core collapse, this must happen after the 
jet head achieves a distance of 5000 km and after a post-bounce time of one second,
perhaps long after.  In fact, from the analysis of our results viewed collectively,
we see no reason that relativistic speeds would obtain as long as a protoneutron star
remains at the core and does not collapse into a black hole through the requisite extended 
phase of mass accretion.

\section{Discussion and Conclusions}
\label{conclusion}
 
In this paper, we have presented the first 2D, rotating, multi-group, radiation
magnetohydrodynamics (RMHD) simulations of supernova core collapse, bounce, and explosion.
These calculations cover the longest stretch of physical time and incorporate 
the most physics ever attempted for MHD simulations in the supernova context.
Moreover, many of the features and results we find are new. Our focus
has been on the evolution, strength, and role of magnetic stresses and on the 
creation and propagation of MHD jets in the context of rapid rotation.
We find that a quasi-steady state can be quickly established during which 
a well-collimated MHD jet is maintained by continuous pumping of power from
the differentially rotating core.  If the initial spin period of the progenitor
core is $\sim$2 seconds, the free energy reservoir in the secularly evolving PNS is more 
than adequate to power a supernova explosion, and may be enough for a hypernova.    
We speculate that general relativity (GR), by making the core more compact and
thereby leading to more rapid post-bounce spin rates, will amplify the 
power and energy of the MHD jet beyond what we see in our Newtonian simulations. 
GR may also enable progenitors with initial spin periods 
slightly higher than $\sim$2 seconds to generate MHD jets with supernova potential. 
Given this, initial core periods up to $\sim$3 seconds may lead to strong
MHD-powered supernova.  Including the effects of small-scale dissipation of heat by the MRI 
(TQB) might relax this period bound somewhat, enabling progenitors 
with higher initial periods near 4$-$5 seconds to launch strong jets 
with supernova energies.  However, we infer from our suite of simulations that
the MHD jet phenomenon is a very stiff function of $\Omega_0$.  
Note that, due to the strong gravitational wave bursts that are
signatures of the bounce of rapidly rotating cores, the associated supernovae  
may be centrally relevant to ongoing efforts to detect 
gravitational waves (Ott et al. 2004,2007).

We find that the jet emerges first along a core around the rotation axis,
when the magnetic stresses behind the shock approach the local gas pressures.
It is the power density, not the total power, that is important in overcoming
the accretion ram and punching through the local accreta. The jet is well collimated 
by the infalling material and magnetic hoop stresses,
and maintains a small opening angle during our simulations.  Hence, 
it does not experience the Blanford-Payne centrifugal effect 
to any great degree.  Furthermore, we see evidence of sausage/pinch 
instabilities in the emerging jet stream, but otherwise the jet settles into
a very steady pattern.

Neutrino heating is sub-dominant in the rapidly rotating models we explored,
but can contribute 10$-$25\% to the final explosion energy.  Hence, we surmise
that for a range of rapid rotation rates a hybrid magneto-neutrino mechanism is 
possible.  Moreover, and intriguingly, even though as in models M15B11DP4A1H
and M15B10DP2A1H the MHD jet might be weak and the supernova explosion 
might have a different origin (neutrino, acoustic ?), our simulations 
suggest that the supernova explosion can be followed by a secondary MHD jet
event, itself a weak explosion.  Such a secondary explosion, a reasonable 
consequence of modest, but not rapid, rotation, may accompany most supernovae,
but due to the anticipated weakness when the rotation rates are low, to date this
feature may have gone unnoticed.  However, the jet/counter-jet structure
seen in Cas A (Hwang et al. 2004) may be such a secondary explosion.  

A theme that emerges from our simulations with rapid rotation is that
for all cases in which either the ``Millisecond magnetar model" 
(Usov 1992; Thompson,~Chang,~\&~Quataert 2004; Bucciantini et al. 2006; Metzger, Thompson, 
\& Quataert 2006) or the collapsar model (MacFadyan \& Woosley 1999) of GRBs or X-ray flashes 
may be relevant, since both these scenarios require rapid rotation, 
the first manifestation of such rotation will be the emergence of an MHD jet  
during the earlier PNS phase we have addressed in this paper.  
This implies, for instance, that the relativistic jets from the 
later collapsar/black-hole phase must be preceded by a precursor 
jet of some significance (Dessart et al., in preparation; 
see also Wheeler, Meier, \& Wilson 2002 and Wang \& Meszaros 2007).  This delay could be seconds and
is the time it takes the inner core to accrete to the critical,
GR-unstable configuration that dynamically gives birth to a black hole.  
Before the formation of the black hole, there is more than enough spin
angular momentum and free energy of differential rotation in the PNS 
to generate and power a magnetic tower explosion. Such a jet may or may not
be more energetic in aggregate than the GRB jet that follows it later, 
but it won't, if we are correct, contribute to the ultrarelativistic 
component of stellar death which may be the central feature of GRBs.

We have included in our model suite two simulations, M15B10DP2A1F 
(180$^{\circ}$) and M15B10DP2A1H (90$^{\circ}$), which allow us 
to gauge the effect of rapid rotation on the SASI. We find that 
rapid rotation tends to mute it; the SASI takes longer to develop
and achieves smaller amplitudes when it does.  The magnetic 
fields in these ``B10" runs were too small to effect the SASI 
more than rotation, given  $P_0 = 2$ seconds.  However, these
conclusions should be considered preliminary until 3D simulations
with good input physics and sophisticated radiative transfer  
are performed. 

Our simulations and estimates suggest that whatever the initial rotational and magnetic
conditions in the progenitor core, the MRI, dynamo action, and rotational
winding alone should completely alter the magnitude and multipolarity structure
of the pulsar or magnetar B-field that is formed in the supernova context.
Even if the spin rate is low, convection and the SASI will drag and 
twist the flux-frozen field. Mere flux conservation on collapse, what 
has been the default explanation for pulsar fields for decades, can't 
be the whole, or, perhaps, even a major part of the story.  There is too much 
dynamics during the PNS/supernova phase not to have left a profound 
mark on what remains.  

\begin{figure}
 \includegraphics[height=.30\textheight]{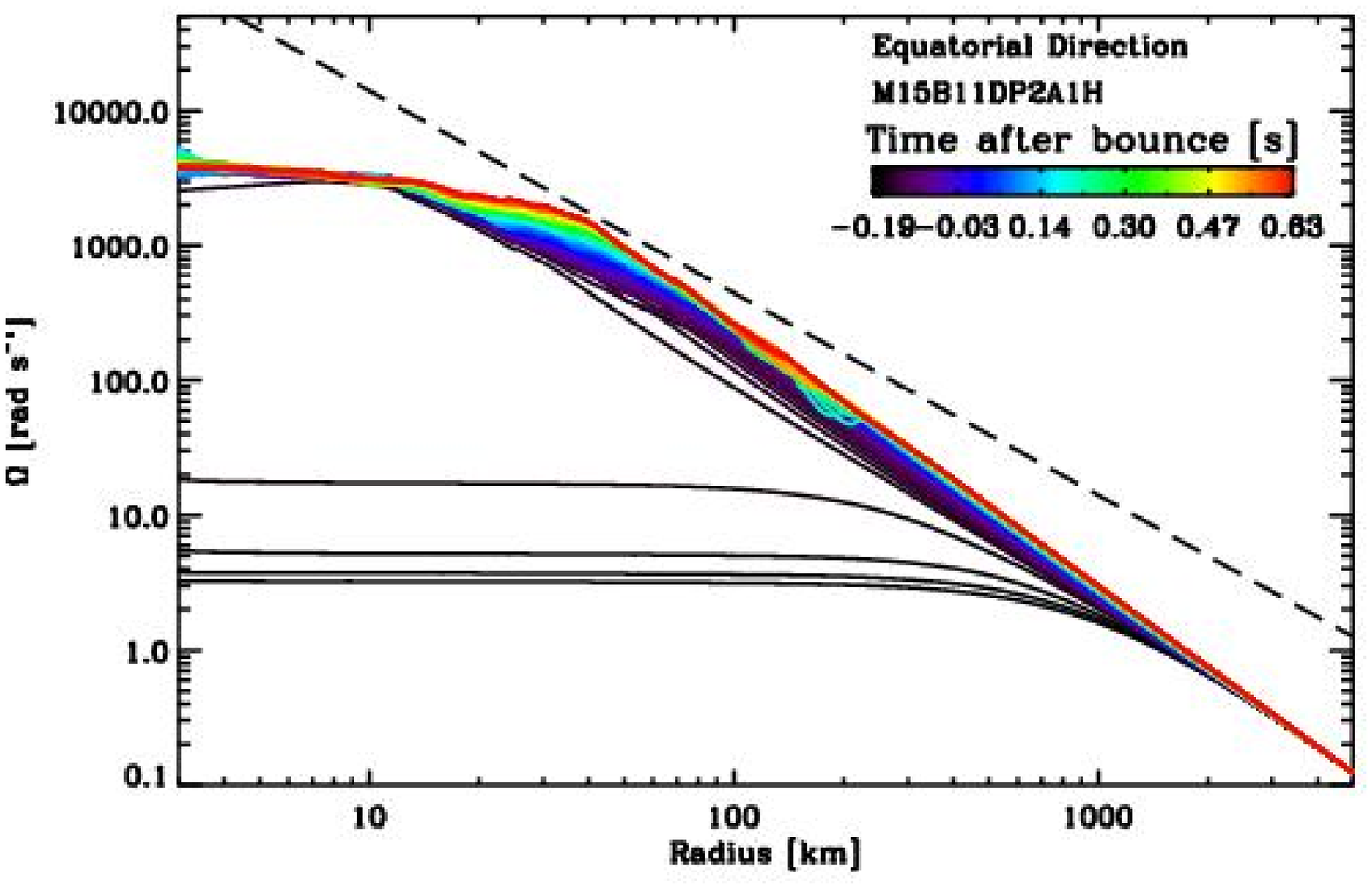}
 \includegraphics[height=.30\textheight]{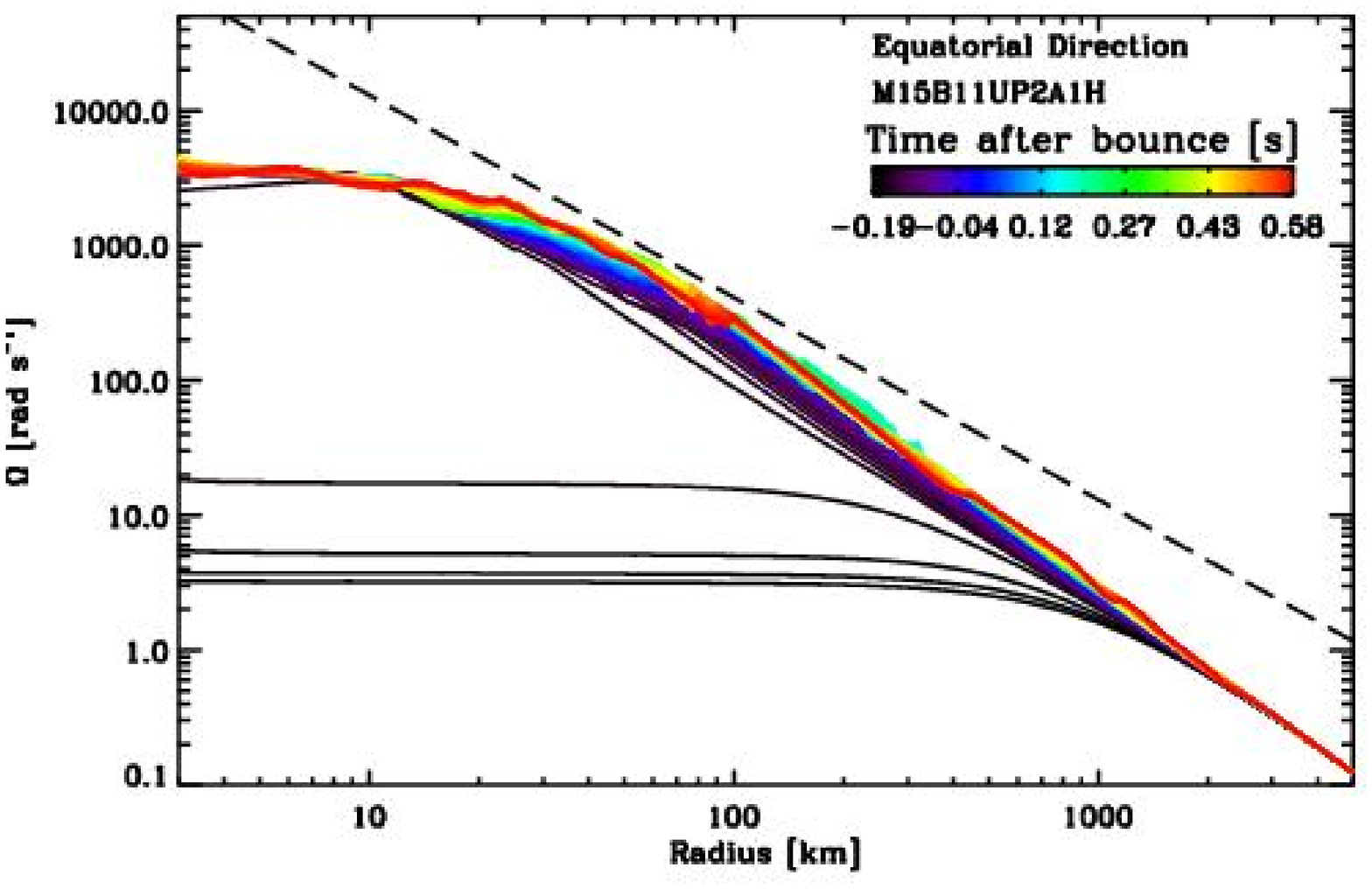}
 \includegraphics[height=.30\textheight]{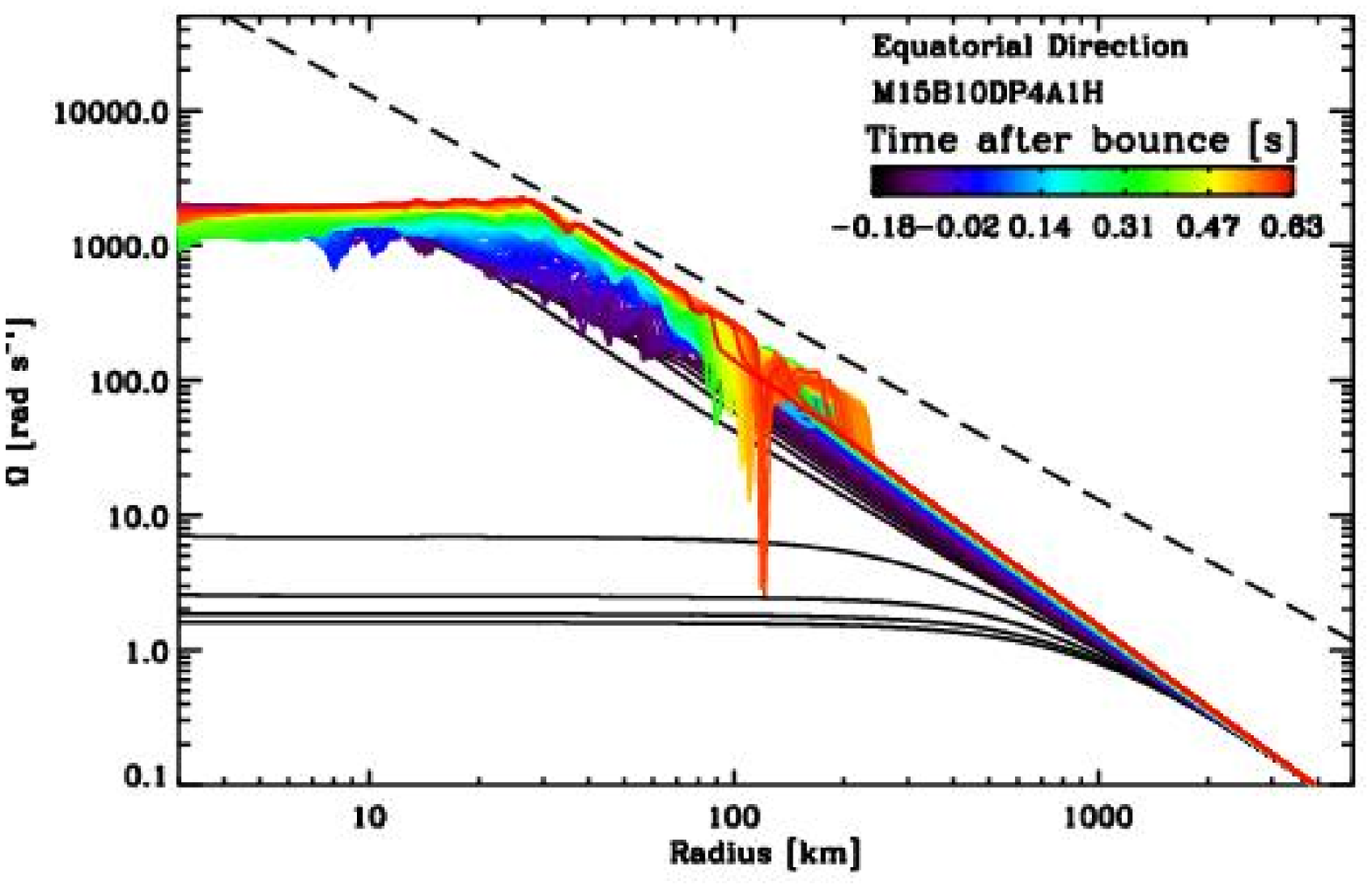}
\caption{The time evolution (see colorbar) of the equatorial angular velocity
for models M15B11DP2A1H (top left),  M15B11UP2A1H (top right), and M15B11DP4A1H (bottom). For comparison, we 
also overplot in each panel, as a dashed line, the local Keplerian velocity for the last
time.  Notice the solid-body rotation inside a radius of $\sim$20-30 \,km, with 
a monotonic spin-up at all radii as time progresses. For model M15B11DP4A1H, the 
angular velocity evolution reflects primarily the collapse of the progenitor
envelope, with lesser effects due to the weak explosion and modest mass loss.
See text for a discussion.
}
\label{mod2p_omega}
\end{figure}

That said, it may be that the initial inner core field is amplified by
flux freezing on collapse and is then afterwards preserved by the fact that this inner
core is not in strong differential motion (Fig. \ref{mod2p_omega}). The
outer region between $\sim$10 km and $\sim$150 km may then experience 
all the winding, twisting and MRI or dynamo amplification during a transient
supernova phase, after which, with the establishment of solid-body rotation
for the whole core, the field, in what is now the outer crust of the young
neutron star, assumes more modest values.  These may be lower than the
values frozen in the inner core, may be higher, or much higher (magnetar?).
Furthermore, the field in the outer crust will experience ohmic dissipation on timescales,
which might be short (days to years), determined in part by the temperatures there. 
These temperatures, and their profiles, are determined by the dynamics
during the collapse and supernova phases. So, one is prompted to ask: Can there 
be a transient phase of high B-field/rotation that is responsible 
for a supernova, that as the rotational shear subsides and the core spins down
transitions to canonical pulsar B-field and spin values?  Can the high B-fields decay
due to ohmic dissipation in the crust, after being generated during 
the hot cool-down phase?  What is the progenitor dependence of all this?
What determines whether a canonical pulsar or a magnetar is left?  Whatever the answers
to these questions, they are determined in no small measure during the 
dynamical supernova phase.  And that phase, if the progenitor core is
rapidly rotating, involves MHD jets and bipolar explosions. 

To summarize, the new features, phenomena, and possibilities we have identified 
in the context of rapidly rotating RMHD core collapse are:

\begin{enumerate}

\item That the presence of the stalled shock changes the evolution
to explosion dramatically.  The confinement it enforces delays explosion
and enables the B-field to grow to explosive conditions;  

\begin{itemize}
\item That the above also makes the accreting matter responsible 
for collimating the MHD jet, and that hoop stresses are sub-dominant 
in this regard.

\item And, moveover, that the Blandford-Payne centrifugal 
phenomenon is suppressed by the accreta, leading to a 
``magnetic tower" explosion that is little aided
by the centrifugal whip effect.
\end{itemize}

\item  The condition on the ratio of the magnetic to gas pressures
behind the shock along the rotation axis that leads to jet explosion.
When this ratio reaches $\sim$1, a polar jet can emerge;

\item Estimates of the range of initial spin periods that may be necessary
to ``ignite" explosions with energies of supernova or hypernova magnitudes;

\item  That accretion and the associated advection of rotational kinetic
energy into the core during the engine phase can maintain the MHD jet 
in a quasi-steady state for a very long time ($\sgreat$1 second), in 
fact for as long as accretion lasts;

\item In the case when MHD jets predominate, the fractional contribution
of neutrino heating to the total supernova/hypernova powers and energies;

\item That even though MHD-jet powered explosions always start along the poles,
they can later explode in an equatorial skirt as well, halting accretion;

\item That if the initial rotation rates are modest, but not
fast enough to dominate the explosion mechanism itself, secondary 
MHD jet explosions may still accompany supernovae powered 
by other means;

\item That rotation partially suppresses the SASI, and that its suppressive effect 
generally trumps that due to MHD stresses, at least in the first hundreds 
of milliseconds.  


\end{enumerate}

The numerical aspects of our simulations that distinguish them 
from those of the past and that allowed us to identify these new features
include:

\begin{itemize}

\item 2D, multi-group flux-limited neutrino radiation transport. No previous simulation,
except for that of LeBlanc \& Wilson (1970), included even a 
crude variant of gray transport;

\item The capacity to simulate for a long stretch of physical time 
($\sim$ 1 second).  This is in contrast to previous simulations,
which generally lasted only a few tens of milliseconds;

\item The use of a realistic nuclear equation of state (EOS).  Almost
all previous work employed a gamma-law EOS;

\item The use of realistic massive-star progenitor density, electron fraction, and 
temperature profiles, though the initial B-fields and rotation profiles 
we employed are to be considered only demonstrational.
Many previous simulations used polytropes or simplified initial cores;

\item The ability to follow collapse up to and after shock stagnation 
realistically by incorporating detailed microphysics, including electron capture,
and neutrino transport. Most previous MHD models that followed collapse 
witnessed a prompt explosion that did not allow the stalled shock and 
post-bounce accretion to play an important role in the build-up of the 
core fields and in the generation and geometry of the resulting MHD jet.

\end{itemize} 

The major limitations of our simulations are:

\begin{itemize}

\item That they are not 3D;

\item That they are not general-relativistic;

\item That the neutrino transport is not multi-angle.  Such a variant of VULCAN/2D
exists (Livne et al. 2004), but is rather slow;

\item That we don't resolve the MRI.  Higher spatial resolution simulations
should remedy this problem;

\item That we don't know the initial B-field and rotation states of the collapsing
progenitor cores.  This is a drawback of all previous work as well.

\end{itemize}

Theorists now have a menu of choices with which to explode a supernova.
Neutrino heating seems to work, if underenergetically, for both the 
lowest-mass massive stars (Kitaura et al. 2006; Buras et al. 2006b)
and accretion-induced collapse (Dessart et al. 2006a). If 3D effects 
are shown to be important, the neutrino mechanism may work in the generic case.  
If no other mechanism aborts it earlier, the acoustic mechanism (Burrows et al. 2006,2007)
seems to obtain after a relatively long delay. As we have presented here, 
rapid rotation leads to strong MHD jets and bipolar explosions.  
In principle, MHD jet explosions with hypernova-scale energies
can emerge from rapidly rotating protoneutron stars, 
without the need to create a rapidly rotating black hole.  Whether such explosions
from a PNS can have a significant relativistic component is as yet unknown.

How rare are rapid rotators?  Woosley \& Heger (2006) suggest that they 
might be more common among low-metallicity massive stars.  However rare or common they are,
it is extremely unlikely that they are absent from the family of 
exploding massive stars. The multi-D, multi-group RMHD simulations
we have describe here are a start along the path towards 
more realistic simulations of the possible role of magnetic fields 
in some of Nature's most dramatic events.



\acknowledgments

We thank Martin Pessah, Todd Thompson and Stan Woosley,
for fruitful discussions and
their insight. We acknowledge support for this work
from the Scientific Discovery through Advanced Computing
(SciDAC) program of the DOE, under grant numbers DE-FC02-01ER41184 
and DE-FC02-06ER41452, and from the NSF under grant number AST-0504947.
E.L. thanks the Israel Science Foundation
for support under grant \# 805/04, and C.D.O. and J.W.M. thank 
the Joint Institute for Nuclear Astrophysics (JINA) for support under 
NSF grant PHY0216783. This research used resources of the National
Energy Research Scientific Computing Center, which is supported by the
Office of Science of the U.S. Department of Energy under Contract No. DE-AC03-76SF00098.
We are happy to acknowledge the National Center for Computational Sciences
at Oak Ridge for an allocation of computer time on Jaguar.
We thank Don Fisher for his help generating both color stills and movies associated with this work
and Jeff Fookson and Neal Lauver of the Steward Computer Support Group
for their invaluable help with the local Beowulf cluster Grendel.
Movies of some of the simulations presented in this paper are 
available from the first author upon request.

\clearpage 

\begin{deluxetable}{lcccccccccccc}
\tablewidth{16cm}
\tabletypesize{\scriptsize}
\tablecaption{Properties of Models
\label{tab_model}}
\tablehead{
\colhead{Name}&
\colhead{Mass}&
\colhead{$B_{\rm poloidal}$}&
\colhead{Field}&
\colhead{$P_0$}&
\colhead{$A_0$}&
\colhead{$\Delta\theta$}&
\colhead{t$_{\rm explosion}$}&
\colhead{t$_{\rm end}$}&
\colhead{v$_{\rm max}$}&
\colhead{E$_{\rm explosion}$}&
\colhead{Power}&
\colhead{$<$P$>$}
\\
\colhead{}&
\colhead{\mo}&
\colhead{Gauss}&
\colhead{Geometry}&
\colhead{s}&
\colhead{km}& 
\colhead{}&
\colhead{ms}&
\colhead{ms}&
\colhead{km\,s$^{-1}$}&
\colhead{10$^{51}$\,erg}&
\colhead{10$^{51}$\,erg\,s$^{-1}$}&
\colhead{ms}
}
\startdata
M15B0DP2A1H     & 15   & 0          & N/A     & 2   & 1000 & 90$^\circ$   
& ... &  595  &  ...   & ...   & ...    &  3.70  \\
M15B10DP2A1H    & 15   & 10$^{10}$  & Dipole  & 2   & 1000 & 90$^\circ$   
& 550 &  944  &  37000 & 0.03  & 0.155  &  3.14  \\
M15B10DP2A1F    & 15   & 10$^{10}$  & Dipole  & 2   & 1000 & 180$^\circ$  
& 550 &  685  &  37000 & 0.03  & 0.118  &  3.18  \\
M15B11DP2A1H    & 15   & 10$^{11}$  & Dipole  & 2   & 1000 & 90$^\circ$   
& 250 &  636  &  50000 & 0.2   & 0.661  &  6.17  \\
M15B11UP2A1H    & 15   & 10$^{11}$  & Uniform & 2   & 1000 & 90$^\circ$   
& 180 &  585  &  55000 & 2.0   & 6.832  &  3.98  \\
M15B11DP4A1H    & 15   & 10$^{11}$  & Dipole  & 4   & 1000 & 90$^\circ$   
& 170 &  415  &  33000 & 0.005 & 0.050  &  4.21  \\
M15B12DP2A1H    & 15   & 10$^{12}$  & Dipole  & 2   & 1000 & 90$^\circ$   
&  80 &  111  &  36000 & 0.6   & 3.168  & 25.60  \\
\enddata
\tablecomments{
In this table, t$_{\rm explosion}$ is 
the time after bounce of the onset of explosion.
t$_{\rm end}$ is the time after bounce at which the simulation was terminated.
v$_{\rm max}$ is the maximum speed of the ejecta at any time during the simulation.
E$_{\rm explosion}$ is an approximation to the explosion energy of the 
ejecta at time t$_{\rm end}$ and in all cases is still accumulating.
``Power" is the engine power being pumped into the exploding ejecta
at time t$_{\rm end}$ and includes the total Bernoulli and Poynting contributions calculated at a radius of 500\,km.
All models use as the reference structure the 15$-$\mo model m15b6 of 
Heger, Woosley, \& Spruit (2005), but the magnetic field and angular velocity
distributions are set according to simple analytical descriptions. The
initial magnetic field is uniform along $z$ within $A_0$, and then switched to a dipole (D) outside or stays
uniform (U) out to the maximum grid radius. The initial toroidal magnetic field component is chosen to be 3 orders
of magnitude smaller than the polodial component, $B_{\rm poloidal}$. The angular velocity
is such that $\Omega(r) = \Omega_0 / (1 + (r/A_0)^2)$, where $\Omega_0 = 2\pi/P_0$, and $P_0$
is the initial period in the core. $r$ is the cylindrical radius.  For all models, the time of bounce is $\sim$190 ms. 
The average period, $<$P$>$, at the end of each simulation is computed using a
density cut of 10$^{10}$\,g\,cm$^{-3}$ and assuming solid-body rotation, given 
the total angular momentum of the matter above this cut. For example, with our definitions, model
M15B10DP2A1H is a 15\,\mo model with initial polodial field of 10$^{10}$\,G, a dipolar configuration outside of the
inner 1000\,km, with an initial period of 2\,s and a rotation law $\Omega(r)$ characterized as above
with the parameter $A_0=1000$\,km, with the simulation performed over one hemisphere only (``H",
in a 90$^{\circ}$ quadrant). (See text for discussion.)
}
\end{deluxetable}

\clearpage

\clearpage

\clearpage

\clearpage

\clearpage

\begin{figure}
\plottwo{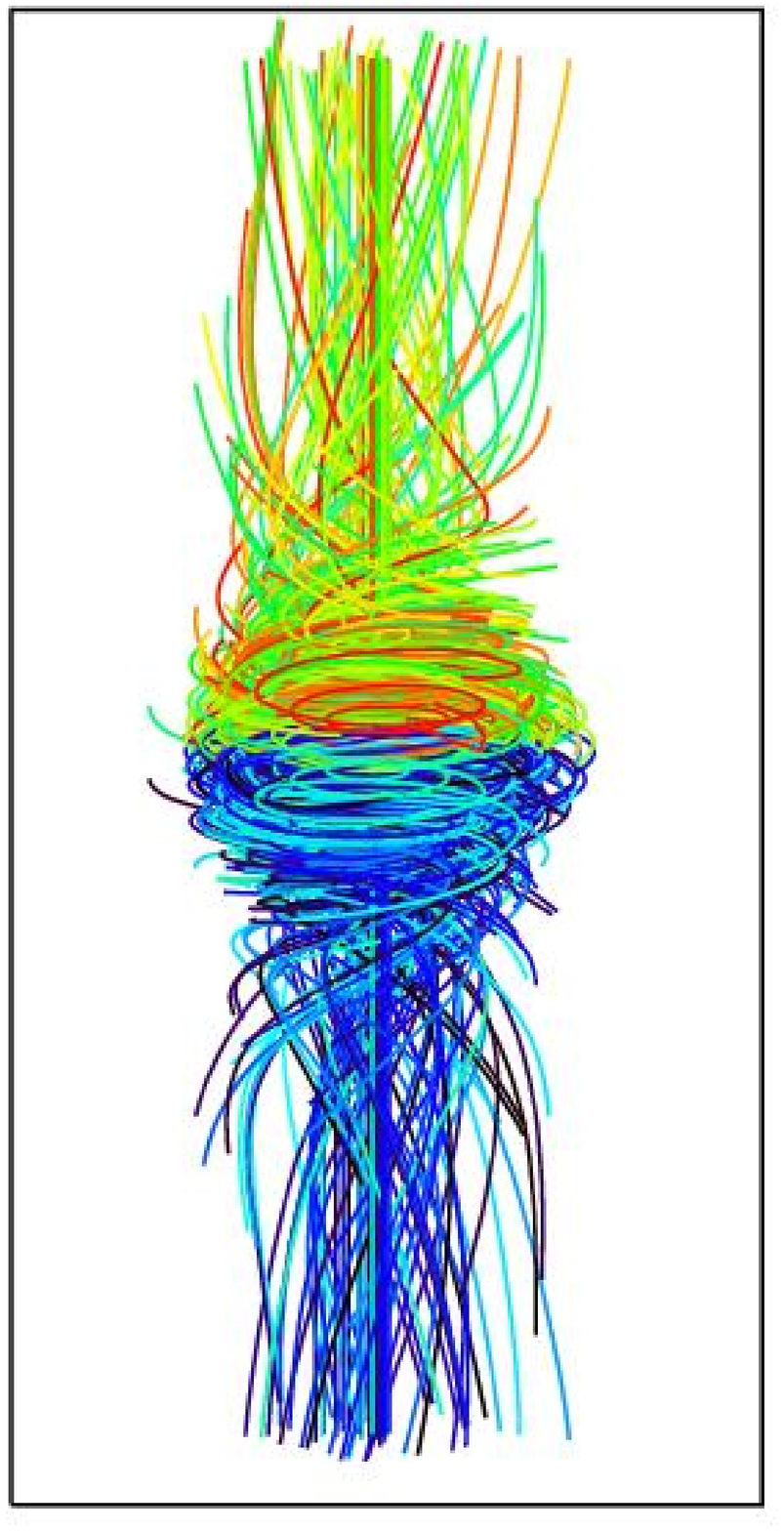}{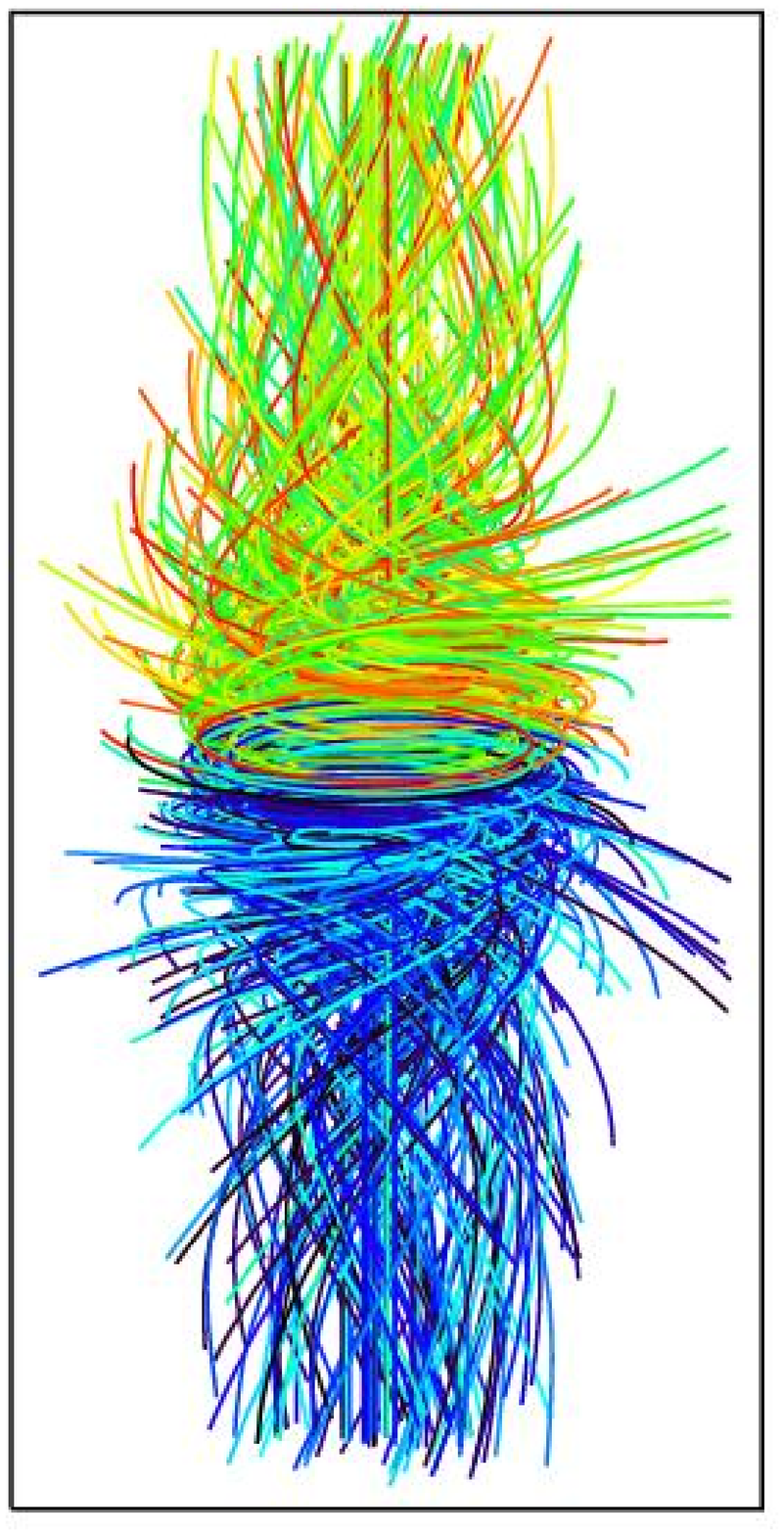}
\caption{
Streaklines of marker particles covering the 200\,ms prior to the end of
the simulation for each model M15B11DP2A1H (left) and  M15B11UP2A1H (right).
Markers are originally disributed uniformly
with radius, using 30 shells between 50 and 200\,km, as well as
with angle, using 30 angular beams between the rotation axis and the
equatorial direction. The marker azimuth is randomly chosen between 0$^{\circ}$
and 360$^\circ$. In its wanderings, if a marker moves closer than 50\,km
from the neutron-star center, we replace it with a new marker located within a
cone, along the rotation axis, with a 10$^\circ$ opening angle and randomly
distributed in radius between 50 and 500\,km.
Here, we show the streaklines for 300 of the 900 particles followed,
the displayed region extending out to 500\,km away from the rotation axis,
and 1000\,km from the equatorial plane.
The viewing angle is 30$^\circ$ with respect to the equatorial plane, with
a color coding randomly allocated, although using warm (cool) colors for the
northern (southern) hemisphere.
}
\label{marker_streaklines_mod2p}
\end{figure}

\clearpage

\clearpage

\clearpage

\clearpage

\clearpage

\clearpage

\clearpage

\clearpage

\clearpage

\clearpage

\clearpage

\clearpage

\clearpage

\end{document}